\documentclass[12pt]{article}
\usepackage[utf8]{inputenc}
\usepackage{graphicx,psfrag,epsf}
\usepackage{booktabs}
\usepackage{textgreek}
\usepackage{threeparttable}
\usepackage{float}
\usepackage{amsmath,amsfonts,amsthm,bm} 
\usepackage{url}
\usepackage{transparent}
\usepackage{adjustbox}
\usepackage{flafter}
\usepackage{lscape}
\usepackage{caption}
\usepackage{subcaption}
\usepackage{graphicx}
\usepackage{geometry}
\usepackage{footnote} 
\makesavenoteenv{tabular} 
\usepackage{verbatim}
\usepackage{natbib}
\usepackage{comment}
\usepackage{rotating}
\usepackage{hyperref}
\usepackage{mdframed}
\usepackage{lipsum}
\usepackage{array}
\newcolumntype{H}{>{\setbox0=\hbox\bgroup}c<{\egroup}@{}}
\usepackage{xcolor}
\usepackage{amsmath}
\usepackage{multirow}
\usepackage{bbm}
\usepackage{algorithmicx}
\usepackage{algorithm,algpseudocode}

\newcommand{\blind}{0}

\begin{document}

\def\spacingset#1{\renewcommand{\baselinestretch}%
{#1}\small\normalsize} \spacingset{1}

\def\spacingset#1{\renewcommand{\baselinestretch}%
{#1}\small\normalsize} \spacingset{1}


\if0\blind
{
  \title{\bf Decoupling Shrinkage and Selection for the Bayesian Quantile Regression\thanks{
    The authors thank Arnab Bhattacharjee, Atanas Christev, Gary Koop, James Mitchell, Mark Schaffer and all participants of the International Symposium on Forecasting (2021) for their invaluable feedback. The usual disclaimers apply.}}
  \author{David Kohns \footnote{Corresponding author: dek1@hw.ac.uk.}\hspace{.2cm}\\
    Department of Economics, Heriot-Watt University\\
    and \\
    Tibor Szendrei \\
    Department of Economics, Heriot-Watt University}
  \maketitle
} \fi

\if1\blind
{
  \bigskip
  \bigskip
  \bigskip
  \begin{center}
    {\LARGE\bf Title}
\end{center}
  \medskip
} \fi

%

\bigskip
\begin{abstract}
\noindent This paper extends the idea of decoupling shrinkage and sparsity for continuous priors to Bayesian Quantile Regression (BQR). The procedure follows two steps: In the first step, we shrink the quantile regression posterior through state of the art continuous priors and in the second step, we sparsify the posterior through an efficient variant of the adaptive lasso, the signal adaptive variable selection (SAVS) algorithm. We propose a new variant of the SAVS which automates the choice of penalisation through quantile specific loss-functions that are valid in high dimensions. We show in large scale simulations that our selection procedure decreases bias irrespective of the true underlying degree of sparsity in the data, compared to the un-sparsified regression posterior. We apply our two-step approach to a high dimensional growth-at-risk (GaR) exercise. The prediction accuracy of the un-sparsified posterior is retained while yielding interpretable quantile specific variable selection results. Our procedure can be used to communicate to policymakers which variables drive downside risk to the macro economy
\end{abstract}


\noindent%
\vfill

\spacingset{1.45} 


\section{Introduction}
While modern day economics, and broadly social science research, is often faced with high dimensional estimation problems in which the number of potential explanatory variables is large, often larger than the number of sample observations, the extant literature for high dimensional methods has focused developments mainly on for conditional mean models. Moving beyond the conditional mean, by estimating quantile regression on the other hand, allows to gauge potentially heterogeneous effects of variables directly across the conditional response distribution. While highly influential in the risk-management and finance literature in calculating risk measures such as VaR (i.e., the loss a portfolio's value incurs at a specific probability level), quantile regression has experienced a recent surge in popularity within the macroeconomic literature to quantify risks and vulnerabilities of output growth in response to summary measures of financial health, aptly named growth-at-risk (GaR) \citep{adrian2019vulnerable,figueres2020vulnerable,adams2020forecasting}. As an important distinction to literature that focuses on forecasting crisis periods directly such as through Markov-switching models \citep{hubrich2015financial,guerin2013markov} or probit models \citep{mccracken2021binary}, GaR instead gives information about the accumulation of risks facing an economy.

Since sources of risk can be numerous, high dimensional quantile problems are becoming ever more pertinent to policy makers and practitioners alike which has spurned methods that deal with variable selection and shrinkage for the quantile regression problem \citep{chernozhukov2010quantile,kohns2020horseshoe,hasenzagl2020financial}. Likewise, recent papers such as \citet{ferrara2020high} and \citet{carriero2020nowcasting} use high dimensional Bayesian quantile regression (BQR) methods to nowcast tail risks in real time. However, most of the previous research has focused on the use of continuous shrinkage priors whose posterior is not easily amenable to interpretation due to lack of exact sparsity. 

In this paper, we propose methods to simultaneously deal with high dimensional quantile estimation and variable selection from a Bayesian decision-theoretic perspective which allows to decouple shrinkage from variable selection in the spirit of \cite{hahn2015decoupling}. We propose easy to implement adaptive sparsification procedures of frontier Bayesian quantile regression priors such as the horseshoe prior \citep{kohns2020horseshoe} that adapt in a data driven manner to the given quantile. Evidence from simulations as well as a high dimensional GaR application to the US shows that our procedure either preserves fit of the unsparsified posterior or even improves both point- as well as - density fit, especially in the left tail which characterises downside risks. Contributing further to investigating sparsity in economic data sets, we find that there is considerable quantile specific sparsity across the conditional distribution which is lost when only focusing on the conditional mean. 


\textbf{The Literature}. The canonical way in which to do Bayesian variable selection for normal likelihood models is through Bayesian model averaging priors \citep{raftery1997bayesian,clyde2004model} in which the model space is discretised and posterior model probabilities allow for posterior probability statements about which is the most likely model, or to provide weights for model averaging. See \citet{alhamzawi2013conjugate} for an adaptation of the related g-prior approach to quantile regression. Due to combinatorial bottlenecks, much of Bayesian research in variable selection in the past decades have proposed a variety of methods (see \citet{polson2010shrink} and \citet{ishwaran2005spike} for excellent reviews), of which the search stochastic variable selection (SSVS) prior, dating back to \citet{george1993variable} and \citet{george1997approaches}, and adapted to the quantile regression by \citet{korobilis2017quantile}, has been notably influential. The SSVS prior is a discrete mixture of normals prior which stays computationally efficient as it stochastically searches among models with highest posterior probability. Although it only approximates the model space, it offers good empirical performance since low probability models are often ignored in the Markov chain.

Instead of approximating the discretised model space, continuous shrinkage priors on the other hand seek to include all variables at all times and instead shrink noise variables' coefficients toward zero so as to minimise their predictive influence. The class of global-local shrinkage priors \citep{polson2010shrink} which include the lasso \citep{park2008bayesian}, horseshoe \citep{carvalho2010horseshoe} and Dirichlet-Laplace (DL) prior \citep{bhattacharya2015dirichlet} increasingly supplant discrete mixture priors due to their excellent theoretical as well as empirical performance \citep{bhadra2019lasso}. Adapting such priors to quantile regression has been alreadyt attempted in the literature: See \citet{li2010bayesian} for lasso type priors and \citet{kohns2020horseshoe} for global-local prior adaptations to the BQR.  The continuous nature of global-local priors allows additionally for fast and efficient computational methods \citep{bhattacharya2016fast,kohns2020horseshoe}. 

Instead of conducting estimation and variable selection in 1 step, \cite{hahn2015decoupling} propose a 2 step procedure in which the posterior is sparsified in an inferentially coherent way through integration over loss-functions which add model size penalties. In particular, they use adaptive lasso type penalisation, akin to \cite{zou2006adaptive}. The idea is to threshold the posterior mean of the coefficients to zero, when they have a negligible effect in terms of squared-error loss to the full model predictions. Piironen et al. (2020) extend this approach to Kullback-Leibler and \cite{kowal2021fast} to more general loss-functions. These approaches have been applied in numerous macroeconomic forecasting articles (see \citet{huber2019inducing,kohns2020developments} among others) which show good forecasting and variable selection properties. 

In section \ref{sec: 2}, we firstly provide a review of the BQR and shrinkage priors considered, followed by a derivation of our proposed decision theoretically motivated sparsification, and details on its implementation. In section \ref{sec: 3}, we conduct a large scale Monte Carlo experiment that tests the proposed methodology's robustness to a variety of high dimensional settings. This is followed in section \ref{sec: 4} by a high dimensional GaR application and lastly we conclude.

\section{Methodology} \label{sec: 2}
\subsection{Bayesian Quantile Regression}
Assuming a linear model such as
\begin{equation}
    y_t = x_t'\beta + \epsilon_t, \; \; t = 1,2,\cdots,T, \label{eq:linmod}
\end{equation}
 where $\{y_t\}_{t=1}^T$ is a scalar response variable and $\{x_t\}_{t=1}^T$ a $K \times 1$ known covariate vector, the objective function of quantile regression can be expressed as the minimised sum of weighted residuals which are zero in expectation for the given quantile $p \in (0,1)$: 

\begin{equation}
\hat{\beta}_p =   \underset{\beta}{min}\sum^n_{t=1}\rho_p(y_t-x_t'\beta), \label{eq:obj}
\end{equation}

whose solution $\hat{\beta}_p$, is a $K \times 1$ quantile specific coefficient vector. Note that the expected quantile $\hat{Q}_p(Y|X) = X\hat{\beta}(p)$ is a consistent estimator of $Q_p(Y|X)$, independent of any parametric assumption about residuals $\{\epsilon\}_{t=1}^T$ \citep{koenker2005}. We will maintain the assumption throughout that paper that the design X is known. The loss function $\rho_{p}(.)$ is often expressed as a tick loss function of the form $\rho_{p}(y)=[p-I(y<0)]y$ where $I(.)$ is an indicator function taking on a value of 0 or 1 depending on whether the condition is satisfied. As noted by \cite{koenker2017handbook}, this loss function is proportional to the negative log  density of the asymmetric laplace distribution. This connection has been used to recast quantile regression as a maximum likelihood solution of model (\ref{eq:linmod}) with an Asymmetric-Laplace distribution, denoted as $\mathcal{ALD}(p,0,\sigma)$, where $\sigma$ denotes the scale of the $\mathcal{ALD}$. Assuming an $\mathcal{ALD}$ error distribution, the working likelihood $f(Y|X,\beta_p,\sigma)$ becomes: 

\begin{equation} \label{likelihood}
    f(Y|\beta,\sigma)=\frac{p^T(1-p)^T}{\sigma^{T}}\prod_{t=1}^T\Big[ e^{-{\rho_p(y_t-x_t'\beta_p)}/{\sigma}} \Big]. 
\end{equation}


As posterior moments with conventional priors are not analytically available with an $\mathcal{ALD}$ working likelihood, it has become standard practice in the literature to use a mixture representation, proposed by \citet{kozumi2011gibbs}, in which the $\mathcal{ALD}$ error process can be recovered as a mixture between an exponentially distributed variable $z_t$, $z_t \sim exp(\sigma)$, and a standard normal variable, $u_t$, $u_t \sim N(0,1)$: 

\begin{equation}
\begin{split}
    \epsilon_t &= \xi z_t + \tau\sqrt{\sigma z_t}u_t \\
    \xi &= \frac{1-2p}{p(1-p)} \\
    \tau^2 &= \frac{2}{p(1-p)}
\end{split}
\end{equation}
    
\noindent where $\xi$ and $\tau$ are deterministic quantile specific parameters. The conditional likelihood stacked over all observations thus becomes: 

\begin{equation} \label{worklik}
    f(Y |X,\beta_p,Z,\sigma) \propto det(\Sigma)^{-\frac{1}{2}} exp \Big\{-\frac{1}{2}[(y-X\beta_p-\xi Z)'\Sigma(y-X\beta_p-\xi Z)] \Big\},
\end{equation}

\noindent where $Y = (y_1,\cdots,y_T)'$, $X = (x_1,\cdots,x_T)'$, $Z = (z_1,\cdots,z_T)'$ and $\Sigma = diag(1/\tau^2z_1\sigma,\cdots,1/\tau^2z_T\sigma)$. Hence, the mixture representation results in a normal kernel for the likelihood which enables analytical solutions for conditional posteriors as shown below.

Throughout the paper, we consider priors on $\beta_p$ that take the following form: 

\begin{equation}
    \beta_p \sim N(\boldsymbol{0}_K,\Lambda_*),
\end{equation} \label{eq:gen_priro}

where a prior mean of zero implies shrinkage toward sparsity and the prior variance parameters, $\Lambda^*$ control the amount of shrinkage toward zero.

By applying independent priors $p(\beta_p,\sigma,Z)=p(\beta_p)p(\sigma)p(Z)$, the conditional posterior for $\beta_p$ is normal:

\begin{equation}\label{eq:post_beta}
\begin{split}
    p(\beta_p|\cdot) &\sim N(\overline{\beta}_p,\overline{\Lambda}_*) \\
    \overline{\beta}_p &= \overline{\Lambda}_*(X'\Sigma(Y-\xi Z)) \\
    \overline{\Lambda}_* &= (X'\Sigma X + \Lambda_*^{-1})^{-1}.
\end{split}
\end{equation} 


\noindent The conditional posterior of the scale parameter is:

\begin{equation}\label{eq:post_sigma} 
\begin{split}
    p(\sigma|\cdot) &\sim IG(\overline{a},\overline{b}) \\
    \overline{a} &= \underline{a} + \frac{3T}{2} \\
    \overline{b} &= \underline{b} + \sum_{t=1}^T \frac{(y_t - x_t'\beta_p-\xi z_t)2}{2z_t+\tau^2} + \sum_{t=1}^T z_t,
\end{split}
\end{equation} 

\noindent where 
IG stands for the inverse-Gamma distribution. Finally, the conditional posterior for $z_t$ is:

\begin{equation}\label{eq:post_zt}
\begin{split}
    p(z_t|\cdot) &\sim 1/iG(\overline{c}_t,\overline{d}_t) \\
    \overline{c}_t &= \frac{\sqrt{\xi^2 + 2\tau^2}}{|y_t-x_t'\beta_p|} \\
    \overline{d}_t &= \frac{\xi^2 + 2\tau^2}{\sigma\tau^2},
\end{split}
\end{equation}

\noindent where iG stands for the inverse Gaussian density with location ($\overline{c}_t$) and rate ($\overline{d}_t$) parameter.

The conditionally conjugate posteriors (\ref{eq:post_beta},\ref{eq:post_sigma},\ref{eq:post_zt}) allow for efficient Gibbs sampling algorithms which for the independent prior setup have been shown to be geometrically ergodic by \citet{khare2012geometric}, independent of any assumptions on X. Hence, X could include more variables than observations, dependent or deterministic variables.

Since taking the inverse of the posterior covariance of the regression coefficients $\overline{\Lambda}_*$ can be computationally demanding in high dimensions, we make use of the fast BQR sampler proposed by \citet{kohns2020horseshoe} that reduces the computational complexity involved in obtaining a draw from $\overline{\beta}_p$ from $\mathcal{O}(K^3)$ to $\mathcal{O}(T^2K)$. This algorithm is particularly suitable for macroeconomic data sets in which K is typically much larger than T. Details of the exact sampling steps for all priors considered are given in the appendix. 

\subsection{Shrinkage Priors} \label{sec: 3}
Shrinkage priors, both for conditional mean and quantile models, can be understood as a Bayesian generalisation to frequentist penalised regression, where the penalisation is a function of the number of active coefficients:

\begin{equation} \label{eq:gen_shrink}
    \beta = \underset{\title{\tilde{\beta}}}{\mathrm{argmin}} \; \underset{t}{\sum} h(y_t,x_t,\tilde{\beta}) + \delta Q(\tilde{\beta}). 
\end{equation}

\noindent $h$ and $Q$ are two positive functions, and $\delta$ controls the amount of penalisation. Choosing $h(\bullet)$ to be the negative log-likelihood of the data and $Q(\tilde{\beta})$ to be the $\ell_1$-norm, $|| \tilde{\beta}||_1$, (\ref{eq:gen_shrink}) yields sparse optimal solutions for $\beta$ which are maximum likelihood equivalents the popular lasso estimator of \citet{tibshirani1996regression} when the likelihood is Gaussian. Likewise, using $\ell_1$-norm penalisation and setting $h(\bullet)$ to the negative log of the $\mathcal{ALD}$ likelihood,  (\ref{eq:gen_shrink}) recovers the maximum likelihood equivalent of the quantile lasso of \citet{chernozhukov2010quantile}. From a Bayesian perspective, $Q(\beta)$ instead can be understood as the negative log prior distribution imposed on $\beta$, which renders (\ref{eq:gen_shrink}) the negative log-posterior. The Bayesian paradigm has the added advantage of being able to set a prior also for the amount of penalisation $p(\delta)$ which gives a probabilistic way to conduct shrinkage. Different prior forms of $p(\delta)$ will result in different shrinkage properties. \\

\textbf{Lasso Prior}. Fistly introduced by \citet{park2008bayesian}, the lasso prior generalises the frequentist lasso by placing a mixture exponential prior on $\delta$. Adapted to the BQR by \citet{li2010bayesian}, the prior takes the following form:

\begin{equation} \label{eq:b_lasso}
    \begin{split}
        \beta_j|\phi & \sim N(0,\lambda_j), \\
        \lambda_j | \phi & \sim exp(\frac{\phi}{2}) \\
        \phi & \sim G(a_1,b_1)
    \end{split}
\end{equation}

\noindent where $\Lambda_* = diag(\lambda_1,\cdots,\lambda_K)$. The conditional posteriors for the hyperparameters are standard:

\begin{equation} \label{eq:post_lasso}
    \begin{split}
        p(\lambda^{-1}_j|\bullet) & \sim iG(\sqrt{\frac{\phi}{\beta_{j,p}^2}},\phi) \\
        p(\phi|\bullet) & \sim G(K+a_1, \frac{1}{2}\sum_{j=1}^K\lambda_j + b_1).
    \end{split}
\end{equation}

\noindent If instead $\phi$ is allowed to vary with the same distribution as in (\ref{eq:b_lasso}) for each j, this prior will become the adaptive lasso prior of \citet{alhamzawi2012bayesian}\footnote{Since in simulations and the application we found there to be no significant difference in terms of performance between the Bayesian lasso and adaptive lasso BQR, we report only resuls for the Bayesian lasso. See \citet{kohns2020horseshoe} for a thorough investigation of the differences in performance between these priors.}.\\

\textbf{Horseshoe Prior}. Similar to the standard result as to why the $\ell_1$-norm penalty in a frequentist approach to solving (\ref{eq:gen_shrink}) tends to over shrink signals in lasso regression \citep{zou2006adaptive}, the Bayesian lasso over shrinks large coefficients as well\footnote{\citet{carvalho2010horseshoe} and \citet{piironen2017sparsity} formalise this argument through implied shrinkage distributions for the normal means and normal regression model respectively.}. The general class of global-local priors seek to solve this problem by adding another shrinkage scale hierarchy, which, with suitably chosen priors, approximate the ideal behaviour of shrinking noise variables aggressively toward zero, while leaving signal variables untouched. Global-local priors take the following form:
\begin{equation} \label{eq:GL}
\begin{split}
    \beta_j | \lambda_j^2, \nu^2 & \sim N(0,\lambda_j^2\nu^2), j \in (1,\cdots, K) \\
    \lambda_j^2 & \sim \pi(\lambda_j^2)d\lambda_j^2, j \in (1,\cdots, K) \\
    \nu^2 & \sim \pi(\nu^2)d\nu^2.
\end{split}
\end{equation}

The horseshoe prior of \citet{carvalho2010horseshoe} employes two independently distributed half Cauchy distributions on the positive support for both the global and local scales:

 \begin{equation}  \label{prior3}
     \lambda_j \sim C_+(0,1)
 \end{equation}
 \begin{equation}  \label{prior4}
    \nu \sim C_+(0,1),     
 \end{equation}
 
\noindent where $\Lambda^* = \nu^2diag(\lambda_1^2,\cdots,\lambda_K)$. Due to the assumption of independence on the scales $(\lambda,\nu)$, it is straightforward to show that their posteriors follow independent Cauchy distributions. To sample from their posteriors, we make use of the slice sampler used for the Horseshoe-BQR in \citet{kohns2020horseshoe}.

\textbf{SSVS Prior}. While the lasso and horseshoe prior are continuous shrinkage priors, the SSVS prior, dating back to \citet{mitchell1988bayesian} and \citet{george1993variable,george1997approaches}, discretises the model space akin to Bayesian model averaging priors \citep{raftery1997bayesian,clyde2004model}. Since evaluating each model to compare marginal likelihoods becomes quickly infeasible in high dimensions, the SSVS saves computation time by exploring models through its Markov chain with highest posterior probability. This is achieved by modeling $\beta$ by a mixture prior, where coefficients are sorted into two groups, the "spike" and the "slab". When sorted into the first group, the value of the coefficient is shrunk heavily toward zero, while in the second it is modelled through a disperse normal prior. We follow \citet{george2008bayesian}'s implementation, which has been adapted to the BQR by \citet{korobilis2017quantile}: 
\begin{equation} \label{eq:ssvs_prior}
    \begin{split}
        \beta_{j,p} | \gamma_j,\lambda_j & \sim (1-\gamma_j)N(0,c\lambda_j^2) + \gamma_jN(0,\lambda_j^2) \forall j \in \{1,\cdots,K\} \\
        \lambda_j^2 & \sim G(a_2,b_2) \\
        \gamma_j|\pi_0 & \sim Bern(\pi_0) \\
        \pi_0 & \sim B(a_3,b_3), 
    \end{split}
\end{equation}
where $Bern(\bullet)$ stands for the Bernoulli distribution, $B(\bullet)$ for the Beta distribution and $c = 10^{-5}$, which effectively shrinks the spike group of coefficients to 0. $\pi_0$ controls the probability of inclusion into the slab group. $\Lambda^*$ for the SSVS priors becomes $diag(\lambda^2_1,\cdots,\lambda^2_K)$ for all $j$ if $\gamma_j=1$ and $diag(c\lambda^2_1,\cdots,c\lambda^2_K)$ if $\gamma_j = 0$. Under prior (\ref{eq:ssvs_prior}), the conditional posteriors are: 

\begin{equation}
    \begin{split}
        \lambda^2_j | \bullet & \sim G(a_2 + \frac{1}{2}, \frac{\beta^2_{j,p}}{2}+b_2) \\
        \gamma_j | \bullet & \sim Bern(\frac{\pi_0N(\beta_{j,p},\lambda_j^2)}{\pi_0N(\beta_{j,p},\lambda_j^2 + (1-\pi_0)N(\beta_{j,p},c\lambda_j^2)}) \\
        \pi_0|\bullet & \sim B(1+a_3, k-1+b_3),
    \end{split}
\end{equation}
where $k$ denotes the size of the slab group. Due to the very strong shrinkage implied by the scalar c, we treat $\mathbbm{1}'\gamma_j^s|\bullet$ as a posterior estimate of the model size on an iteration basis. Further, throughout all simulation and the empirical application, we set $a_3=b_3=1$, which embeds the assumption that a-priori all model sizes are equally likely, and thus allows for dense as well as sparse models as recommended by \citet{giannone2017economic}.

\subsection{Decoupling Shrinkage and Sparsity}
While continuous shrinkage priors such as global-local priors in (\ref{eq:GL}) yield good forecasting performance \citep{cross2020macroeconomic,huber2019inducing,kohns2020developments}, interpretation of forecasts based on the posterior are impeded by the fact that the marginal posterior $p(\beta_p|Y)$ is continuous on $\mathbb{R}^K$. To aid interpretability, the assumption of sparsity is often employed, which forces small coefficients to zero when they have small enough effects on the target, $Y$. The Bayesian approach to enforcing sparsity can be seen as an optimal action that minimises an expected loss function that embeds the preference of sparsity as: 

\begin{equation} \label{eq:loss_hd}
    \mathcal{L}(\Tilde{Y},\alpha) = \phi ||\alpha||_0 + T^{-1}||X\alpha-\Tilde{Y}||^2_2,
\end{equation}

\noindent which was proposed by \citet{hahn2015decoupling}, and $||\bullet||_0$ refers to the $\ell_0$-norm. $\Tilde{Y}$ refers to a realisation of the predictive distribution with density: 

\begin{equation} \label{eq:predictive_distribution}
    p(\Tilde{Y}|Y) = \int p(Y|\Tilde{X},\theta)p(\theta|Y,\Tilde{X})d\theta.
\end{equation}

\noindent In (\ref{eq:predictive_distribution}), we have collected all unknown parameters into $\theta$ for the sake of brevity. For simplicity, we focus here on in-sample predictions which renders $\Tilde{X}$ known and equal to X.\footnote{Note that in general, this need not be the case and X might contain an entirely different subset to X. See \citet{hahn2015decoupling} and \citet{piironen2020projective} for more discussion on that.} Note that in the following discussion, we omit the condition on $\Tilde{X}$ for readability. Since $\Tilde{Y}$ is a latent quantity, the expectation of (\ref{eq:loss_hd}) requires integration both with the respect to $\Tilde{Y}$ and $\theta$: 

\begin{equation} \label{eq:expected_loss}
    \mathcal{L}(\alpha) = \int \int \phi||\alpha||_0 + T^{-1} ||X\alpha-\Tilde{Y}||^2_2 \; p(\Tilde{Y}|\theta)d\Tilde{Y} \; p(\theta|\Tilde{Y})d\theta.
\end{equation}
\textbf{Theorem 1}: Assume the observation likelihood is the ALD, given by (\ref{likelihood}), and that the posterior for $\theta$, $p(\theta|Y)$, given by (\ref{eq:post_beta}), (\ref{eq:post_sigma}), and (\ref{eq:post_zt}), has been obtained through posterior sampling, then the expected loss (\ref{eq:expected_loss}) is given by: 
\begin{equation} \label{eq:intloss_adl}
    \mathcal{L}(\alpha) \propto \phi||\alpha||_0 + T^{-1}||X\alpha-X\overline{\beta}_p||_2^2 + tr(X'X\overline{\Sigma}_{\beta_p}) - T^{-1}\alpha'X'\xi\overline{Z}_p,
\end{equation}

\noindent where $\overline{\beta}$ and $\overline{\Sigma}_{\beta_p}$ refer to the posterior mean and covariance of $p(\beta_p|Y)$, and $\overline{Z}_p$ is defined as $\overline{Z}_p = \Big( \frac{|y_1-x_1'\beta_p|}{\sqrt{\xi^2+2\tau^2}} + \frac{\sigma\tau^2}{\xi^2 + 2\tau^2}, \cdots, \frac{|y_T-x_T'\beta_p|}{\sqrt{\xi^2+2\tau^2}} + \frac{\sigma\tau^2}{\xi^2 + 2\tau^2} \Big)'$. The proof is provided in the appendix.
\\~\\
\noindent\textbf{Remark on Theorem 1}: the integrated loss function (\ref{eq:intloss_adl}) is remarkably similar to the normal likelihood case derived in \citet{hahn2015decoupling}, with the difference being a term involving the posterior means of Z, $\overline{Z}_p$, which appear due to the mixture representation of the ALD. Since we are interested in sparsifying the vector $\beta_p|\bullet$, so as to minimise the Euclidean distance to the expected quantile $\hat{Q}_p = X\beta_{p|\bullet}$, instead of the expected location of the ALD directly, we proceed from this analysis by neglecting the term involving $\overline{Z}_p$. While this may seem a strong simplification, notice that 1) the contributions to this term become very small with increasing T as the entries of $\overline{Z}_p$ are multiplied one covariate at a time, which will become clear from the coordinate descent algorithm employed below; and 2) for $p=0.5$, the term vanishes completely due to $\xi=0$. Thus central quantiles are virtually unaffected. Lastly, ignoring the terms involving $\overline{Z}_p$ can also be understood from the perspective that the ALD in the BQR is often only treated as a working likelihood rather than the true data generating process of $Y$ in order to retrieve posterior estimates of the quantile regression coefficients \citep{yang2016posterior}. To convince on the negligible effect of dropping the last term in (\ref{eq:intloss_adl}), we provide simulation evidence below. 

\noindent Proceeding by dropping all constant terms, the objective function becomes:

\begin{equation} \label{eq:obj_pen}
    \alpha^*_{p} = \underset{\alpha}{\mathrm{argmin}} \;\; \phi||\alpha||_0 + T^{-1}||X\alpha-X\overline{\beta}_p||^2_2.
\end{equation}
At this point, it is interesting to note the difference in the objective function to frequentist penalised quantile regression such as the quantile lasso of \citet{chernozhukov2010quantile}. While \citet{chernozhukov2010quantile} compute the expected quantiles from observations $Y$, (\ref{eq:obj_pen}) finds a sparsified vector $\alpha^*_p$ by minimising the squared differences to the expected quantile directly. Hence, due to the availability of the posterior for $\beta_p|\bullet$, $\overline{\beta}_p$, the quantile can be treated as observed. To verify that indeed squared error loss applied in (\ref{eq:obj_pen}) results in recovering the desired quantile coefficients, we provide Lemma 1 below:
\\~\\
\noindent\textbf{Lemma 1}: Let $\beta_p^* \in \mathbb{R}^K$ be the true quantile regression parameter vector, and X be of full column rank. Then, the following minimisation problem recovers the true quantile coefficients: 
\begin{equation} \label{eq:lemma_1}
    \hat{\beta}_p = \underset{\Tilde{\beta}}{\mathrm{argmin}} \;\; ||X\beta^*_p-X\Tilde{\beta}||_2^2/T.
\end{equation}
\noindent\textbf{Proof}: Expanding (\ref{eq:lemma_1}) and solving the first order conditions yields: $\hat{\beta}_p = (X'X)^{-1}X'X\beta^*_p = \beta^*_p$.

\subsection{Signal Adaptive Variable Selection for the BQR} \label{sec: SAVS}
Following \citet{hahn2015decoupling} and \citet{ray2018signal}, we make three modifications to the objective function (\ref{eq:obj_pen}). Firstly, we make use of the $\ell_1$-norm instead of $\ell_0$-norm penalisation to obtain a convex objective function whose solution is computable with standard techniques such as the coordinate descent algorithm of \citet{friedman2010regularization}. Secondly, it is well known that the $\ell_1$-norm penalisation as an approximation to the $\ell_0$ norm overshrinks signals which might hurt predictive performance. To tackle this we make use of adaptive penalisation akin to \citet{zou2006adaptive}: 

\begin{equation} \label{eq:obj_bqrsavs}
    \alpha^*_p = \underset{\alpha}{\mathrm{argmin}} \;\; \frac{1}{2}||X\overline{\beta}_p - X\alpha||_2^2 + \sum^K_{j=1}\phi_j|\alpha|_j.
\end{equation}
\noindent Lastly, \citet{ray2018signal} have observed that starting the coordinate descent algorithm with the posterior means of $\beta|\bullet$ results in convergence after the first iteration by setting $\phi_j = \frac{1}{|\overline{\beta}_j|^{\kappa}}$ and fixing $\kappa = 2$. Hence, the solution to (\ref{eq:obj_bqrsavs}) with one iteration of the coordinate descent algorithm simplifies to:

\begin{equation} \label{eq:savsbqr}
    \hat{\alpha}_{p,j}^* = sign(\overline{\beta}_{p,j})||X_j||^{-2}(|\overline{\beta}_{p,j}|\cdot ||X_j||^2 - \phi_j)_+,
\end{equation}
where $sign(y) = 1 $ if $y\geq0$ and -1 otherwise, and let $y_+ = max\{y,0\}$. See the appendix for a full derivation of the coordinate descent algorithm.

While the original SAVS procedure proposed by \citet{ray2018signal} provides good sparsification properties by fixing $\kappa_j = 2$ across all coefficients for conditional mean models, a uniform value of $\kappa$ across quantiles can be inappropriate if the degree of sparsity varies across the conditional distribution. Quantile specific sparsity can arise when the coefficients $\beta_p$ have a quantile profile which crosses the zero line for any $p \in (0,1)$. Hence, to accommodate this, we make two contributions to the SAVS algorithm which we call the $BQR_{SAVS}$ algorithm. Firstly, we treat $\kappa_j$ as a parameter to be estimated from the data. And secondly, since standard score functions such as the log-score employed in projective variable selection \citep{piironen2020projective} or other functionals \citep{kowal2021fast} would entail variable selection consistent with conditional mean models, we use quantile specific score functions. In particular, we make use of the quantile BIC (qBIC) of \citet{lee2014model} and choose $\hat{\kappa}_{p}$ such that it minimises the qBIC criterion:

\begin{equation} \label{eq:qBIV}
    \begin{split} 
        \hat{\kappa}_p & = \underset{\kappa}{\mathrm{argmin}} \;\; qBIC(\kappa) \\
        & = \underset{\kappa}{\mathrm{argmin}} \;\; log(\sum_{t=1}^T \rho_p(y_t - x_t'\overline{\beta}_{p,\kappa})) + |\hat{\mathcal{S}}_{\kappa}|\frac{log(T)}{2T}C_K,
    \end{split}
\end{equation}
where $|\hat{\mathcal{S}}_{\kappa}|$ is the cardinality of sparsified $\overline{\beta}_{p,\kappa}$ conditional on parameter $\hat{\kappa}_p$, and $C_T$, is a positive constant which diverges to infinity with K. We follow \citet{lee2014model} by setting $C_T = log(p)$. \citet{lee2014model} show that estimating the level of penalisation in $\ell_1$ penalised quantile regression through the qBIC, results in model selection consistency under expanding K and T.\footnote{Model selection consistency is also dependent on specifying an upper bound, s, for $\mathcal{S}$ when $K >> T$. Not specifying s still leads to model selection consistency, however, with mild extra assumptions on the rate of K.} While model selection consistency is hard to prove in the Bayesian paradigm and is an active field of investigation, it stands to reason that we get close to model selection consistency in the frequentist sense when $\overline{\beta}_p$ is consistently estimated. We leave this proof for future research.

\begin{figure}[!h]
    \centering
    \includegraphics[width=\textwidth]{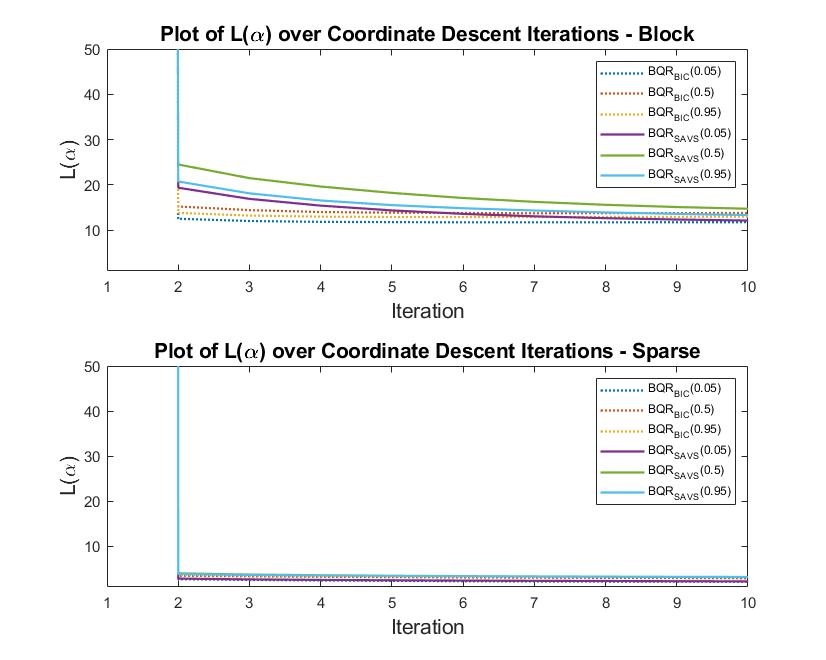}
    \caption{Coordinate Descent paths for objective function (\ref{eq:obj_bqrsavs}) for a block-sparsity and sparse DGP taken from the Monte Carlo simulations below. The convergence paths are an average over 50 simulations with 100 in-sample observations with 100 covaraites, not including the constant.}
    \label{fig:Coord_descent}
\end{figure}

Alternatively to finding $\hat{\kappa}_p$ through minimisation of (\ref{eq:qBIV}), one can embed a decision rule that incorporates uncertainty in $\hat{\kappa}_p$, such as choosing the smallest $\hat{\kappa}_p$ (and therefore the smallest model size), that is no worse as measured by the qBIC, to the un-sparsified posterior with probability $\zeta$. Obtaining a standard error estimate for the difference between the qBICs of the sparsified and un-sparsified posteriors are readily available from the the output of the Markov chain. We found, however, throughout simulations as well as the application that either method gave similar results, such that we use the minimum qBIC procedure as in (\ref{eq:qBIV}). We will denote the $BQR_{SAVS}$ using $\hat{\kappa}_p$ that minimises the qBIC as $BQR_{BIC}$.

The importance of selecting $\hat{\kappa}_p$ in a data informed manner, even for simple data generating processes without quantile varying sparsity, can be gauged from the convergence path of the coordinate descent algorithms, shown in figure (\ref{fig:Coord_descent}).

Taking average solution paths over 50 simulations based on a simple high dimensional block and sparse DGP, which are further elaborated in section 3, figure (\ref{fig:Coord_descent}) shows clearly that convergence is reached for the qBIC penalised SAVS objective function after only one iteration, while the plain SAVS takes usually more than 10 iterations on DGPs which are non-sparse. The level of the objective function is additionally much lower for qBIC compared to SAVS. Hence, stopping the coordinate descent algorithm with the plain SAVS bears the risk of inefficient penalisation.

Finally, since with correlated designs, the mean of the posterior distribution of $\beta$, can perform badly in terms of forecasting due to multimodality, we implement sparsification (\ref{eq:savsbqr}) as well as penalisation parameter choice (\ref{eq:qBIV}) on an iteration basis. This bears the added advantage of being able to quantify variable selection uncertainty through the frequency of how often a coefficient is selected into model $\mathcal{M}^s$, where $\mathcal{M}^s$ denotes a quantile regression model indexed by a vector of binary indicators of dimension K per iteration s. The percentage of times coefficient j appears in $\mathcal{M}^s$, can thus be interpreted akin to posterior variable inclusion probabilities such as $\mathbb{E}_{\gamma}(\gamma_j|\bullet)$ . Alongside posterior credible intervals for $\beta_p$, this gives an interpretable, Bayesian decision theoretically motivated way to conduction variable selection for any continuous prior for the BQR framework.

\section{Monte Carlo Experiment} \label{sec: 3}
\subsection{Monte Carlo Setup}


Simulated experiments are conducted in order to investigate the potential benefits of sparsification for the BQR in terms of its variable selection properties, coefficient bias and robustness to different data generating processes (DGPs). Another purpose of this simulation experiment is to compare variable selection of the proposed methods to the standard way of conducting Bayesian variable selection, namely via the SSVS prior. In particular, we are interested in the $BQR_{SAVS}$' and $BQR_{BIC}$'s ability to adapt to different degrees of sparsity and error distributions. We generate data from the following model:

\begin{equation} \label{eq:rand_mod}
    y_t = \beta_0 + x_{t,c}'\beta_c + x_{t,q}'\beta_q + (x_t'\vartheta)u_t, \; \; t = 1,\cdots,T,
\end{equation}
where $\{u_t\}_{t=1}^T$ are assumed to follow some otherwise unspecified cumulative density function (CDF), F, and the subscripts c and q refer to covariate groups whose regression coefficients don't vary and vary across quantiles respectively. From (\ref{eq:rand_mod}), one can see that the quantile profile of the quantile varying coefficients is in turn dictated by the correlation between the covariate and error process $u_t$ which is enforced through $\vartheta$ being a K-dimensional binary vector which is non-zero if $x_{t,i} \in x_{t,q}$ (as well as for the constant coefficient). Assuming a tick-loss function and solving for quantile regression coefficients $\beta(p)$ for any p, \citet{koenker2005} has shown that quantile regression coefficients have a random coefficient interpretation, where the quantile profile is proportional to the quantiles of the error process, with its centre on $\beta_c$ and $\beta_q$ respectively:

\begin{equation}\label{eq:beta_decomp}
\begin{split}
    \beta_0(p)&=\beta_0+\vartheta_0 F^{-1}(p) \\
    \beta_{q}(p) & = \beta_q + \vartheta_q F^{-1}(p)\footnote{Footnote to explain that we set $\vartheta$ to 0 decrease the variance due to large tail effects.} \\
    \beta_c(p) & = \beta_c 
\end{split}
\end{equation}

The constant coefficient $\beta_0$ by default always has a quantile profile so as to enforce location effects. 
Solution (\ref{eq:beta_decomp}) to model (\ref{eq:rand_mod}) therefore motivates the investigation of sparsity detection for both the $\beta_c$ and $\beta_q$ vectors, and also whether the assumption of the error distribution effects this.

In particular, consider a homoskedastic design in which $\vartheta$ is populated by zeros such that $\beta(p) = \beta \; \forall p $ where $\beta = (\beta_c',\beta_q')'$. These offer baseline DGPs in which, as per (\ref{eq:beta_decomp}), only the constant has a non-constant quantile function. Sparsity patterns considered are:
\begin{itemize}
    \item $\beta_{sparse} = (1,1.5,1,0.5,0.33,0.25, 0_{1 \times (K-4)})$
    \item $\beta_{block}=(1,0.5_{1 \times \frac{K}{5}},0_{1 \times 2\frac{K}{5}},0.5_{1 \times \frac{K}{5}})$
\end{itemize}
To investigate how well the $BQR_{SAVS}$ and $BQR_{BIC}$ detect sparsity specific to a given quantile, we consider additionally a heteroskedastic set of DGPs where $\beta_q$ is zero only for some p. In particular, we let the constant coefficient be zero at the median and an additional covariate be zero only for $0.15 < p < 0.85$.\footnote{Centering the quantile function of $\beta_0$ on 0 helps identifying quantile profiles on other covariates on account of tails of coefficients being 'aligned'} This is equivalent to making the $\vartheta$ be quantile varying as well. In particular $\vartheta_q$ for the quantile specific sparsity is defined as:

\begin{equation}
    \vartheta^*_q=\vartheta_q[I(u_t\leq F^{-1}(0.15))+I(u_t\geq F^{-1}(0.85))]
\end{equation}

Sparsity patterns that are considered for the quantile specific sparse Monte Carlo designs are:
\begin{itemize}
    \item $\beta_{sparse,c} = (0,0.5,0.33,0.25, 0_{1 \times (K-4)})$, $\beta_{q} = 0$ and $\beta_0 = 0$
    \item $\beta_{block,c}=(0,0.5_{1 \times \frac{K}{5}},0_{1 \times 2\frac{K}{5}},0.5_{1 \times \frac{K}{5}})$, $\beta_{q} = 0$ and $\beta_0 = 0$
\end{itemize}


\begin{figure}[b!]
    \centering
    \includegraphics[width=\textwidth]{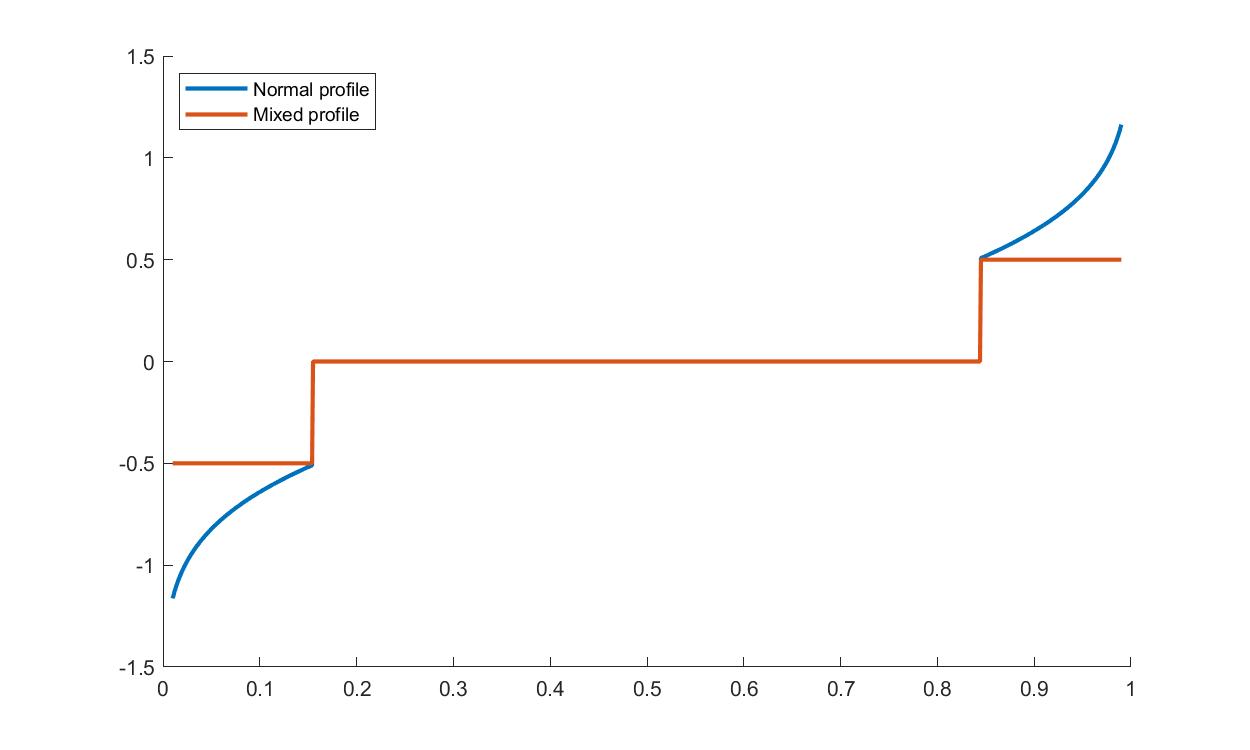}
    \caption{Quantile specific sparsity profiles}
    \label{fig:quantprof}
\end{figure}

For the homoskedastic DGPs, we consider standard normal and student-t-distributed errors with 3 degrees of freedom. For heteroskedastic DGPs, we also consider standard normal error and one further variant. Since it may also be plausible to find that there is sparsity across p, but with constant non-zero coefficients for non-sparse quantiles, we consider as a final heteroskedastic design the quantile varying coefficient to be -0.5 for $p \in (0,0.15]$ and 0.5 for $p \in [0.85,1)$. For clarity, the quantile profiles for the quantile specific sparse coefficient are presented in figure (\ref{fig:quantprof}). The quantile profile of the quantile varying beta coefficients for the 4 designs are summarised in table (\ref{tab:simsetups}). 

\begin{table}[t]
\centering
\begin{tabular}{c|l}
\textbf{Design}    &  \textbf{$\beta$ profiles}\\
\hline
$y_1$ & $\beta_0(p)=\beta_0+F^{-1}_{N(0,1)}(p)$ \\
\hline
$y_2$ & $\beta_0(p)=\beta_0+F^{-1}_{T(3)}(p)$\\
\hline
\multirow{2}{*}{$y_3$} & $\beta_0(p)=\beta_0+F^{-1}_{N(0,1)}(p)$ \\
  & $\beta_1(p)=\beta_1+[I(u_t\leq F^{-1}_{N(0,1)}(0.15))+I(u_t\geq F^{-1}_{N(0,1)}(0.85))]F^{-1}_{N(0,1)}(p)$ \\
\hline
\multirow{2}{*}{$y_4$} & $\beta_0(p)=\beta_0+F^{-1}_{N(0,1)}(p)$ \\
& $\beta_1(p)=\beta_1+[-\frac{1}{2}I(u_t\leq F^{-1}_{N(0,1)}(0.15))+\frac{1}{2}I(u_t\geq F^{-1}_{N(0,1)}(0.85))]$
\end{tabular}
\caption{Summary of simulation setups} \label{tab:simsetups}
\end{table}

For all DGPs, we generate 50 synthetic data sets and set $K=100$ (constant not included). The design matrix, X, is simulated using a multivariate Gaussian with zero mean and Toeplitz covariance structure, $\Omega$, where $\Omega_{i,j} = \rho^{|i-j|}$ and $\rho = 0.5$. There are two sample sizes considered $T_i \in \{100,500\}$.

Performance of the priors are gauged in terms of coefficient bias and measures of correct variable selection. Coefficient bias is calculated as the average (over Monte Carlo iterations) root mean deviation to the true quantile regression coefficients:

\begin{equation}
    \frac{1}{iter} \sqrt{|| \hat{\beta}(p) - \beta(p)||_2^2}
\end{equation}

\noindent where $iter$ is the number of Monte Carlo iterations, and $\hat{\beta}(p)$ is the mean posterior vector of the estimator.

Variable selection will be evaluated with Matthew's correlation coefficient (MCC) and the hit rate. While the MCC gives an overall measure of goodness of variable selection based on the true and false positives and negatives, the hit rate tells us how well we identify the true positives. The MCC and the hit rate can be calculated from the confusion matrix with the following formulas:

\begin{align}
    MCC &= \frac{TP \times TN - FP \times FN}{\sqrt{(TP + FP)(TP + FN)(TN + FP)(TN + FN)}} \\
    Hit~rate &= \frac{TP}{TP+FN}
\end{align}

\noindent where T stands for "True", F stands for "False", N stands for "Negative", and P stands for "Positive". The MCC and hit rate are commonly used for binary classification problems but high-dimensional problems can be easily amended to fit the framework: non-zero coefficients have a value of 1 and zero coefficients have a value of 0. The same way coefficients shrunk to 0 receive a value of 0, while coefficients that are not 0 get a value of 1. This way a confusion matrix can be constructed from which the measures can be calculated. The MCC is between -1 (worst) and +1 (best), while the hit rate will range between 0 (worst) and 1 (best).

\subsection{Monte Carlo Results} \label{sec:MCresults}
The coefficient bias results of the designs for a selection of quantiles are presented in table (\ref{tab:coeffbias}), the results for the MCC are shown in table (\ref{tab:MCC}), and the results of for the hit rates can be seen in table (\ref{tab:Hitrate}) \footnote{Note that the results presented in tables (\ref{tab:coeffbias})-(\ref{tab:Hitrate}) do not include the ALD adjustment term $\overline{Z}_p$.}. The tables tell a very clear story: sparsification via the $BQR_{SAVS}$ and $BQR_{BIC}$ improves or yields similar coefficient bias compared to the unsparsified posteriors, while yielding more often than not better variable selection accuracy than the $SSVSBQR$, independent of sparsity, in-sample length and error-distribution. The $BQR_{BIC}$ procedure stands out particularly in the sparse DGPs in which this sparsification procedure can reduce bias close to half for some quantiles in the larger sample setting compared to the unsparsified posterior and offers further bias reduction compared to the $BQR_{SAVS}$. 
Results including the adjustment term $\overline{Z}_p$, presented in the appendix shows that the results are largely unaffected as expected. 

\subsubsection{Baseline DGPs}
Focusing first on bias for the unsparsified posterior, one can see from the first two columns of table (\ref{tab:coeffbias}), that the $HSBQR$ and $LBQR$ do similarly well in the sparse DGPs, while clearly outperforming the $SSVSBQR$. Although, all priors reduce bias with more data, the overall tendency doesn't change in the larger T=500 DGP. The largest difference among the SSVS, Lasso and horseshoe priors manifests itself in the block DGP's where the $HSBQR$ does consistently better, especially in the tails and in the smaller sample size of T=100. These findings corroborate the findings of \citet{kohns2020horseshoe} that the adaptive shrinkage from global local priors are able to capture different sparsity blocks among the covariates and adapt to low data settings.

\begin{landscape}
\begin{table}[]
\centering
\resizebox{1\textwidth}{!}{
\begin{tabular}{lrccccc|ccccc|ccccc|ccccc}
 \hline
       & \multicolumn{1}{l}{}     & 0.05  & 0.25  & 0.50  & 0.75  & 0.95  & 0.05  & 0.25  & 0.50  & 0.75  & 0.95  & 0.05  & 0.25  & 0.50  & 0.75  & 0.95  & 0.05  & 0.25  & 0.50  & 0.75  & 0.95  \\ 
       \hline  \hline
       & \multicolumn{1}{l}{}     & \multicolumn{5}{c|}{$y_1$}            & \multicolumn{5}{c|}{$y_2$}            & \multicolumn{5}{c|}{$y_3$}            & \multicolumn{5}{c}{$y_4$}             \\
\multicolumn{2}{l}{\textbf{T=100}}&       &       &       &       &       &       &       &       &       &       &       &       &       &       &       &       &       &       &       &       \\
\multicolumn{2}{l}{Sparse}        &       &       &       &       &       &       &       &       &       &       &       &       &       &       &       &       &       &       &       &       \\
       & $SSVSBQR$                   & 0.206 & 0.210 & 0.217 & 0.212 & 0.221 & 0.227 & 0.209 & 0.195 & 0.201 & 0.238 & 0.235 & 0.241 & 0.227 & 0.253 & 0.243 & 0.193 & 0.219 & 0.223 & 0.222 & 0.194 \\
       & $LBQR$                   & 0.132 & 0.079 & 0.068 & 0.084 & 0.143 & 0.173 & 0.082 & 0.067 & 0.084 & 0.164 & 0.158 & 0.081 & 0.063 & 0.087 & 0.153 & 0.133 & 0.084 & 0.065 & 0.087 & 0.140 \\
       & $LBQR_{SAVS}$            & 0.132 & 0.076 & 0.065 & 0.082 & 0.143 & 0.173 & 0.080 & 0.065 & 0.082 & 0.164 & 0.158 & 0.082 & 0.061 & 0.088 & 0.154 & 0.133 & 0.086 & 0.062 & 0.087 & 0.141 \\
       & $LBQR_{BIC}$             & 0.132 & 0.076 & 0.065 & 0.082 & 0.143 & 0.173 & 0.080 & 0.065 & 0.081 & 0.164 & 0.158 & 0.081 & 0.064 & 0.088 & 0.153 & 0.133 & 0.085 & 0.062 & 0.086 & 0.141 \\
       & $HSBQR$                  & 0.119 & 0.094 & 0.085 & 0.097 & 0.127 & 0.172 & 0.099 & 0.087 & 0.097 & 0.176 & 0.165 & 0.104 & 0.075 & 0.100 & 0.157 & 0.131 & 0.099 & 0.077 & 0.099 & 0.131 \\
       & $HSBQR_{SAVS}$           & 0.108 & 0.074 & 0.064 & 0.080 & 0.117 & 0.164 & 0.080 & 0.066 & 0.081 & 0.169 & 0.157 & 0.095 & 0.061 & 0.090 & 0.148 & 0.123 & 0.089 & 0.061 & 0.088 & 0.124 \\
       & $HSBQR_{BIC}$            & 0.101 & 0.063 & 0.058 & 0.072 & 0.109 & 0.159 & 0.067 & 0.056 & 0.073 & 0.164 & 0.149 & 0.096 & 0.060 & 0.090 & 0.140 & 0.119 & 0.090 & 0.060 & 0.088 & 0.119 \\
\multicolumn{2}{l}{Block}         &       &       &       &       &       &       &       &       &       &       &       &       &       &       &       &       &       &       &       &       \\
       & $SSVSBQR$                   & 0.320 & 0.325 & 0.325 & 0.336 & 0.336 & 0.339 & 0.333 & 0.322 & 0.328 & 0.360 & 0.358 & 0.371 & 0.362 & 0.359 & 0.353 & 0.337 & 0.340 & 0.353 & 0.339 & 0.332 \\
       & $LBQR$                   & 0.369 & 0.328 & 0.327 & 0.323 & 0.382 & 0.338 & 0.329 & 0.326 & 0.328 & 0.342 & 0.414 & 0.352 & 0.339 & 0.355 & 0.419 & 0.410 & 0.348 & 0.342 & 0.346 & 0.411 \\
       & $LBQR_{SAVS}$            & 0.368 & 0.326 & 0.325 & 0.321 & 0.381 & 0.337 & 0.327 & 0.324 & 0.326 & 0.341 & 0.413 & 0.349 & 0.336 & 0.353 & 0.418 & 0.410 & 0.346 & 0.339 & 0.344 & 0.411 \\
       & $LBQR_{BIC}$             & 0.368 & 0.326 & 0.326 & 0.321 & 0.381 & 0.337 & 0.327 & 0.324 & 0.326 & 0.341 & 0.413 & 0.349 & 0.336 & 0.352 & 0.418 & 0.410 & 0.346 & 0.339 & 0.344 & 0.410 \\
       & $HSBQR$                  & 0.250 & 0.248 & 0.251 & 0.253 & 0.244 & 0.260 & 0.240 & 0.244 & 0.249 & 0.272 & 0.288 & 0.293 & 0.290 & 0.286 & 0.278 & 0.258 & 0.267 & 0.269 & 0.263 & 0.255 \\
       & $HSBQR_{SAVS}$           & 0.252 & 0.244 & 0.244 & 0.247 & 0.243 & 0.263 & 0.236 & 0.239 & 0.244 & 0.271 & 0.289 & 0.287 & 0.282 & 0.280 & 0.277 & 0.261 & 0.263 & 0.262 & 0.258 & 0.256 \\
       & $HSBQR_{BIC}$            & 0.259 & 0.246 & 0.246 & 0.247 & 0.247 & 0.271 & 0.241 & 0.244 & 0.247 & 0.275 & 0.287 & 0.287 & 0.282 & 0.280 & 0.277 & 0.258 & 0.263 & 0.263 & 0.259 & 0.254 \\ 
\multicolumn{2}{l}{\textbf{T=500}}&       &       &       &       &       &       &       &       &       &       &       &       &       &       &       &       &       &       &       &       \\
\multicolumn{2}{l}{Sparse}        &       &       &       &       &       &       &       &       &       &       &       &       &       &       &       &       &       &       &       &       \\
       & $SSVSBQR$                   & 0.133 & 0.095 & 0.087 & 0.095 & 0.131 & 0.165 & 0.096 & 0.083 & 0.095 & 0.170 & 0.159 & 0.084 & 0.069 & 0.087 & 0.156 & 0.134 & 0.088 & 0.073 & 0.083 & 0.132 \\
       & $LBQR$                   & 0.092 & 0.053 & 0.042 & 0.054 & 0.091 & 0.138 & 0.060 & 0.044 & 0.060 & 0.147 & 0.119 & 0.056 & 0.039 & 0.056 & 0.119 & 0.102 & 0.053 & 0.036 & 0.055 & 0.105 \\
       & $LBQR_{SAVS}$            & 0.092 & 0.046 & 0.033 & 0.049 & 0.090 & 0.139 & 0.053 & 0.035 & 0.056 & 0.147 & 0.119 & 0.056 & 0.029 & 0.056 & 0.119 & 0.102 & 0.052 & 0.027 & 0.055 & 0.105 \\
       & $LBQR_{BIC}$             & 0.092 & 0.041 & 0.029 & 0.046 & 0.090 & 0.138 & 0.048 & 0.032 & 0.053 & 0.147 & 0.119 & 0.056 & 0.031 & 0.056 & 0.119 & 0.101 & 0.051 & 0.030 & 0.054 & 0.105 \\
       & $HSBQR$                  & 0.099 & 0.072 & 0.062 & 0.074 & 0.103 & 0.152 & 0.074 & 0.061 & 0.078 & 0.153 & 0.137 & 0.066 & 0.047 & 0.068 & 0.134 & 0.110 & 0.065 & 0.047 & 0.067 & 0.113 \\
       & $HSBQR_{SAVS}$           & 0.086 & 0.051 & 0.038 & 0.054 & 0.089 & 0.145 & 0.057 & 0.037 & 0.060 & 0.145 & 0.125 & 0.057 & 0.025 & 0.059 & 0.122 & 0.096 & 0.056 & 0.026 & 0.058 & 0.101 \\
       & $HSBQR_{BIC}$            & 0.074 & 0.043 & 0.029 & 0.047 & 0.076 & 0.139 & 0.050 & 0.030 & 0.053 & 0.138 & 0.106 & 0.063 & 0.030 & 0.064 & 0.107 & 0.083 & 0.060 & 0.030 & 0.062 & 0.090 \\
\multicolumn{2}{l}{Block}         &       &       &       &       &       &       &       &       &       &       &       &       &       &       &       &       &       &       &       &       \\
       & $SSVSBQR$                   & 0.150 & 0.101 & 0.094 & 0.104 & 0.145 & 0.177 & 0.101 & 0.091 & 0.101 & 0.180 & 0.190 & 0.097 & 0.079 & 0.098 & 0.180 & 0.162 & 0.097 & 0.084 & 0.099 & 0.156 \\
       & $LBQR$                   & 0.305 & 0.100 & 0.079 & 0.094 & 0.306 & 0.290 & 0.097 & 0.079 & 0.099 & 0.288 & 0.340 & 0.102 & 0.074 & 0.098 & 0.327 & 0.322 & 0.100 & 0.073 & 0.098 & 0.320 \\
       & $LBQR_{SAVS}$            & 0.304 & 0.101 & 0.080 & 0.096 & 0.305 & 0.289 & 0.099 & 0.081 & 0.100 & 0.287 & 0.340 & 0.104 & 0.074 & 0.100 & 0.326 & 0.321 & 0.103 & 0.073 & 0.100 & 0.319 \\
       & $LBQR_{BIC}$             & 0.304 & 0.100 & 0.079 & 0.094 & 0.305 & 0.289 & 0.097 & 0.079 & 0.099 & 0.287 & 0.340 & 0.102 & 0.073 & 0.098 & 0.326 & 0.321 & 0.100 & 0.073 & 0.098 & 0.319 \\
       & $HSBQR$                  & 0.142 & 0.104 & 0.094 & 0.105 & 0.143 & 0.173 & 0.103 & 0.091 & 0.104 & 0.175 & 0.189 & 0.099 & 0.083 & 0.097 & 0.184 & 0.155 & 0.100 & 0.084 & 0.099 & 0.154 \\
       & $HSBQR_{SAVS}$           & 0.137 & 0.092 & 0.082 & 0.093 & 0.137 & 0.172 & 0.092 & 0.078 & 0.093 & 0.171 & 0.186 & 0.090 & 0.072 & 0.089 & 0.182 & 0.151 & 0.091 & 0.073 & 0.091 & 0.150 \\
       & $HSBQR_{BIC}$            & 0.138 & 0.095 & 0.085 & 0.096 & 0.137 & 0.173 & 0.094 & 0.080 & 0.095 & 0.171 & 0.187 & 0.090 & 0.072 & 0.088 & 0.182 & 0.153 & 0.091 & 0.073 & 0.090 & 0.152 \\
       \hline
\end{tabular}}
\caption{Root Mean Coefficient Bias Results}
\label{tab:coeffbias}
\end{table}
\end{landscape}

\begin{landscape}
\begin{table}[]
\centering
\resizebox{1.25\textwidth}{!}{
\begin{tabular}{lrccccc|ccccc|ccccc|ccccc}
 \hline
       & \multicolumn{1}{l}{}     & 0.05  & 0.25  & 0.50  & 0.75  & 0.95  & 0.05  & 0.25  & 0.50  & 0.75  & 0.95  & 0.05  & 0.25  & 0.50  & 0.75  & 0.95  & 0.05  & 0.25  & 0.50  & 0.75  & 0.95  \\ 
       \hline  \hline
       & \multicolumn{1}{l}{}     & \multicolumn{5}{c|}{$y_1$}            & \multicolumn{5}{c|}{$y_2$}            & \multicolumn{5}{c|}{$y_3$}            & \multicolumn{5}{c}{$y_4$}             \\
\multicolumn{2}{l}{\textbf{T=100}}&       &       &       &       &       &       &       &       &       &       &       &       &       &       &       &       &       &       &       &       \\
\multicolumn{2}{l}{Sparse}        &       &       &       &       &       &       &       &       &       &       &       &       &       &       &       &       &       &       &       &       \\
            & $SSVSBQR$              & 0.243 & 0.275 & 0.269 & 0.292 & 0.325 & 0.266 & 0.273 & 0.305 & 0.304 & 0.353 & 0.225 & 0.188 & 0.180 & 0.184 & 0.217 & 0.262 & 0.210 & 0.172 & 0.205 & 0.266 \\
            & $LBQR_{SAVS}$       & 0.589 & 0.710 & 0.789 & 0.781 & 0.660 & 0.570 & 0.712 & 0.797 & 0.791 & 0.680 & 0.428 & 0.581 & 0.558 & 0.546 & 0.432 & 0.442 & 0.551 & 0.545 & 0.538 & 0.399 \\
            & $LBQR_{BIC}$        & 0.589 & 0.720 & 0.829 & 0.814 & 0.656 & 0.574 & 0.717 & 0.824 & 0.821 & 0.680 & 0.433 & 0.586 & 0.518 & 0.528 & 0.436 & 0.439 & 0.548 & 0.572 & 0.548 & 0.404 \\
            & $HSBQR_{SAVS}$      & 0.390 & 0.532 & 0.560 & 0.550 & 0.513 & 0.382 & 0.523 & 0.554 & 0.556 & 0.503 & 0.350 & 0.422 & 0.442 & 0.415 & 0.353 & 0.394 & 0.426 & 0.429 & 0.411 & 0.396 \\
            & $HSBQR_{BIC}$       & 0.552 & 0.720 & 0.782 & 0.772 & 0.695 & 0.541 & 0.722 & 0.793 & 0.789 & 0.689 & 0.468 & 0.495 & 0.542 & 0.488 & 0.472 & 0.510 & 0.504 & 0.539 & 0.493 & 0.506 \\
\multicolumn{2}{l}{Block}         &       &       &       &       &       &       &       &       &       &       &       &       &       &       &       &       &       &       &       &       \\
            & $SSVSBQR$              & 0.420 & 0.386 & 0.381 & 0.384 & 0.405 & 0.416 & 0.375 & 0.392 & 0.382 & 0.423 & 0.374 & 0.319 & 0.314 & 0.330 & 0.394 & 0.410 & 0.352 & 0.311 & 0.376 & 0.417  \\
            & $LBQR_{SAVS}$       & 0.255 & 0.429 & 0.488 & 0.473 & 0.261 & 0.252 & 0.437 & 0.496 & 0.471 & 0.256 & 0.230 & 0.415 & 0.439 & 0.410 & 0.217 & 0.232 & 0.424 & 0.427 & 0.426 & 0.222 \\
            & $LBQR_{BIC}$        & 0.255 & 0.424 & 0.478 & 0.471 & 0.261 & 0.254 & 0.437 & 0.492 & 0.471 & 0.256 & 0.229 & 0.411 & 0.436 & 0.409 & 0.217 & 0.232 & 0.421 & 0.424 & 0.423 & 0.222 \\
            & $HSBQR_{SAVS}$      & 0.387 & 0.408 & 0.421 & 0.410 & 0.393 & 0.406 & 0.430 & 0.440 & 0.420 & 0.401 & 0.337 & 0.342 & 0.343 & 0.351 & 0.344 & 0.383 & 0.391 & 0.385 & 0.395 & 0.383 \\
            & $HSBQR_{BIC}$       & 0.423 & 0.442 & 0.459 & 0.452 & 0.436 & 0.436 & 0.458 & 0.471 & 0.462 & 0.444 & 0.251 & 0.327 & 0.350 & 0.333 & 0.255 & 0.276 & 0.352 & 0.369 & 0.355 & 0.272 \\ 
\multicolumn{2}{l}{\textbf{T=500}}&       &       &       &       &       &       &       &       &       &       &       &       &       &       &       &       &       &       &       &       \\
\multicolumn{2}{l}{Sparse}        &       &       &       &       &       &       &       &       &       &       &       &       &       &       &       &       &       &       &       &       \\
            & $SSVSBQR$              & 0.275 & 0.308 & 0.305 & 0.309 & 0.314 & 0.294 & 0.311 & 0.320 & 0.324 & 0.318 & 0.284 & 0.331 & 0.304 & 0.334 & 0.293 & 0.301 & 0.316 & 0.291 & 0.348 & 0.325 \\
            & $LBQR_{SAVS}$       & 0.704 & 0.745 & 0.755 & 0.789 & 0.790 & 0.740 & 0.707 & 0.758 & 0.792 & 0.809 & 0.571 & 0.797 & 0.718 & 0.773 & 0.593 & 0.589 & 0.783 & 0.757 & 0.776 & 0.571 \\
            & $LBQR_{BIC}$        & 0.708 & 0.903 & 0.924 & 0.926 & 0.805 & 0.745 & 0.878 & 0.907 & 0.924 & 0.820 & 0.594 & 0.836 & 0.835 & 0.839 & 0.603 & 0.598 & 0.848 & 0.852 & 0.823 & 0.590 \\
            & $HSBQR_{SAVS}$      & 0.401 & 0.538 & 0.563 & 0.539 & 0.470 & 0.393 & 0.570 & 0.579 & 0.555 & 0.470 & 0.395 & 0.665 & 0.635 & 0.642 & 0.416 & 0.422 & 0.658 & 0.634 & 0.636 & 0.441 \\
            & $HSBQR_{BIC}$       & 0.711 & 0.899 & 0.899 & 0.901 & 0.817 & 0.713 & 0.909 & 0.901 & 0.900 & 0.837 & 0.699 & 0.810 & 0.847 & 0.805 & 0.722 & 0.724 & 0.819 & 0.845 & 0.797 & 0.747 \\
\multicolumn{2}{l}{Block}         &       &       &       &       &       &       &       &       &       &       &       &       &       &       &       &       &       &       &       &       \\
            & $SSVSBQR$              & 0.682 & 0.725 & 0.711 & 0.713 & 0.696 & 0.694 & 0.730 & 0.722 & 0.726 & 0.708 & 0.638 & 0.753 & 0.768 & 0.768 & 0.657 & 0.679 & 0.760 & 0.744 & 0.753 & 0.707 \\
            & $LBQR_{SAVS}$       & 0.584 & 0.937 & 0.959 & 0.959 & 0.595 & 0.595 & 0.951 & 0.963 & 0.955 & 0.591 & 0.552 & 0.935 & 0.956 & 0.945 & 0.549 & 0.556 & 0.941 & 0.958 & 0.937 & 0.563 \\
            & $LBQR_{BIC}$        & 0.584 & 0.930 & 0.956 & 0.950 & 0.595 & 0.595 & 0.937 & 0.957 & 0.946 & 0.591 & 0.552 & 0.925 & 0.952 & 0.938 & 0.549 & 0.556 & 0.935 & 0.950 & 0.930 & 0.563 \\
            & $HSBQR_{SAVS}$      & 0.691 & 0.800 & 0.810 & 0.799 & 0.696 & 0.708 & 0.804 & 0.818 & 0.812 & 0.712 & 0.612 & 0.830 & 0.863 & 0.847 & 0.616 & 0.679 & 0.823 & 0.855 & 0.843 & 0.677 \\
            & $HSBQR_{BIC}$       & 0.745 & 0.929 & 0.947 & 0.930 & 0.766 & 0.749 & 0.934 & 0.953 & 0.941 & 0.783 & 0.433 & 0.867 & 0.906 & 0.880 & 0.441 & 0.439 & 0.833 & 0.887 & 0.847 & 0.449 \\ 
            \hline
\end{tabular}}
\caption{MCC Results}
\label{tab:MCC}
\end{table}
\end{landscape}

\begin{landscape}
\begin{table}[]
\centering
\resizebox{1.25\textwidth}{!}{
\begin{tabular}{lrccccc|ccccc|ccccc|ccccc}
 \hline
       & \multicolumn{1}{l}{}     & 0.05  & 0.25  & 0.50  & 0.75  & 0.95  & 0.05  & 0.25  & 0.50  & 0.75  & 0.95  & 0.05  & 0.25  & 0.50  & 0.75  & 0.95  & 0.05  & 0.25  & 0.50  & 0.75  & 0.95  \\ 
       \hline  \hline
       & \multicolumn{1}{l}{}     & \multicolumn{5}{c|}{$y_1$}            & \multicolumn{5}{c|}{$y_2$}            & \multicolumn{5}{c|}{$y_3$}            & \multicolumn{5}{c}{$y_4$}             \\
\multicolumn{2}{l}{\textbf{T=100}}&       &       &       &       &       &       &       &       &       &       &       &       &       &       &       &       &       &       &       &       \\
\multicolumn{2}{l}{Sparse}        &       &       &       &       &       &       &       &       &       &       &       &       &       &       &       &       &       &       &       &       \\
            & $SSVSBQR$              & 0.622 & 0.804 & 0.854 & 0.848 & 0.818 & 0.637 & 0.796 & 0.867 & 0.862 & 0.831 & 0.573 & 0.650 & 0.688 & 0.676 & 0.570 & 0.591 & 0.695 & 0.676 & 0.687 & 0.581 \\
            & $LBQR_{SAVS}$       & 0.399 & 0.647 & 0.788 & 0.742 & 0.512 & 0.362 & 0.646 & 0.772 & 0.768 & 0.524 & 0.226 & 0.445 & 0.501 & 0.424 & 0.233 & 0.240 & 0.428 & 0.458 & 0.417 & 0.211 \\
            & $LBQR_{BIC}$        & 0.400 & 0.568 & 0.759 & 0.712 & 0.509 & 0.364 & 0.597 & 0.747 & 0.755 & 0.524 & 0.227 & 0.410 & 0.333 & 0.353 & 0.233 & 0.235 & 0.385 & 0.392 & 0.378 & 0.214 \\
            & $HSBQR_{SAVS}$      & 0.576 & 0.814 & 0.844 & 0.835 & 0.769 & 0.567 & 0.792 & 0.827 & 0.822 & 0.759 & 0.530 & 0.589 & 0.611 & 0.573 & 0.550 & 0.544 & 0.600 & 0.609 & 0.578 & 0.558 \\
            & $HSBQR_{BIC}$       & 0.483 & 0.638 & 0.706 & 0.706 & 0.700 & 0.505 & 0.645 & 0.727 & 0.725 & 0.695 & 0.437 & 0.367 & 0.385 & 0.353 & 0.449 & 0.455 & 0.382 & 0.384 & 0.360 & 0.473 \\
\multicolumn{2}{l}{Block}         &       &       &       &       &       &       &       &       &       &       &       &       &       &       &       &       &       &       &       &       \\
            & $SSVSBQR$              & 0.575 & 0.652 & 0.682 & 0.666 & 0.593 & 0.570 & 0.650 & 0.699 & 0.669 & 0.601 & 0.557 & 0.620 & 0.638 & 0.634 & 0.557 & 0.573 & 0.645 & 0.652 & 0.660 & 0.571 \\
            & $LBQR_{SAVS}$       & 0.120 & 0.325 & 0.387 & 0.372 & 0.129 & 0.124 & 0.333 & 0.406 & 0.369 & 0.122 & 0.113 & 0.332 & 0.370 & 0.329 & 0.105 & 0.121 & 0.334 & 0.359 & 0.340 & 0.111 \\
            & $LBQR_{BIC}$        & 0.121 & 0.318 & 0.371 & 0.366 & 0.129 & 0.123 & 0.332 & 0.400 & 0.368 & 0.122 & 0.112 & 0.326 & 0.359 & 0.324 & 0.105 & 0.121 & 0.330 & 0.349 & 0.334 & 0.111 \\
            & $HSBQR_{SAVS}$      & 0.554 & 0.651 & 0.686 & 0.672 & 0.583 & 0.569 & 0.665 & 0.692 & 0.671 & 0.585 & 0.548 & 0.631 & 0.649 & 0.639 & 0.555 & 0.566 & 0.648 & 0.666 & 0.655 & 0.568 \\
            & $HSBQR_{BIC}$       & 0.362 & 0.414 & 0.443 & 0.439 & 0.389 & 0.369 & 0.423 & 0.445 & 0.436 & 0.387 & 0.686 & 0.654 & 0.632 & 0.664 & 0.694 & 0.718 & 0.702 & 0.685 & 0.709 & 0.721 \\ 
\multicolumn{2}{l}{\textbf{T=500}}&       &       &       &       &       &       &       &       &       &       &       &       &       &       &       &       &       &       &       &       \\
\multicolumn{2}{l}{Sparse}        &       &       &       &       &       &       &       &       &       &       &       &       &       &       &       &       &       &       &       &       \\
            & $SSVSBQR$              & 0.875 & 0.973 & 0.984 & 0.973 & 0.964 & 0.905 & 0.980 & 0.989 & 0.984 & 0.949 & 0.857 & 0.963 & 0.990 & 0.976 & 0.874 & 0.845 & 0.968 & 0.966 & 0.985 & 0.881 \\
            & $LBQR_{SAVS}$       & 0.569 & 0.919 & 0.950 & 0.938 & 0.698 & 0.608 & 0.880 & 0.937 & 0.929 & 0.712 & 0.413 & 0.886 & 0.918 & 0.869 & 0.442 & 0.427 & 0.872 & 0.931 & 0.872 & 0.430 \\
            & $LBQR_{BIC}$        & 0.573 & 0.851 & 0.886 & 0.896 & 0.700 & 0.611 & 0.819 & 0.874 & 0.888 & 0.712 & 0.410 & 0.848 & 0.733 & 0.838 & 0.428 & 0.417 & 0.858 & 0.786 & 0.852 & 0.430 \\
            & $HSBQR_{SAVS}$      & 0.773 & 0.950 & 0.962 & 0.952 & 0.926 & 0.752 & 0.959 & 0.958 & 0.955 & 0.917 & 0.824 & 0.927 & 0.945 & 0.908 & 0.840 & 0.846 & 0.934 & 0.947 & 0.925 & 0.856 \\
            & $HSBQR_{BIC}$       & 0.637 & 0.845 & 0.844 & 0.855 & 0.813 & 0.648 & 0.864 & 0.851 & 0.855 & 0.827 & 0.648 & 0.752 & 0.755 & 0.746 & 0.664 & 0.652 & 0.796 & 0.768 & 0.768 & 0.674 \\
\multicolumn{2}{l}{Block}         &       &       &       &       &       &       &       &       &       &       &       &       &       &       &       &       &       &       &       &       \\
            & $SSVSBQR$              & 0.969 & 1.000 & 1.000 & 0.999 & 0.983 & 0.966 & 0.999 & 1.000 & 0.999 & 0.980 & 0.938 & 0.991 & 0.998 & 0.992 & 0.938 & 0.957 & 0.992 & 0.998 & 0.992 & 0.962 \\
            & $LBQR_{SAVS}$       & 0.500 & 0.974 & 0.996 & 0.986 & 0.499 & 0.509 & 0.980 & 0.993 & 0.984 & 0.500 & 0.473 & 0.963 & 0.991 & 0.964 & 0.476 & 0.482 & 0.964 & 0.992 & 0.967 & 0.487 \\
            & $LBQR_{BIC}$        & 0.500 & 0.975 & 0.996 & 0.987 & 0.499 & 0.509 & 0.981 & 0.993 & 0.985 & 0.500 & 0.474 & 0.965 & 0.990 & 0.968 & 0.476 & 0.482 & 0.968 & 0.993 & 0.970 & 0.487 \\
            & $HSBQR_{SAVS}$      & 0.964 & 0.999 & 0.999 & 0.997 & 0.971 & 0.964 & 0.999 & 0.999 & 0.998 & 0.972 & 0.932 & 0.986 & 0.995 & 0.989 & 0.927 & 0.955 & 0.988 & 0.995 & 0.989 & 0.950 \\
            & $HSBQR_{BIC}$       & 0.951 & 0.986 & 0.991 & 0.985 & 0.956 & 0.955 & 0.989 & 0.993 & 0.990 & 0.954 & 0.968 & 0.984 & 0.994 & 0.986 & 0.965 & 0.983 & 0.988 & 0.994 & 0.989 & 0.980 \\ 
            \hline
\end{tabular}}
\caption{Hit Rate Results}
\label{tab:Hitrate}
\end{table}
\end{landscape}

With sparsification, one sees generally that the $BQR_{SAVS}$ and $BQR_{BIC}$ reduce bias nearly uniformly across DGPs. However, while the $LBQR$ and $HSBQR$ had similar bias in sparse DGPs, the $HSBQR_{BIC}$ has the lowest bias among all models, while trading blows with the $HSBQR_{SAVS}$ in the block DGPs. The $HSBQR$ coming ahead with sparsification indicates that signals are better identified through the horseshoe than the lasso prior.

The variable selection results in the first two columns of ($\ref{tab:MCC}$) and (\ref{tab:Hitrate}), mirror the bias results, 
while highlighting that the reason for the better performance from the $BQR_{BIC}$ models comes from lower false positive rates. This can be deduced from the fact that while the hit rates in table (\ref{tab:Hitrate}) are higher for the other models, the MCC are lower. This also holds for the SSVS prior. Hence, both the $SSVSBQR$ and $BQR_{SAVS}$ models tend to select too many noise variables. 

\subsubsection{Quantile specific Sparse DGPs}
Similar to the baseline DGPs, the bias tendencies carry mostly also over to the quantile specific simulations, with the difference that the $HSBQR$ does exceptionally well in the tails of the block DGPs. This can be understood from the fact that adaptive shrinkage plays an even larger role with quantile specific sparsity, particularly in the more extreme tails which are data sparse \citep{koenker2005}.    

Adding sparsification, the $BQR_{BIC}$ does better than $BQR_{SAVS}$ in sparse DGPs, however, similarly to the baseline results, does not offer a systematic improvement over the $BQR_{SAVS}$ in the block DGPs. This makes sense in the high-dimensional settings considered where capturing the quantile profile of the single covariate has a larger impact on fit in the sparse DGP than the block DGP, where many more covariates have non-zero coefficients. 

In terms of variable selection, we find that the continuous priors considered outperform the $SSVSBQR$ in sparse DGPs and the $HSBQR_{BIC}$ does especially well in the tails, consistently having the highest MCC. The results in the block design deviate from the baseline results: here we see that the $SSVSBQR$ and $BQR_{SAVS}$ models tend to do better than the $HSBQR_{BIC}$ in terms of the MCC, yet have similar or higher bias. Considering the hit rates, this difference indicates that the $BQR_{BIC}$ now offers lower penalisation in order to capture the quantile effect from the quantile specific variable which the tick-loss function heavily tilts the predicted quantile towards. Allowing for proper ALD correction by inlcuding $\overline{Z}_p$ as in (\ref{eq:intloss_adl}) for these DGPs, however, allows the $HSBQR_{BIC}$ to again outperform in the tails or come close to the best performer in terms of variable selection. Looking at the hit rates, it is suggestive that the $\overline{Z}_p$ correction implies some additional penalisation particularly in the tails which increases sparsity and decreases the amount of false positives from the sparse parts. Nevertheless, in terms of bias, the correction induces very small differences 
and vanishes with more data.


All in all, the simulations have shown that sparsification often significantly lowers bias, or at worst holds bias stable, while improving variable selection compared to the SSVS. The $BQR_{BIC}$ outperforms the $BQR_{SAVS}$, especially in sparse designs and in the tails where the extra flexibility of determining the penalisation from the data allows to capture the quantile profile. The results have also shown that although sparsification often helps the priors' performance, care should be taken in selecting the shrinkage prior apriori.

\section{Growth at Risk application} \label{sec: 4}

The initial motivation for growth-at-risk, as introduced by \citet{adrian2019vulnerable}, is to measure the downside risks, or vulnerabilities, to real GDP growth, which are captured as the response of forecasted conditional quantiles to changes in summary measures of financial conditions. Such models provide useful information for policymakers, as it has been empirically and theoretically shown that crises stemming from the financial sector carry the risk of creating negative feedback loops between the real economy and the financial system \citep{jorda2015leveraged}. However, as seen from recent macroeconomic developments, vulnerabilities to GDP growth may stem from various, possibly a priori unknown parts of the economy which entails that a larger information set is needed. Recent contributions to the GaR literature indeed conduct real time nowcasting exercises \citep{carriero2020nowcasting,ferrara2020high} and forecast density combination \citep{kohns2020horseshoe,korobilis2017quantile} using larger macro data sets in combination with shrinkage priors in order to deal with parameter proliferation. 

However, shrinkage priors, such as the very popular global-local priors, are continuous and are therefore hard to interpret in higher dimensions. For this application, similar to \citet{kohns2020horseshoe}, we are interested in creating quantile and combined density forecasts for real GDP growth based on the entire \citet{mccracken2020fred} data base. The logic of combining forecasts of multiple quantiles is to allow for the full heterogeneity of the data's effects to permeate to the conditional forecast distribution, which has been shown in the previous literature to often outperform models focused on only location effects \citep{korobilis2017quantile} \footnote{This step should also be natural from the standpoint that conditional quantiles characterise empirical CDFs which in turn provide direct links to the latent forecasted probability density function.}. However, unlike previous contributions, we apply the introduced sparsification methods to understand which variables drive the forecasting results in an effort to better communicate which subsets of the data set explain particular parts of the forecast distribution.


The \citet{mccracken2020fred} database\footnote{\url{https://research.stlouisfed.org/econ/mccracken/fred-databases/}}, including real GDP,  consists of 248 macroeconomic time series, starting from 1959Q1 at a quarterly frequency and is updated monthly. We take the quarter-on-quarter growth rate of annualized real GDP as our dependent variable and all others as independent covariates. These variables include a wide variety of macroeconomic effects which cover real, financial as well as national accounts data. Since not all series start at 1959Q1, for the growth at risk application, only variables that are available from 1970Q1 on wards were considered which gives 229 explanatory variables.

Predictions for our quantile methodology are generated from a direct forecasting model:
\begin{equation}
    y_{t+h} = x_t'\beta + \epsilon_{t+h}
\end{equation}

for $t=1,\cdots,T-h$, where h refers to the forecast horizon. We consider one- to four-quarter ahead forecast horizons ($h=1,\cdots,4$). Following \cite{yu2001bayesian}, the quantile specific predictive distribution can be obtained by marginalising out the uncertainty of the quantile regression parameter posterior $p(\beta_p|\bullet)$:

\begin{equation} \label{eq:quant_pred1}
    p(y_{t+h,p}|y_{1:t}) = \int p(y_{1:t}|\beta_p)p(\beta_p|y_{1:t})d\beta_p,
\end{equation}
where in (\ref{eq:quant_pred1}) $y_{1:t}$ refers to in-sample observations for $(y_1, \cdots, y_t)$. (\ref{eq:quant_pred1}) is conveniently approximated through Monte Carlo integration, as posterior draws $\beta_p^s$ are available from MCMC algorithms detailed in the appendix:

\begin{equation} \label{eq:quant_pred2}
    p(y_{t+h,p}|y_{1:t}) \approx \frac{1}{S}\sum_{s=1}^S \; p(y_{1:t}|\beta_p^s).
\end{equation}

Note that for each MCMC iteration $p(\beta_p|\bullet)$, $\beta_p^s$, we obtain a sparsified vector through the $BQR_{SAVS}$ and $BQR_{BIC}$ procedure respectively, which are used again in (\ref{eq:quant_pred2}) to generate separate predictive distributions. Predictive distributions are estimated for plain, $BQR_{SAVS}$ sparsified and $BQR_{BIC}$ sparsified posteriors of the $HSBQR$ and $LBQR$ as described in section (\ref{sec: SAVS}) and the $SSVSBQR$ 
serves as a benchmark for the large data models. To further contrast our results to the well established 'vulnerable growth' model of \citet{adrian2019vulnerable}, we estimate a Bayesian quantile regression with a fairly uninformative standard normal prior on the regression coefficients of only lagged real GDP and the NFCI index. 
This model is abbreviated in the tables below as $ABG_{BQR}$. Finally, to obtain a combined predictive density, we follow \citet{gaglianone2012constructing} and \citet{korobilis2017quantile} by estimating quantile predictive distributions for 19 equidistant quantiles, $p \in [0.05,0.10,\cdots, 0.90,0.95]$, sort the stacked $vec(S \times 19)$ vector, 
and smooth this stacked vector via a Gaussian kernel\footnote{This can be implemented for example via the "kdensity" function in Matlab}. Although stacking quantile predictive distributions will inevitably create some quantile crossing, the combination of multiple neighbouring quantiles helps in providing probability density in tail regions, which, due to the tick-loss function in the likelihood, are statistically data sparse relative to the median \citep{koenker2005}. 


To provide direct reference also to the original work of \cite{adrian2019vulnerable}, we construct quantile and density forecasts based on their two-step estimation procedure where conditional quantiles are smoothed via the skew-t-distribution \citep{azzalini2003distributions}.\footnote{Results for the skew-t big data models are reported in the appendix} Forecasts for all models are computed on an expanding window basis where the initial in-sample period uses the first 50 observations of the sample, which makes for 149-h forecast windows.\footnote{The ABG model contains 135-h rolling forecasts due to the availability of the NFCI index.}  


To evaluate forecasts, we employ both measures of overall density accuracy and quantile specific fit. For the overall density, we consider two strictly proper scoring rules in the sense of \citet{gneiting2007strictly}, the average log-predictive density score (LPDS) and average cumulative rank probability score (CRPS). Define $T_0$ as the initial in-sample period, taken to be (1970Q1-1982Q2), and T as the final forecast period (Q42019), then the LPDS and CRPS are calculated as follows:

\begin{equation}
    \begin{split}
        LPDS & = \frac{1}{T-T_0-h+1}\sum^{T-T_0-h+1}_{t=1} \; log \Big( \frac{1}{S}\sum^S_{t=1} \; p(y_{T_0+t+h-1}|y_{1:T_0+t}) \Big) \\
        CRPS & = \frac{1}{T-T_0-h+1}\sum^{T-T_0-h+1}_{t=1} \; |y_{T_0+t+h-1} - \hat{y}^A_{T_0+t+h-1}| - \frac{1}{2}|\hat{y}^A_{T_0+t+h-1}-\hat{y}^B_{T_0+t+h-1}|,
    \end{split}
\end{equation}
where $\hat{y}^A_{T_0+t+h-1}$ and $\hat{y}^B_{T_0+t+h-1}$ are two independent draws from the forecast distribution. To facilitate the discussion, the objective is to maximise the LPDS and minimise the CRPS. 
Since the LPDS and CRPS do not elucidate forecast calibration across different parts of the overall forecast distribution, we further make use of \cite{rossi2019alternative}'s calibration test. If a model is well calibrated, then the forecasted CDF of the probability integral transforms (PITs), $g_{t+h} = \int^{y_{t+h}}_0 = p(u|y_{t+h})du$, should be statistically indistinguishable from a 45 degree line \citep{diebold1998vevaluating}. Following \citet{rossi2019alternative}, we report confidence bands around the 45 degree line to account for sampling uncertainty \footnote{The confidence bands of \cite{rossi2019alternative} should be taken as general guidance, as strictly speaking, they are derived using a rolling window of estimation, while we use an expanding window. As recommended by \cite{rossi2019alternative}, uncertainty intervals for higher order forecasts ($h>1$), are computed using bootstrapping.}.

Lastly, as a measure of local fit, we make use of the quantile weighted CRPS (qwCRPS) to asses tail forecast performance. This metric is based on the quantile score (QS), as proposed by \cite{gneiting2011comparing} and is calculated as follows: let $\hat{y}_{t+h,p}$ denote the expected quantile for real GDP growth h steps ahead, then the QS is computed as: 
\begin{equation}
    QS_{t+h,p} = (y_{t+h}-\hat{y}_{t+h,p}) \times (p-\mathbb{I}(y_{t+h}-\hat{y}_{t+h,p})).
\end{equation}
The qwCRPS is thus given by: 
\begin{equation}
    qwCRPS_{t+h} = \int^1_0 \; w_p QS_{t+h,p}dp,
\end{equation}
where $w_p$ denotes a weighting scheme to evaluate specific parts of the forecast density. Since quantile regression is often used to capture downside risks, we use $w_p = (1-p)^2$, as suggested by \citet{gneiting2011comparing} for left tail evaluation.  
\newcolumntype{C}{>{\centering\arraybackslash}p{1.7cm}}

\begin{table}[t]
\centering
\resizebox{\textwidth}{!}{%
\begin{tabular}{r|CCCC|CCCC}
               & MSFE     & LPDS      & CRPS     & qwCRPS   & MSFE     & LPDS      & CRPS     & qwCRPS \\ \hline \hline
           & \multicolumn{4}{c|}{\textit{h=1}} & \multicolumn{4}{c}{\textit{h=2}} \\
$ABG_{BQR}$    & 0.504    & -11.607   & 0.313    & 1.049    & 0.404    & -7.870    & 0.254    & 0.891 \\
$ABG_{BQR-Skt}$& 0.519    & -1.727    & 0.301    & 1.026    & 0.426    & -4.102    & 0.277    & 0.970 \\
$SSVSBQR$      & 0.660*** & -0.955*** & 0.337*** & 1.492    & 0.433*** & -0.622**  & 0.235*** & 1.106 \\
$HSBQR$        & 0.533*   & -0.858*** & 0.296*** & 0.988*   & 0.411*** & -0.558**  & 0.222*** & 0.804* \\
$HSBQR_{BIC}$  & 0.519    & -0.895*** & 0.291*** & 0.948**  & 0.422*** & -1.064**  & 0.227*** & 0.794* \\
$HSBQR_{SAVS}$ & 0.536    & -0.976*** & 0.296*** & 0.968    & 0.468**  & -2.856*   & 0.254*** & 0.855 \\
$LBQR$         & 0.551*   & -4.603**  & 0.305*** & 1.091    & 0.441*** & -2.522*   & 0.238*** & 0.863 \\
$LBQR_{BIC}$   & 0.552    & -4.628**  & 0.305*** & 1.088    & 0.441*** & -2.517*   & 0.238*** & 0.863 \\
$LBQR_{SAVS}$  & 0.572    & -4.832*** & 0.316*** & 1.085    & 0.467*** & -3.006*   & 0.250*** & 0.859 \\ \hline
           & \multicolumn{4}{c|}{\textit{h=3}} & \multicolumn{4}{c}{\textit{h=4}} \\
$ABG_{BQR}$          & 0.378    & -6.825    & 0.237    & 0.847    & 0.351    & -4.387    & 0.235    & 0.844 \\
$ABG_{BQR-Skt}$ & 0.397   &-1.006	  & 0.249	 & 0.857    & 0.367	   & -1.252	   & 0.251	  & 0.875 \\
$SSVSBQR$      & 0.391*** & -0.797    & 0.219*** & 1.027    & 0.394*** & -0.796    & 0.221*** & 1.038 \\
$HSBQR$        & 0.390*** & -0.649    & 0.215*** & 0.833    & 0.376**  & -0.657    & 0.213*** & 0.773 \\
$HSBQR_{BIC}$  & 0.397*** & -1.589    & 0.219*** & 0.817    & 0.374**  & -0.933    & 0.213*** & 0.761 \\
$HSBQR_{SAVS}$ & 0.435*   & -2.913    & 0.237*** & 0.830    & 0.391    & -3.605    & 0.216*** & 0.757 \\
$LBQR$         & 0.422*** & -2.820    & 0.230*** & 0.872    & 0.382*** & -3.498    & 0.216*** & 0.809 \\
$LBQR_{BIC}$   & 0.423*** & -2.920    & 0.230*** & 0.870    & 0.383*** & -3.603    & 0.216*** & 0.809 \\
$LBQR_{SAVS}$  & 0.442**  & -4.323    & 0.239*** & 0.860    & 0.409**  & -4.949    & 0.230*** & 0.823 \\ \hline
\end{tabular}%
}
\caption{Forecast Evaluation Results}
\label{tab:Forecast Eval}
\end{table}

\subsection{Forecast Density Evaluation Results}
\begin{figure}[b!]
     
         \centering
         \includegraphics[width=\textwidth]{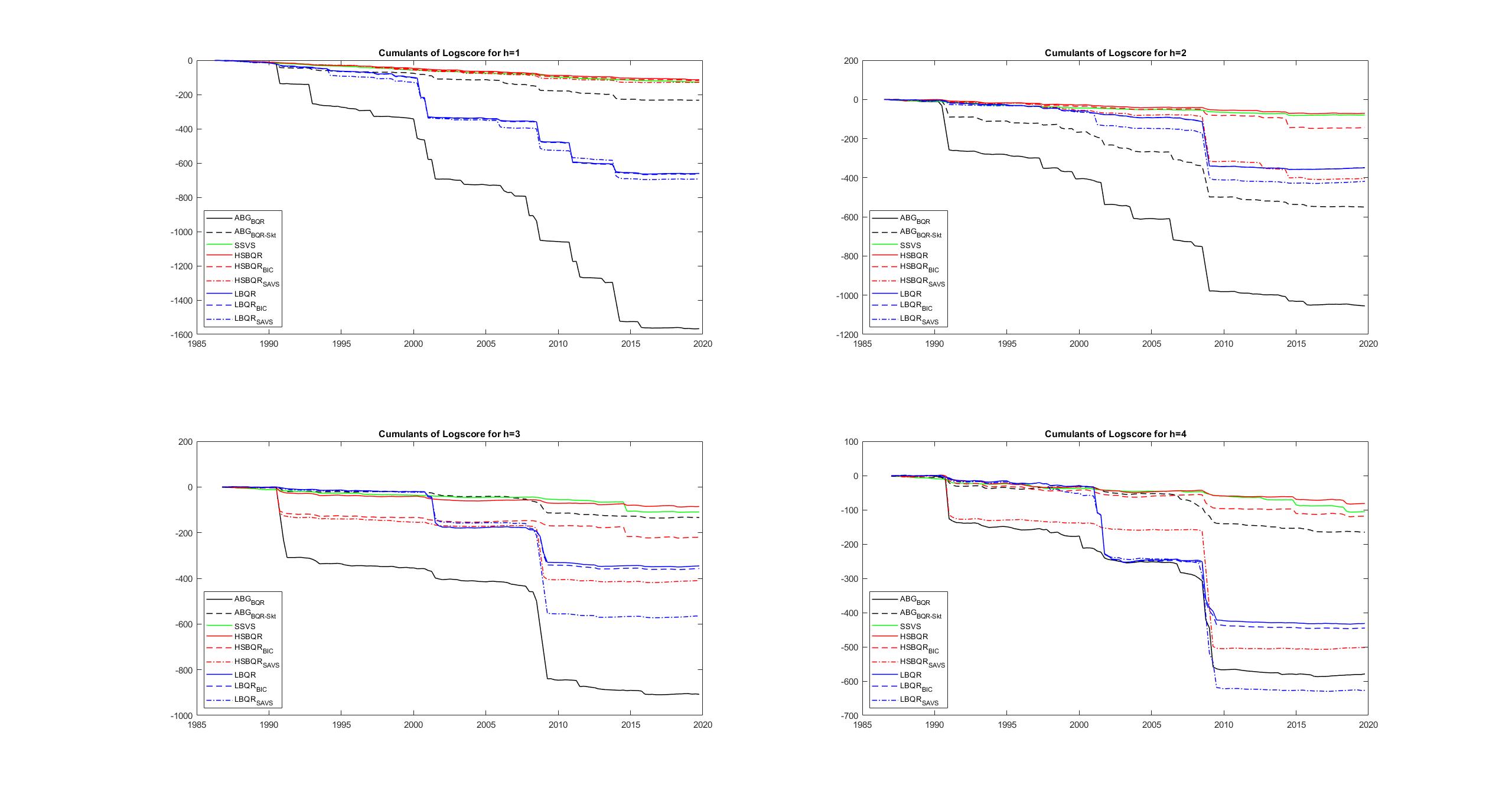}
     
     \caption{Cumulative log-scores.}
     \label{fig:Cum-LPDS}
\end{figure}

\begin{figure}[h]
     \centering
     \begin{subfigure}[b]{0.45\textwidth}
         \centering
         \includegraphics[width=\textwidth]{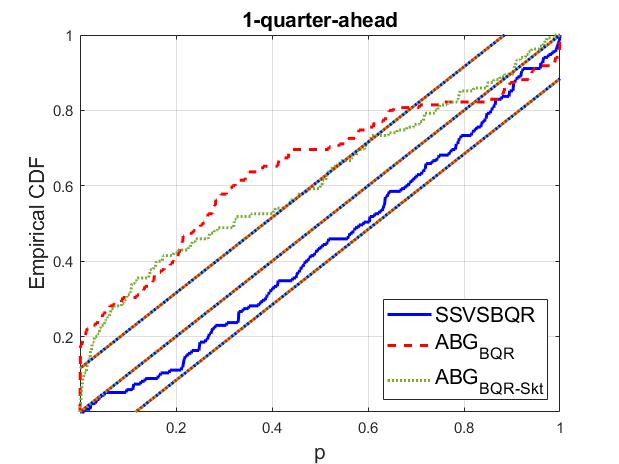}
     \end{subfigure}
     \begin{subfigure}[b]{0.45\textwidth}
         \centering
         \includegraphics[width=\textwidth]{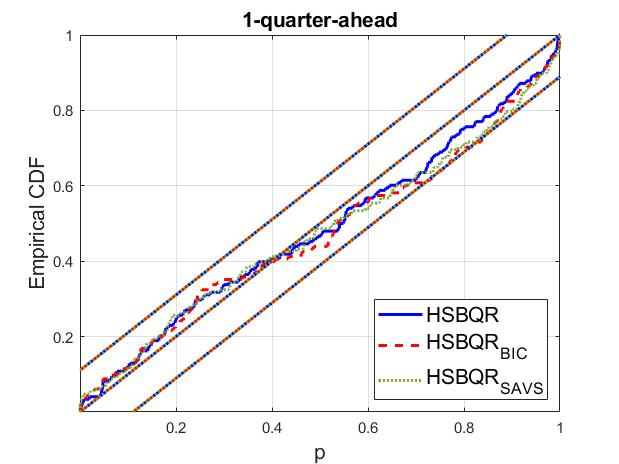}
     \end{subfigure}
      \centering
     \begin{subfigure}[b]{0.45\textwidth}
         \centering
         \includegraphics[width=\textwidth]{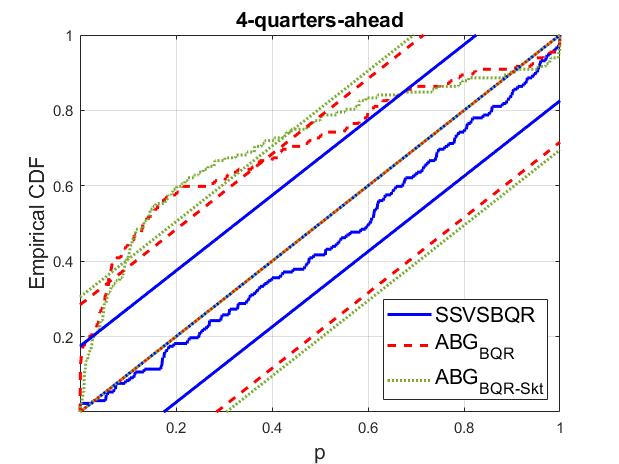}
     \end{subfigure}
      \centering
     \begin{subfigure}[b]{0.45\textwidth}
         \centering
         \includegraphics[width=\textwidth]{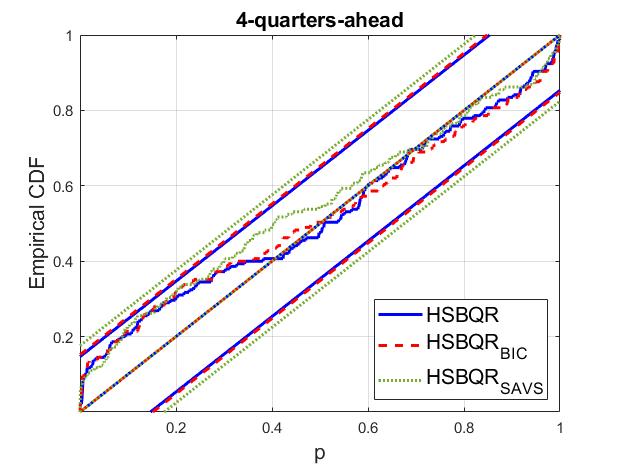}
     \end{subfigure}
     \caption{\citet{rossi2019alternative} calibration test.}
     \label{fig:Calibration}
\end{figure}

Point and density scoring results for all forecast horizons are reported in table (\ref{tab:Forecast Eval}), where stars indicate statistical significance as per the \cite{diebold1998vevaluating} test and calibration results are graphed in figure (\ref{fig:Calibration}). The density scores show generally that there is a clear benefit to big data, as average LPDS are often several order of magnitudes higher, and CRPS significantly lower, especially for the $HSBQR$ compared to either of the ABG models. The cumulative log-scores over time in graph (\ref{fig:Cum-LPDS}) highlight that the vulnerable growth models underperform particularly during the financial crisis where the realisations of GDP growth seem to fall far into the tails of their forecast distributions, thereby being heavily penalised \footnote{ The large performance discrepancy compared to the CRPS, therefore makes sense given that the LPDS penalises low mass event very heavily.} \footnote{Fitting the skew-t distribution to the ABG model, markedly improves overall density scoring results and calibration at the one-quarter-ahead forecast horizon, where gains are largest for the middle to left parts of the density, as seen in figure (\ref{fig:Calibration}). This gives an indication that the low-dimensional information set cannot provide enough information between quantiles, for which the skew-t provides extra spread. This in particular aids log-score penalisation. We find that applying the skew-t to the big-data models, doesn't not change performance qualitatively vis-a-vis ABG models. The table with the skew-t results are provided in the appendix.}. In line with the simulation results, we find that the $HSBQR$ outperforms both the $SSVSBQR$ and $LBQR$, and that the $BQR_{BIC}$ is clearly preferred for sparsification compared to the $BQR_{SAVS}$, where $BQR_{BIC}$ sparsification often increases performance, as gauged by the CRPS, or only slightly worsens it as per the LPDS. Also similar to the simulations, the quality of sparsification is again dependent on the quality of shrinkage. This can be seen from the generally worse density scores of the $LBQR$ models. Interestingly, for all models considered, forecast accuracy increases with the forecast horizon. This may stem the fact that changes in parts of the economy only affect aggregate output with a lag. Similar to much of the previous macro forecasting literature \citep{stock1998comparison,banerjee2006there,edge2010comparison} we also find that in terms of point predictions, simple autoregressive models such as the ABG models are hard to beat.

In terms of capturing downside risks, the qwCRPS results show that the $HSBQR_{BIC}$ consistently produces the best left tail performance, and significantly outperforms the ABG models at one-quarter and two-quarters ahead. The $BQR_{BIC}$ sparsification procedure in fact increases tail performance both for the $HSBQR$ and $LBQR$.

Calibration based on PITs are able to shed more light on the differing performances. Figure (\ref{fig:Calibration}) shows the models' PIT CDFs with confidence intervals computed as suggested by \citet{rossi2019alternative}. Corroborating the good density scores of the $HSBQR$ variants and the $SSVSBQR$, figure (\ref{fig:Calibration}) shows that these models are also well calibrated, especially in the left tail at all horizons. The ABG models in turn tend to underestimate the lower tail which helps explain the bad log-score performance during the financial crisis in figure (\ref{fig:Cum-LPDS}). Among the big-data models, the general tendency from the calibration results is that the $SSVSBQR$ slightly overpredicts the left tail and underpredicts the right tail, while the $LBQR$ tends to underpredict both tails. The $HSBQR$ in turn falls somewhere in the middle. In line with the density scores, sparsification via the $BQR_{BIC}$ is able to preserve calibration, while the $BQR_{SAVS}$ tends to overshrink. This is especially prevalent in the left to middle quantiles of the $HSBQR_{SAVS}$ at one-year ahead predictions, where more observations fall below those quantiles.



\subsection{Quantile specific variable inclusion}
The results from the density and calibration tests have shown that the $BQR_{BIC}$ variable selection procedure is able to preserve the fit of the unsparsified forecast distributions, and even improve left tail accuracy, which gives credibility to the fact that most salient features of the models have been captured despite sparsification. In order to analyse variable importance both over time and across quantiles, we provide in figures (\ref{fig:HSBQR_quanthmap}), (\ref{fig:SAVS_quanthmap}), (\ref{fig:Other_quanthmap}), and (\ref{fig:timevarying}) heatmaps of variable inclusion probability for each prior and sparsification variant. While figure (\ref{fig:HSBQR_quanthmap}), (\ref{fig:SAVS_quanthmap}), and (\ref{fig:Other_quanthmap}) give a general static picture of variable selection across quantiles, averaged over time, figure (\ref{fig:timevarying}) shows variable inclusion probabilities for left tail (average of $5^{th}$-$15^{th}$ quantiles), right tail  (average of $85^{th}$ to $95^{th}$ quantiles) as well as the middle regions (average of $45^{th}$-$55^{th}$ quantiles) across time to showcase time variation in variable selection. Since it is of interest to understand differences in variable selection compared to widely used conditional mean models, we present at the bottom of figures (\ref{fig:HSBQR_quanthmap}) and (\ref{fig:Other_quanthmap}) variable inclusion probabilities with each of the priors considered for the Bayesian quantile regression, however applied to a normal observational likelihood\footnote{Details on the conditional mean models for the different priors is provided in the appendix.}. To make the graphs more easily interpretable, we present the covariates on the horizontal axis in terms of general variable groupings given by the FRED-QD data base. For example, series in the NIPA group refer to "National Income and Product Account" information that spans information such as aggregate consumption, investment and government expenditures. Exact variable definitions for each group can be found in the appendix of \citet{mccracken2020fred}.

\subsubsection{Variable Inclusion Across Quantiles}

\begin{figure}
    \begin{subfigure}[t]{\textwidth}
        \centering
        \includegraphics[width=\linewidth]{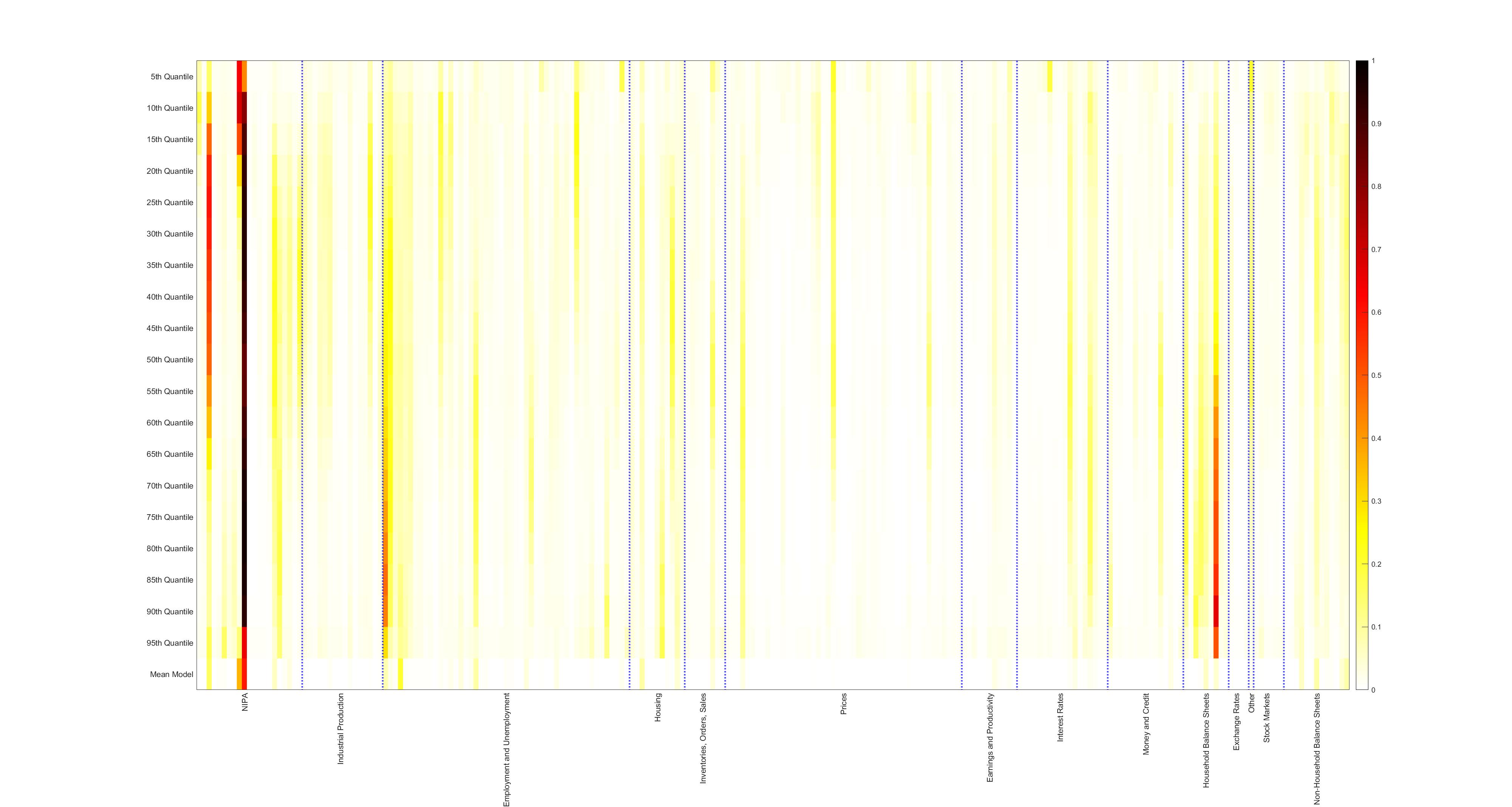}
    \end{subfigure}
    \vfill
    \begin{subfigure}[t]{\textwidth}
        \centering
        \includegraphics[width=\linewidth]{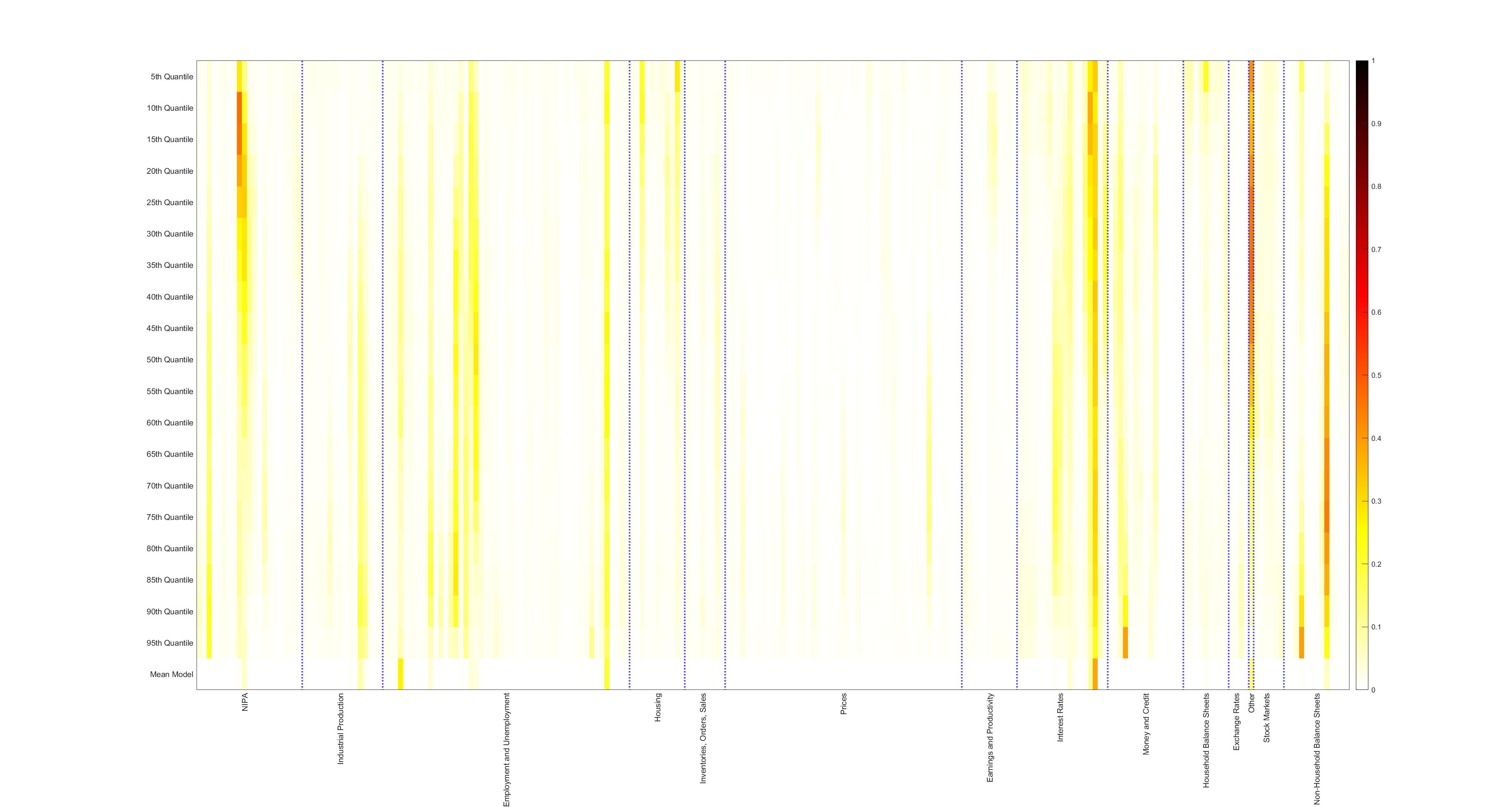}
    \end{subfigure}
    \caption{Average inclusion probability for each quantile of the $HSBQR_{BIC}$ for h=1(top) and h=4 (bottom). HS with SAVS for mean model included for reference}
    \label{fig:HSBQR_quanthmap}
\end{figure}

\begin{figure}
    \begin{subfigure}[t]{\textwidth}
        \centering
        \includegraphics[width=\linewidth]{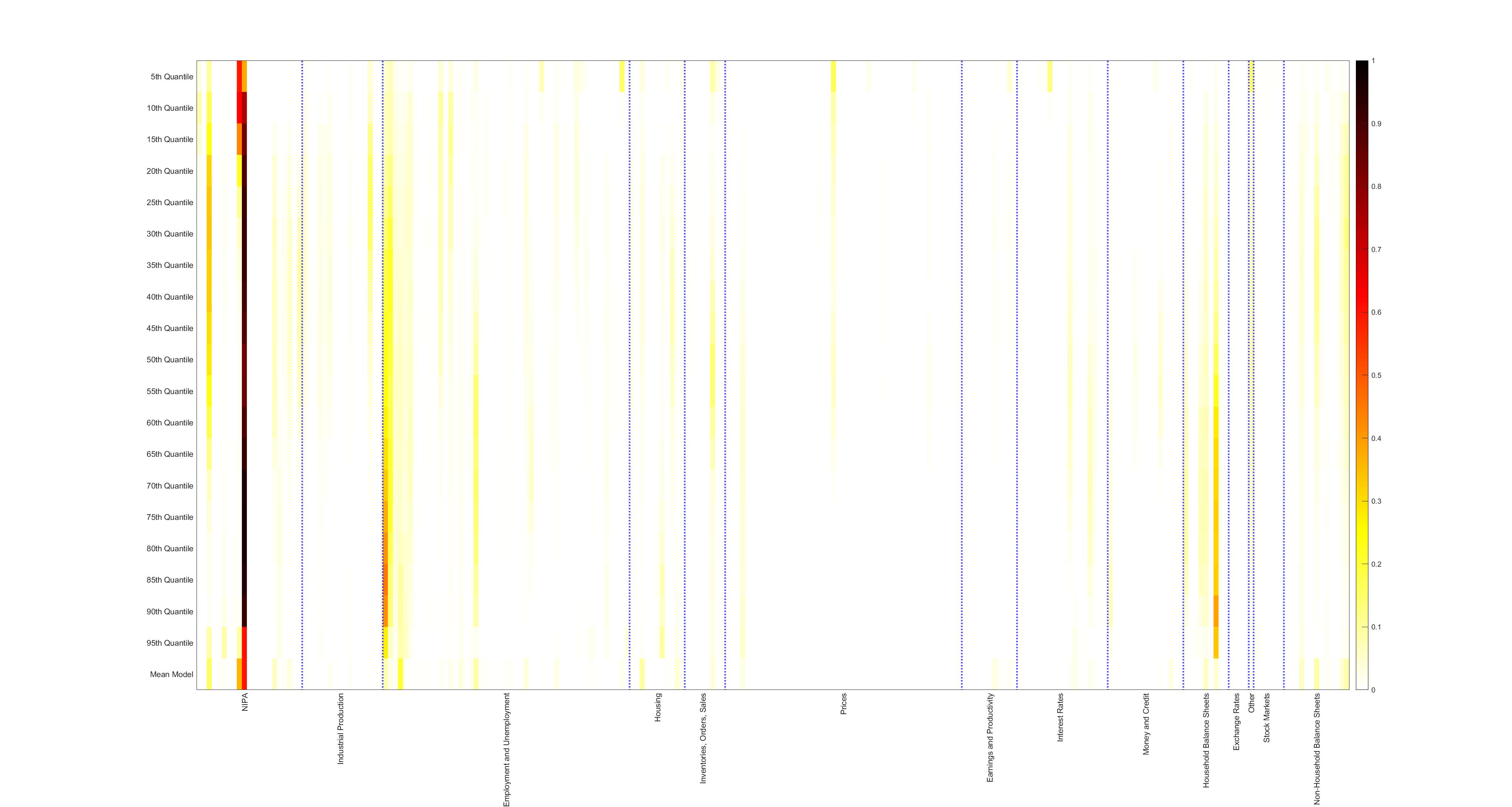}
    \end{subfigure}
    \vfill
    \begin{subfigure}[t]{\textwidth}
        \centering
        \includegraphics[width=\linewidth]{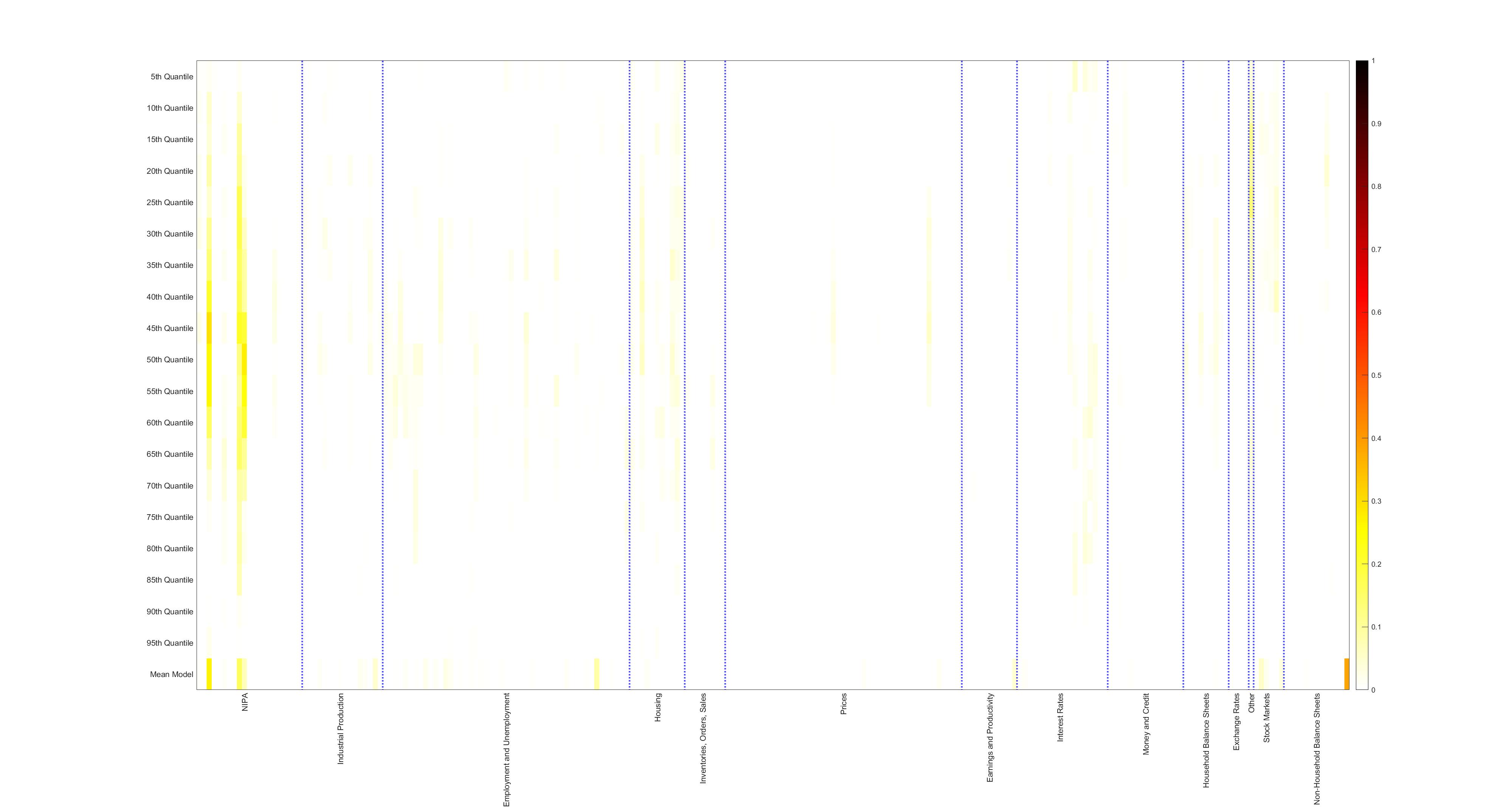}
    \end{subfigure}
    \caption{Average inclusion probability for each quantile of the $HSBQR_{SAVS}$ (top) and the $LBQR_{SAVS}$ (bottom) for h=1}
    \label{fig:SAVS_quanthmap}
\end{figure}

\begin{figure}
    \begin{subfigure}[t]{\textwidth}
        \centering
        \includegraphics[width=\linewidth]{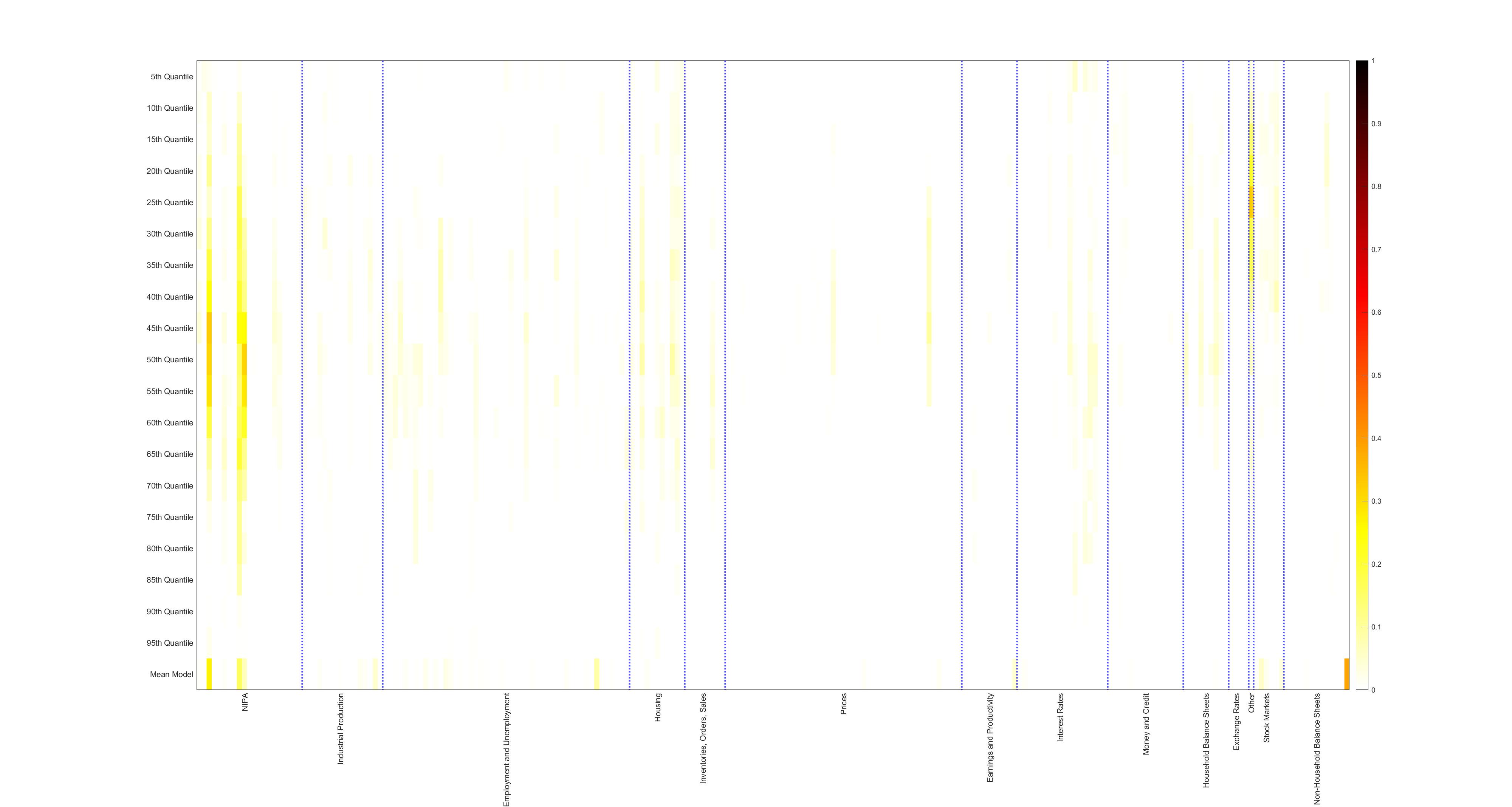}
    \end{subfigure}
    \vfill
    \begin{subfigure}[t]{\textwidth}
        \centering
        \includegraphics[width=\linewidth]{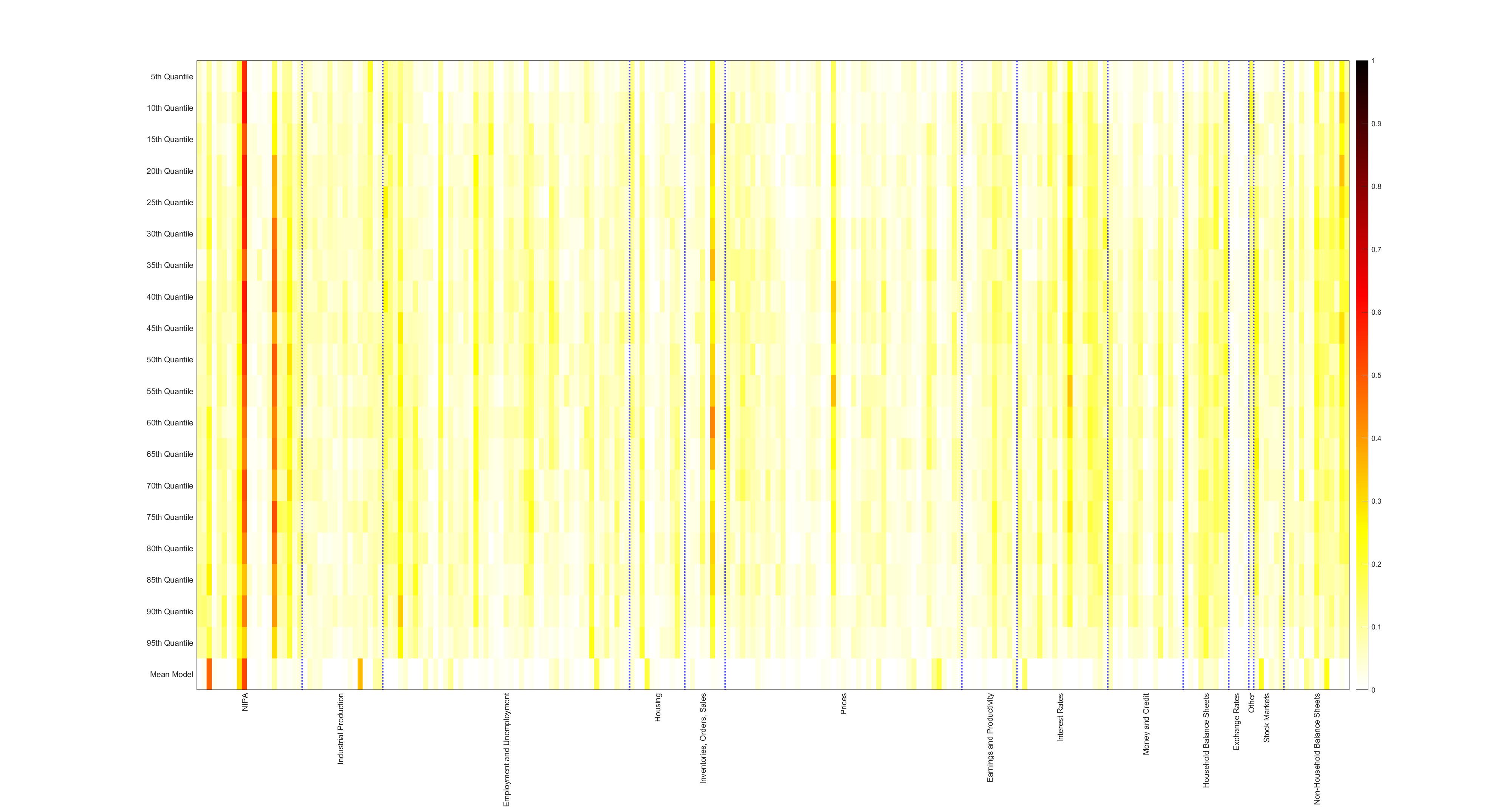}
    \end{subfigure}
    \caption{Average inclusion probability for each quantile of the $LBQR_{BIC}$ (top) and the $SSVSBQR$ (bottom) for h=1}
    \label{fig:Other_quanthmap}
\end{figure}

For all quantile models under consideration, figures (\ref{fig:HSBQR_quanthmap}) - (\ref{fig:Other_quanthmap}) clearly show that there is substantial quantile varying sparsity. For instance, one can see for the $HSBQR_{BIC}$ model in figure (\ref{fig:HSBQR_quanthmap}) at one-quarter ahead, that "Employment and Unemployment" data and "Household Balance Sheets" information are more important for the right tail, while playing only a marginal role in the left tail. At one-year ahead forecasts, variable importance shifts in the left tail toward "Interest Rates" and "Consumer Sentiment", while the right tail has higher inclusion probability for "Money and Credit" and "Non-Household Balance Sheets". The former information can be interpreted as important sectors that predict both the 2001 as well as great financial crisis. As expected, middle quantiles draw information from either tails, displaying variable inclusion patterns which overlap both with the left and right tail.\footnote{ Note that variable groups overlapping need not signify that the fitted quantiles contain the same information since coefficients may vary across quantiles.} 

A commonality among, especially, the one-quarter ahead forecasts is the high inclusion probability of NIPA variables at all quantiles. Since the NIPA variable group contains national account information which gets aggregated via the national account identity to GDP, it is natural to interpret this variable category as capturing information inherent in lagged GDP. In fact taking the first principal component of the four most included NIPA variables by the $HSBQR_{BIC}$ yields a series that is highly correlated with GDP growth
. The fact that these series are included for the majority of the quantiles presents a departure from the previous literature such as \citet{adrian2019vulnerable} who find that current economic conditions, as captured by lagged GDP only forecast the median of the conditional distribution.


\begin{figure}
    \begin{subfigure}[t]{\textwidth}
        \centering
        \includegraphics[width=\linewidth]{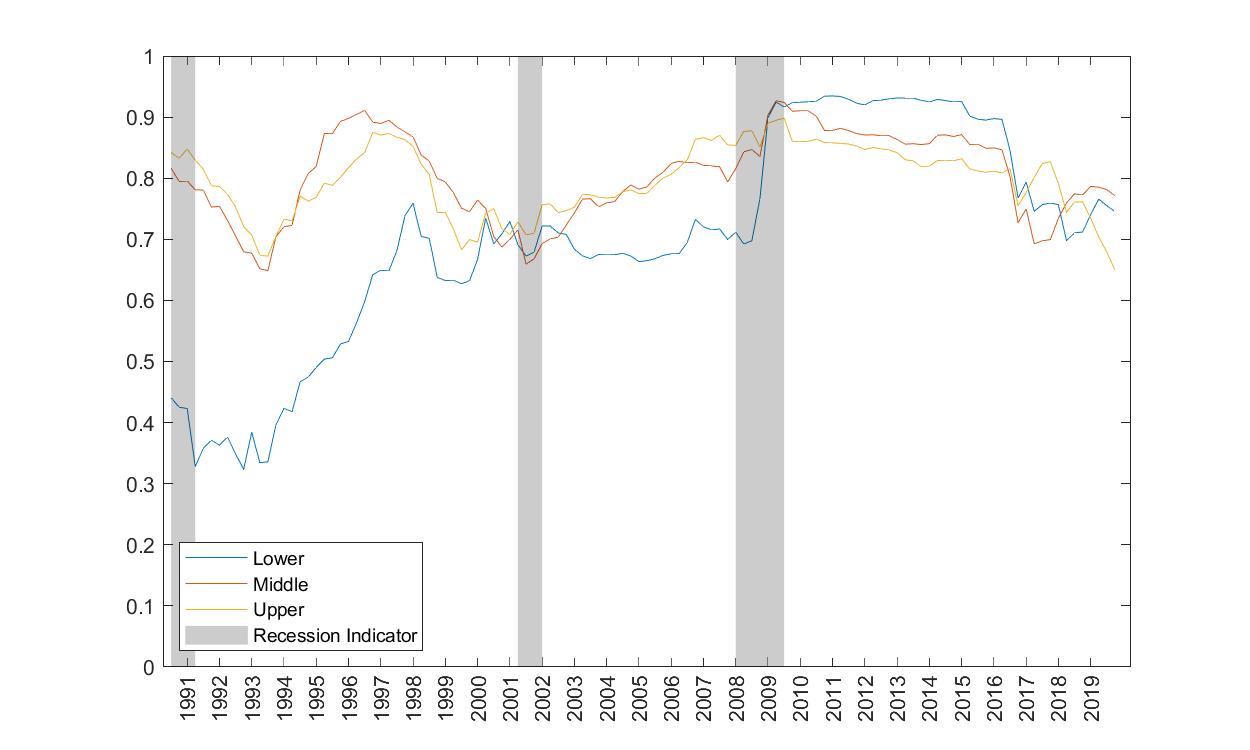}
    \end{subfigure}
    \vfill
    \begin{subfigure}[t]{\textwidth}
        \centering
        \includegraphics[width=\linewidth]{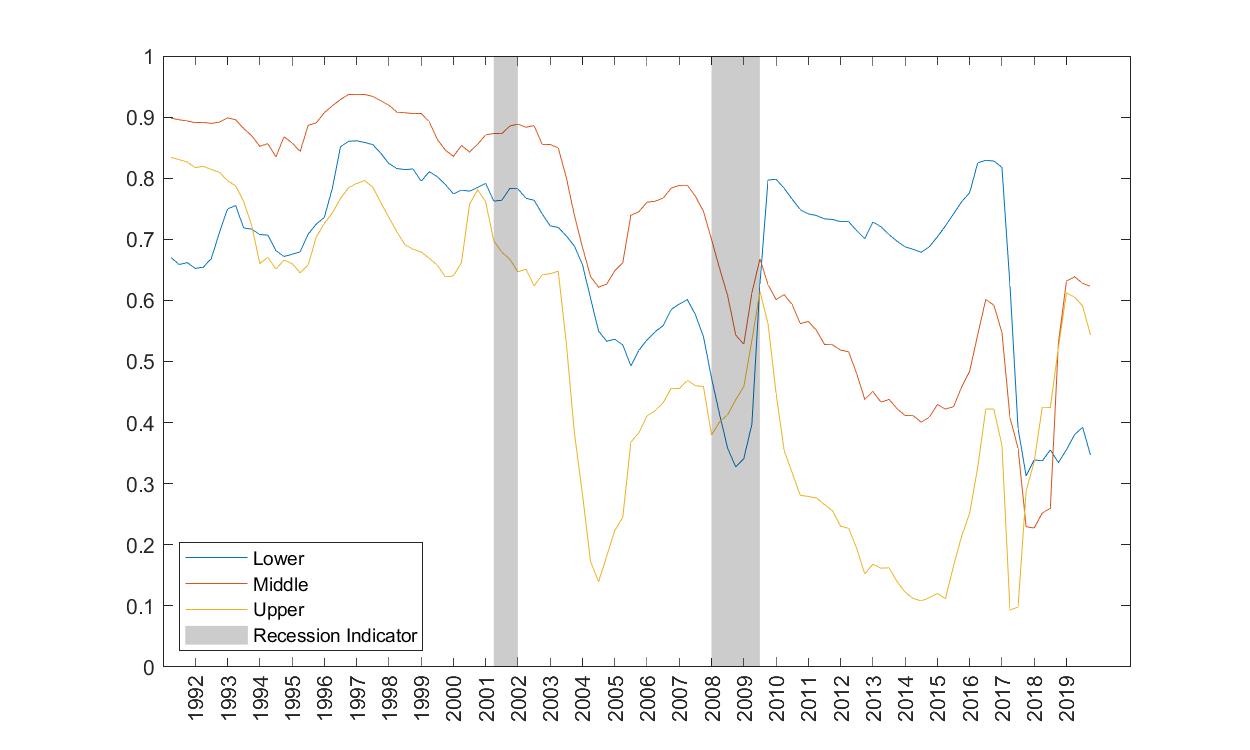}
    \end{subfigure}
    \caption{Correlation coefficients computed on a rolling 30 quarter window}
    \label{fig:corrplot_HSBQR}
\end{figure}

Contrasting these results to conditional mean models, we find that these tend to select variable groups which have high inclusion probability not only in the median, but also the tails. To underscore this, figure (\ref{fig:corrplot_HSBQR}) plots the correlation coefficient between left, right and middle quantiles of the $HSBQR_{BIC}$ in a rolling 30 quarter window with point forecasts of the conditional mean horseshoe prior model with SAVS sparsification. While this figure is not a conclusive test of which quantiles influence the mean the most, it does highlight that correlation is highest with the median and right tail during tranquil times, while in recession periods, as indicated by grey bars, correlation with the lower quantiles increases, hence signifying that information spills over from the left tail to the mean. This figure, also highlights that there are larger differences in correlation between the upper, middle and lower quantiles in the one-year ahead forecasts after the 2001 crisis compared to the one-quarter ahead forecasts. This may reflect the fact that interest rate and credit information propagate only with a lag to aggregated economic activity \citep{romer2018advanced} and therefore are increasingly picked up in higher order forecasts after periods in which monetary policy had changed a lot to counteract the recession. Lastly, in line with the recent macroeconomic forecasting literature such as \citet{giannone2017economic}, we also find that there is evidence for model uncertainty which can be deduced from the amount of yellow shaded areas at all forecasting horizons.

Comparing variable selection among priors and sparsification techniques, some further interesting patterns emerge. First, the $BQR_{SAVS}$ algorithm with fixed penalisation induces much sparser models than under the more adaptive $BQR_{BIC}$ penalisation for both continuous shrinkage priors, and in doing so, eliminates near all model uncertainty, as seen by the large white spaces in figure (\ref{fig:SAVS_quanthmap}). Notably though, variables which are identified as strong signals by the $HSBQR_{BIC}$, tend to also appear with higher inclusion probability in the $HSBQR_{SAVS}$. As seen in the previous section, this heavy sparsification, however, comes at a large penalty on density scores and calibration, hurting particularly tail forecasting performance, as underscored in figure (\ref{fig:Cum-LPDS}) around the financial crisis. Second, in line with the simulation results, we see also in the empirical application that the $LBQR$ tends in general to exert stronger shrinkage than the $HSBQR$, resulting in far sparser models than under the horseshoe prior, while the $SSVSBQR$ tends to select larger models (lower panel of figure \ref{fig:Other_quanthmap}). This is corroborated by table (\ref{tab:modelsize}) which shows average model sizes of selected estimators. Lastly, the $SSVSBQR$ prior also induces far greater variable inclusion uncertainty, as the heatmap shows denser yellow regions compared to the $HSBQR_{BIC}$. Nevertheless, variables with the highest inclusion probabilities tend to overlap with those selected by the $HSBQR_{BIC}$.


\begin{table}[]
\centering
\begin{tabular}{r|ccc|ccc}
         & Left   & Mid    & Right  & Left   & Mid    & Right \\ \hline \hline
         & \multicolumn{3}{c|}{h=1} & \multicolumn{3}{c}{h=4} \\ 
$SSVSBQR$  & 13.023 & 19.195 & 14.668 & 9.063  & 13.024 & 8.505 \\
$HSBQR_{BIC}$ & 9.003  & 9.640  & 8.202  & 7.474  & 7.314  & 7.148 \\
$LBQR_{BIC}$  & 1.541  & 3.055  & 1.107  & 1.439  & 2.748  & 1.119 \\ \hline
\end{tabular}
\caption{Average Model sizes of the different models}
\label{tab:modelsize}
\end{table}

The fact that the $SSVSBQR$ prior increases model uncertainty without outperforming the $HSBQR$ and $HSBQR_{BIC}$ might indicate that the discretisation of the model-space along with the less flexible slab distribution employed by the $SSVSBQR$ prior induces higher uncertainty over variable inclusion, especially with correlated data. We leave exploration of whether global-local priors can illicit different answers to the 'illusion of sparsity' found by \citet{giannone2017economic} for future investigation.    


\subsubsection{Variable Inclusion Across Time}



\begin{figure}
    \begin{subfigure}[b]{0.5\textwidth}
        \centering
        \includegraphics[width=\textwidth]{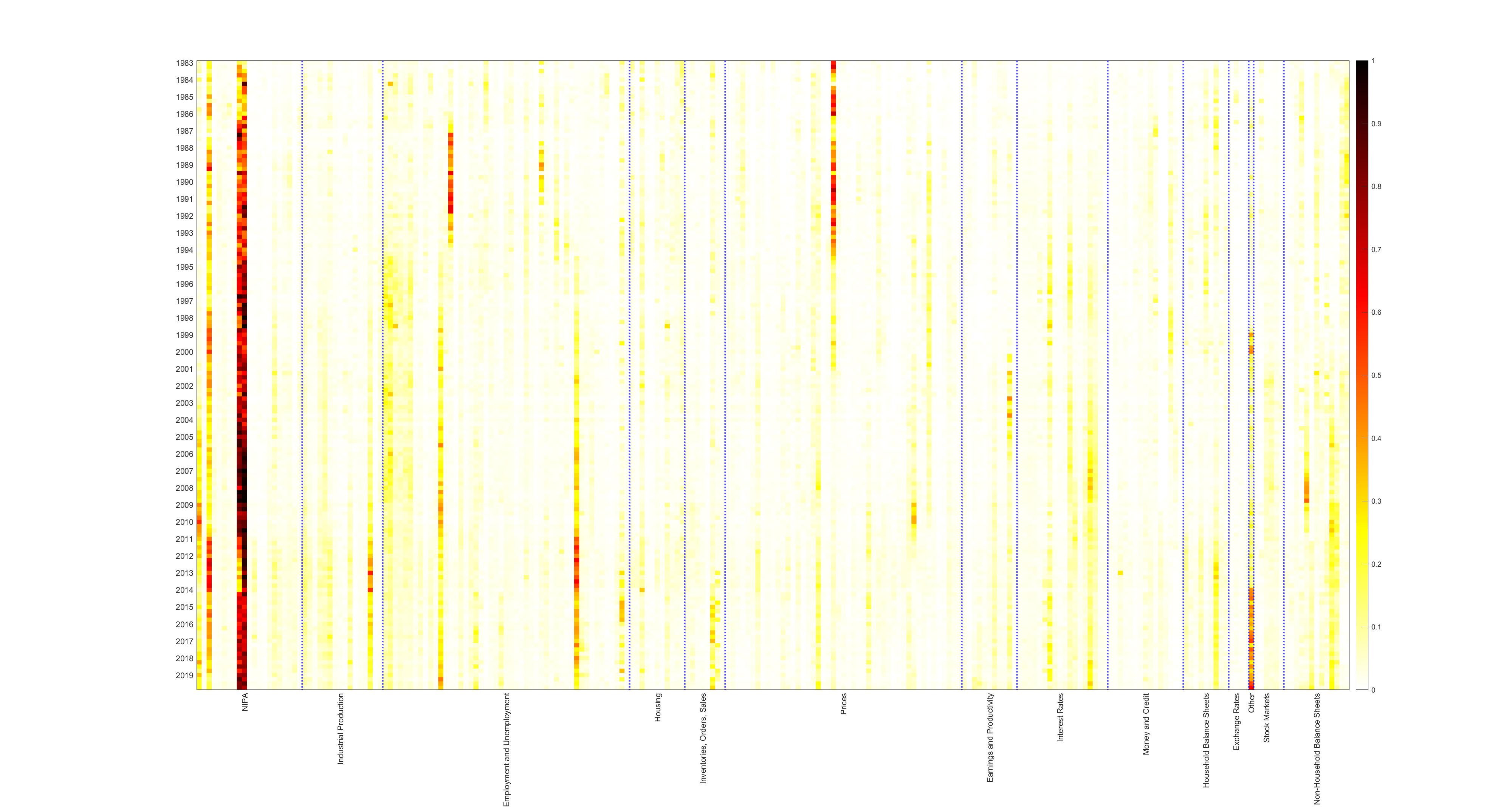}
    \end{subfigure}
    \begin{subfigure}[b]{0.5\textwidth}
        \centering
        \includegraphics[width=\textwidth]{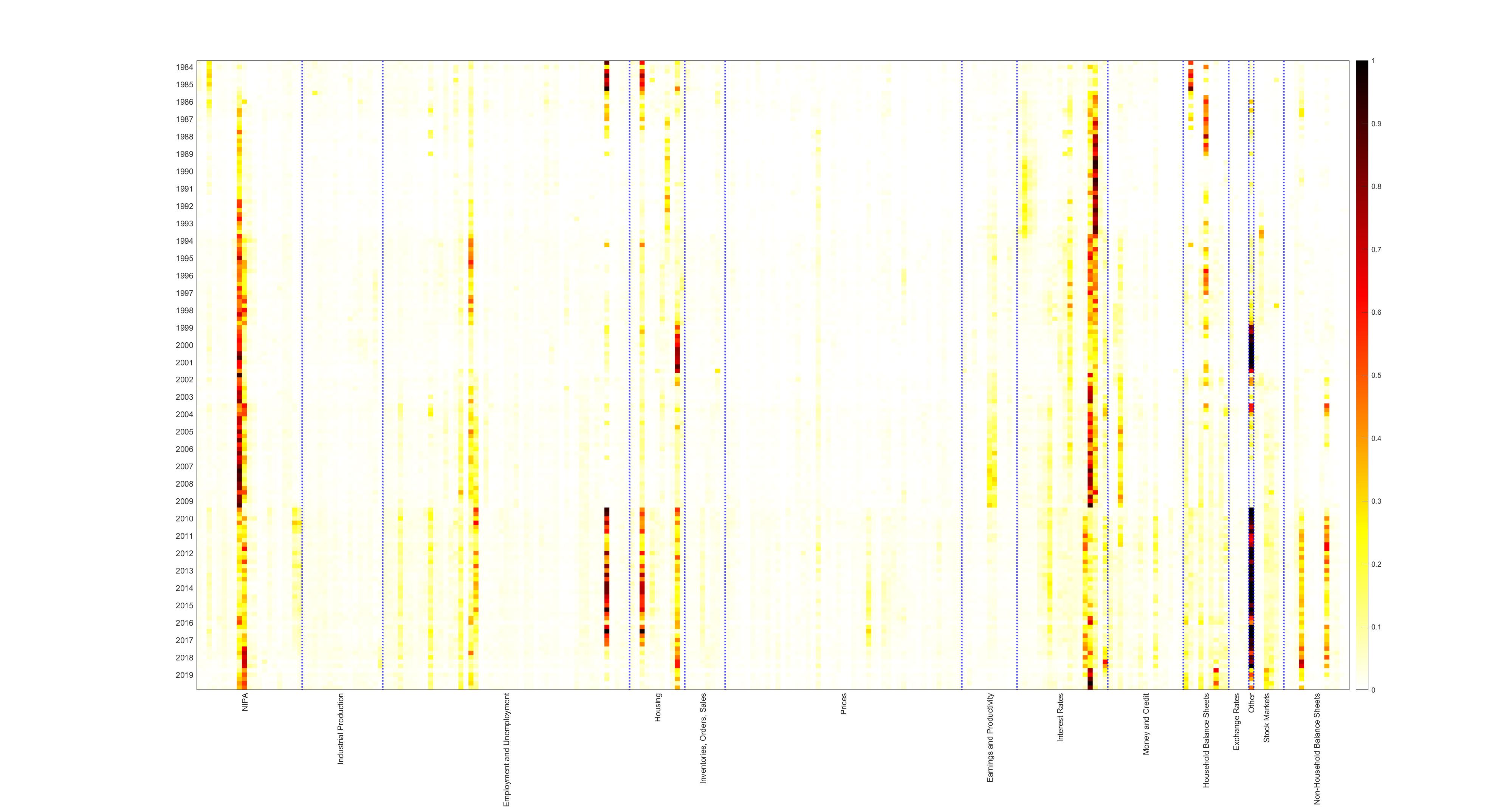}
    \end{subfigure}
    
    \begin{subfigure}[b]{0.5\textwidth}
        \centering
        \includegraphics[width=\linewidth]{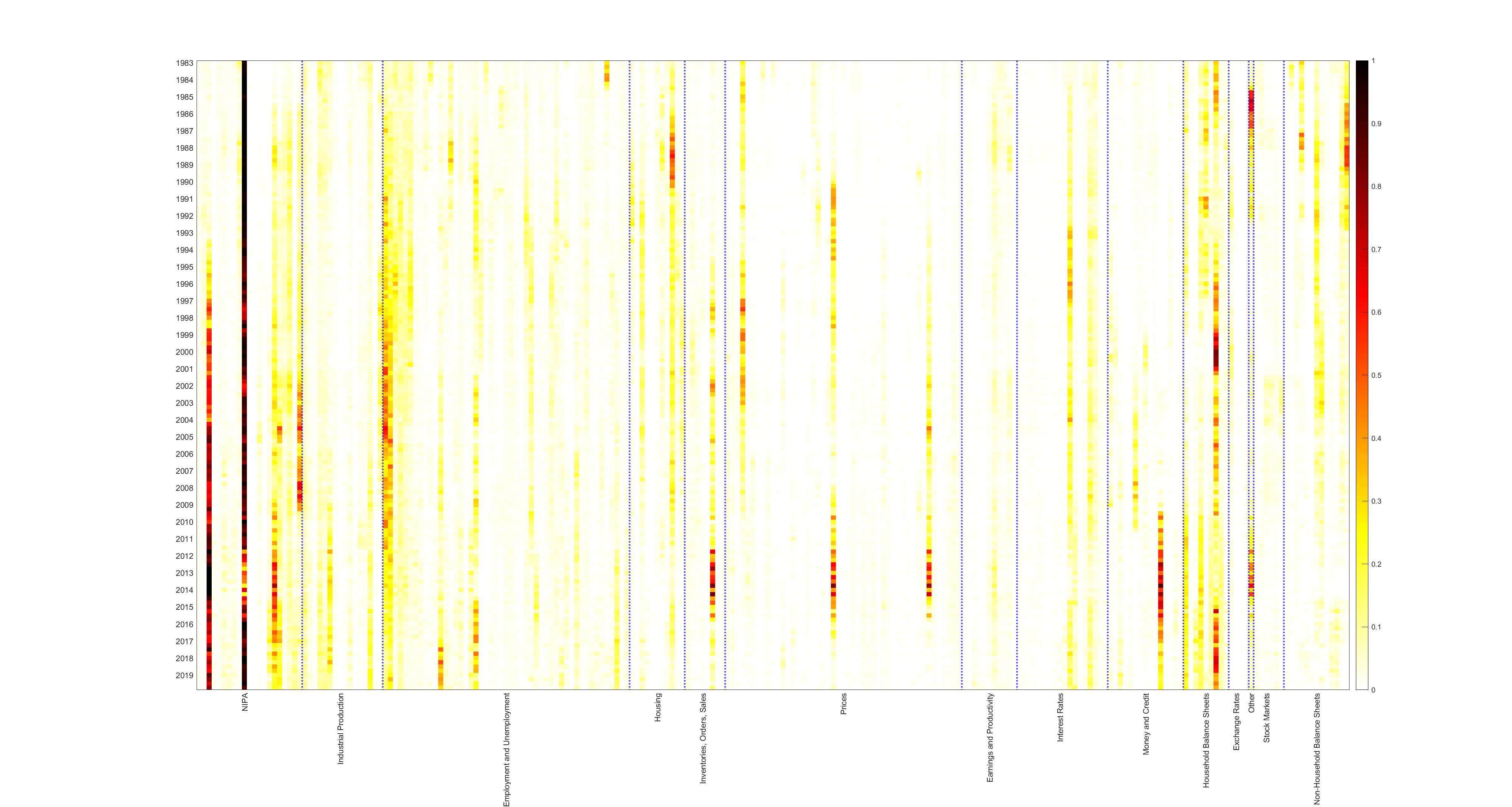}
    \end{subfigure}
    \begin{subfigure}[b]{0.5\textwidth}
        \centering
        \includegraphics[width=\linewidth]{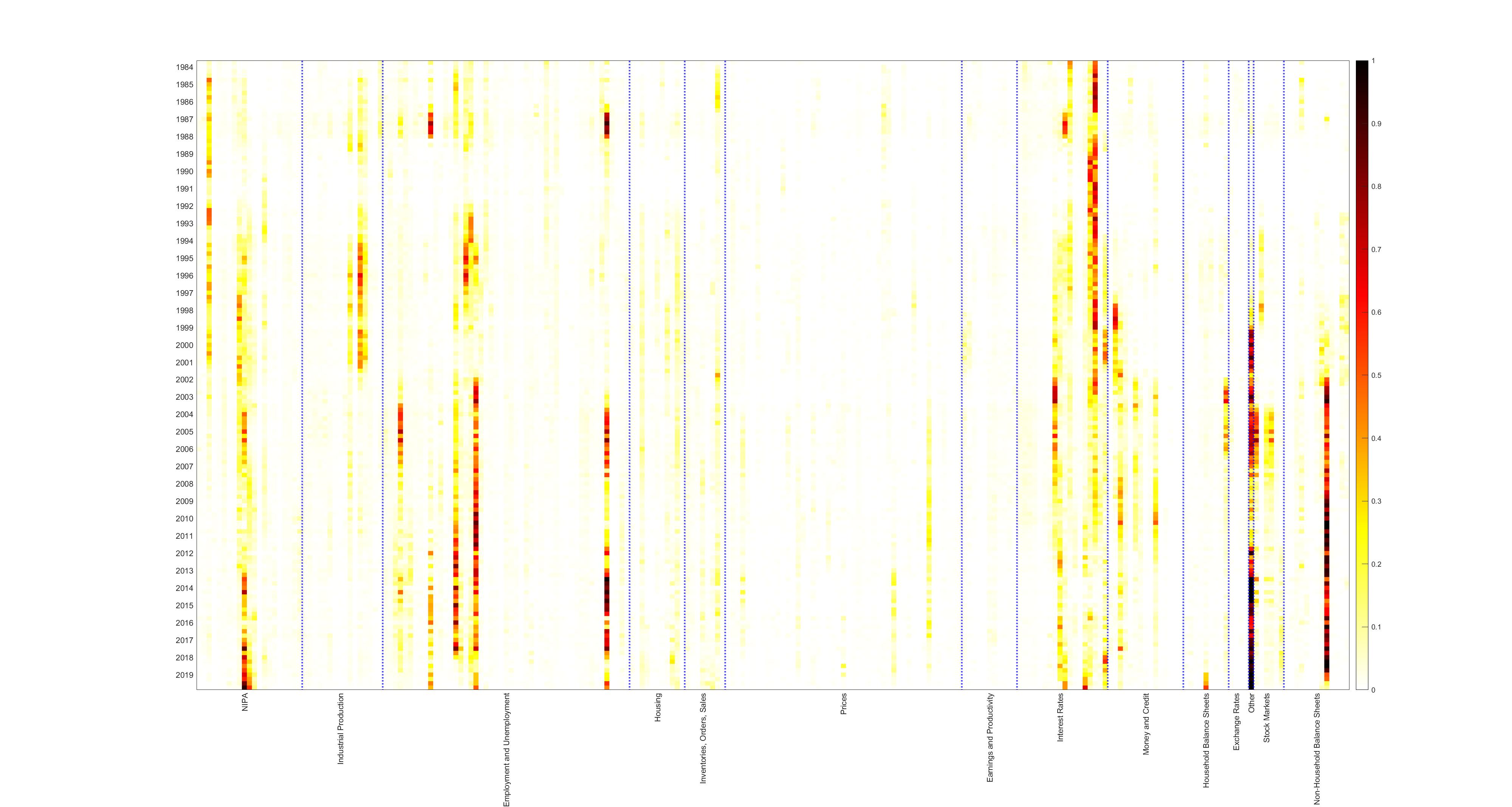}
    \end{subfigure}
    
    \begin{subfigure}[b]{0.5\textwidth}
        \centering
        \includegraphics[width=\textwidth]{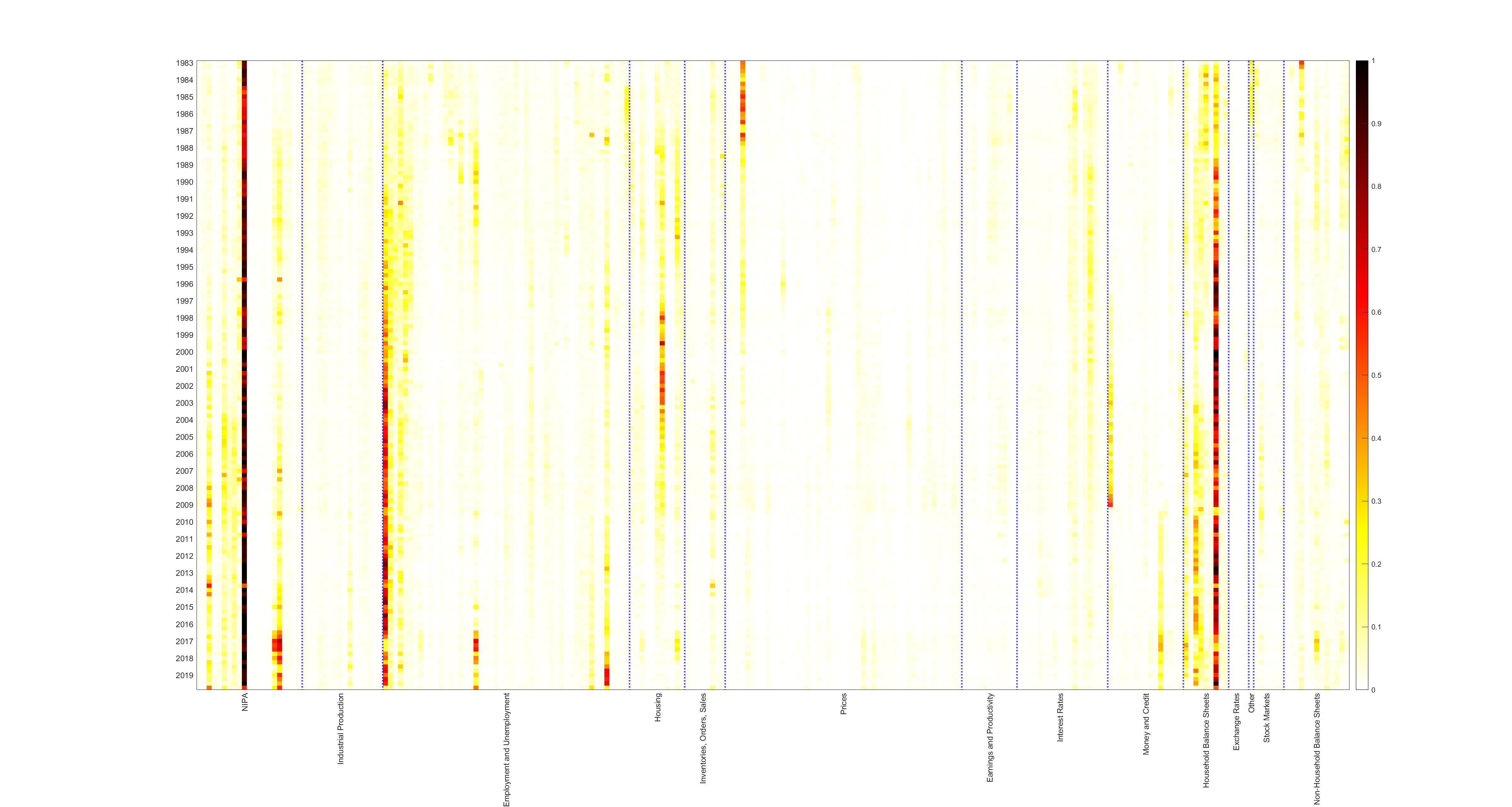}
    \end{subfigure}
    \begin{subfigure}[b]{0.5\textwidth}
        \centering
        \includegraphics[width=\textwidth]{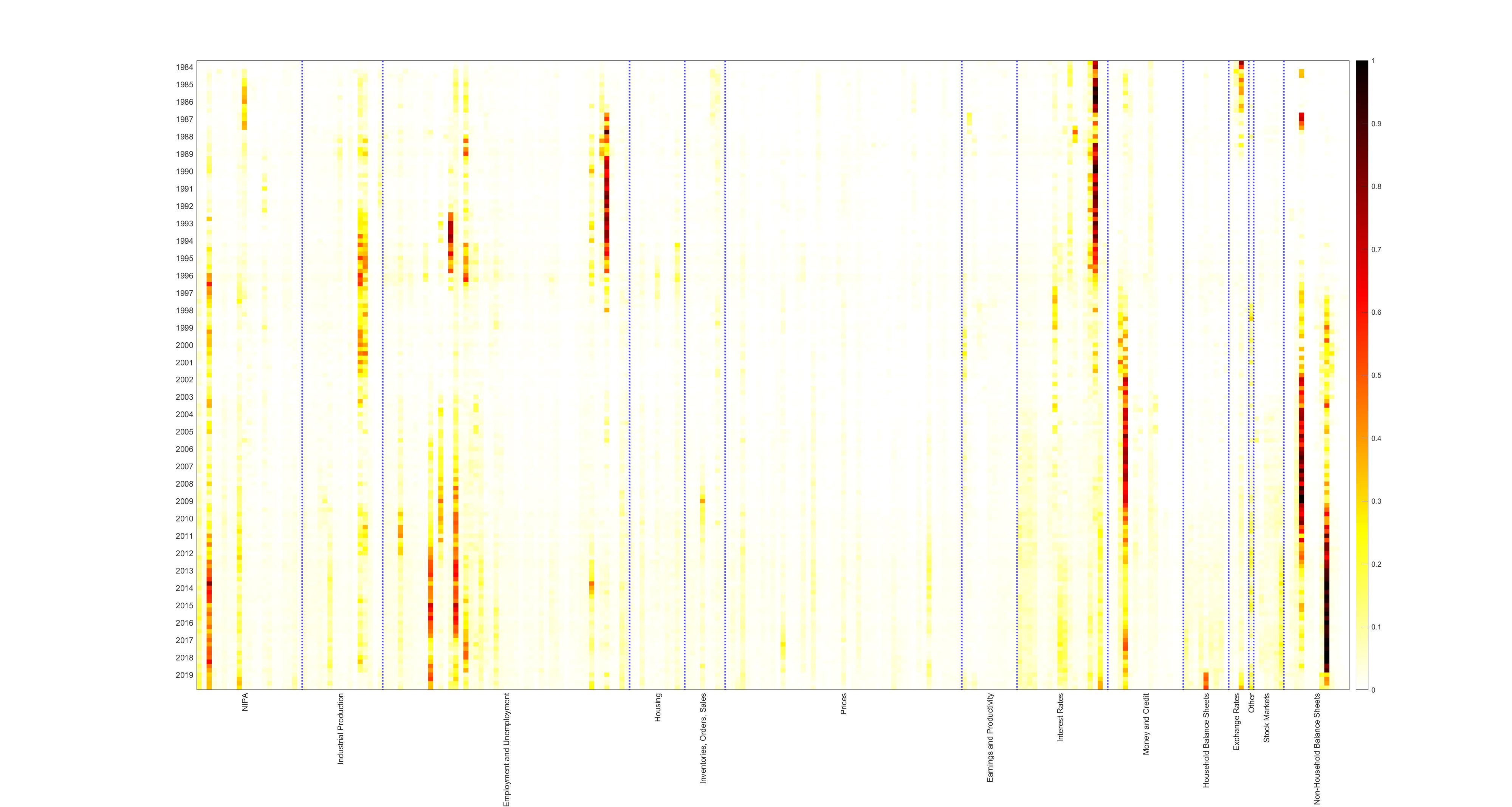}
    \end{subfigure}
    \caption{Average inclusion probability each year for the $HSBQR_{BIC}$ for its 3 left quantiles (top), 3 central quantiles (middle), and 3 right quantiles (bottom) for h=1 (left) and h=4 (right)}
    \label{fig:timevarying}
\end{figure}

\begin{figure}
    \begin{subfigure}[b]{0.5\textwidth}
        \centering
        \includegraphics[width=\textwidth]{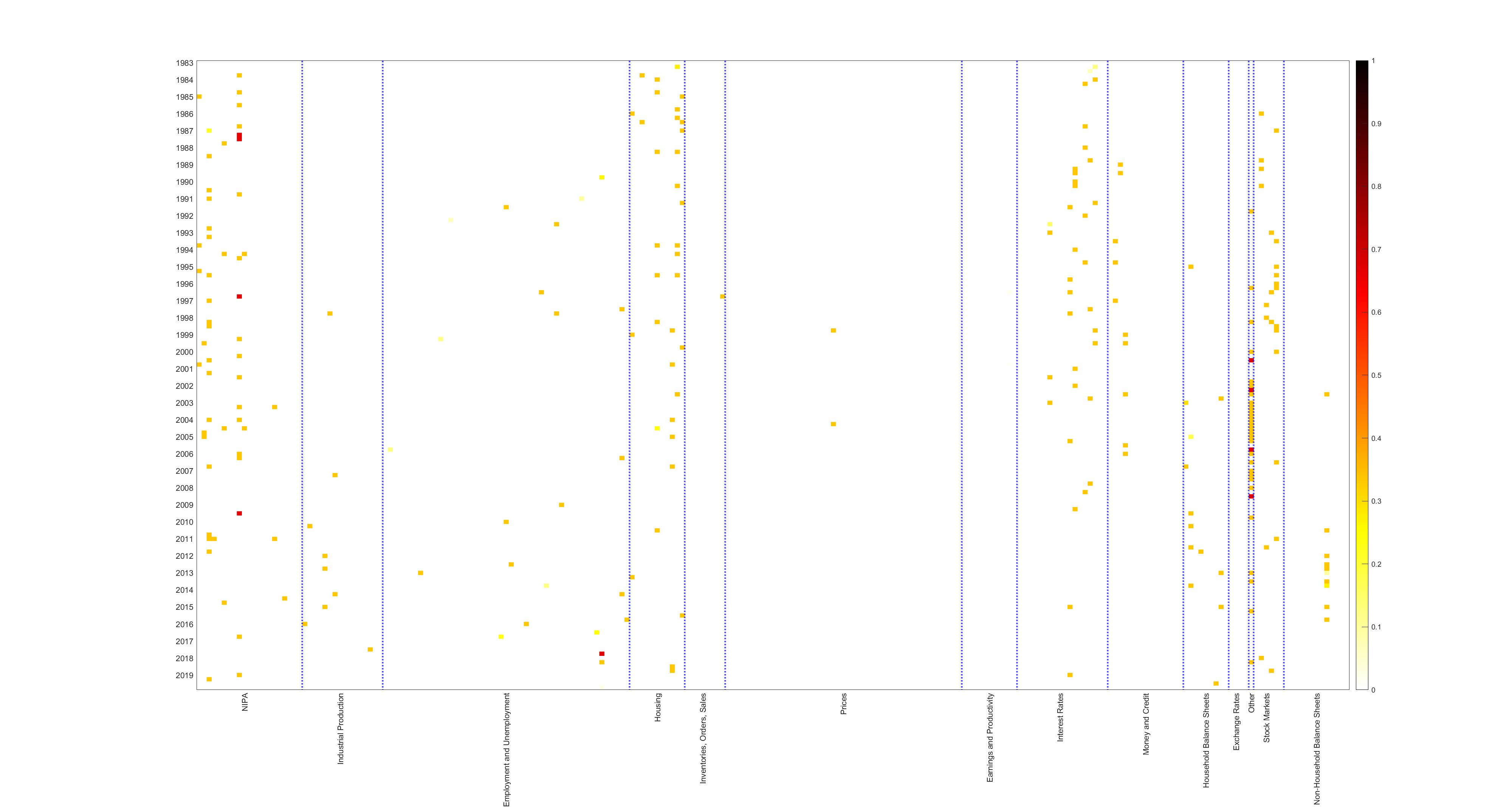}
    \end{subfigure}
    \begin{subfigure}[b]{0.5\textwidth}
        \centering
        \includegraphics[width=\textwidth]{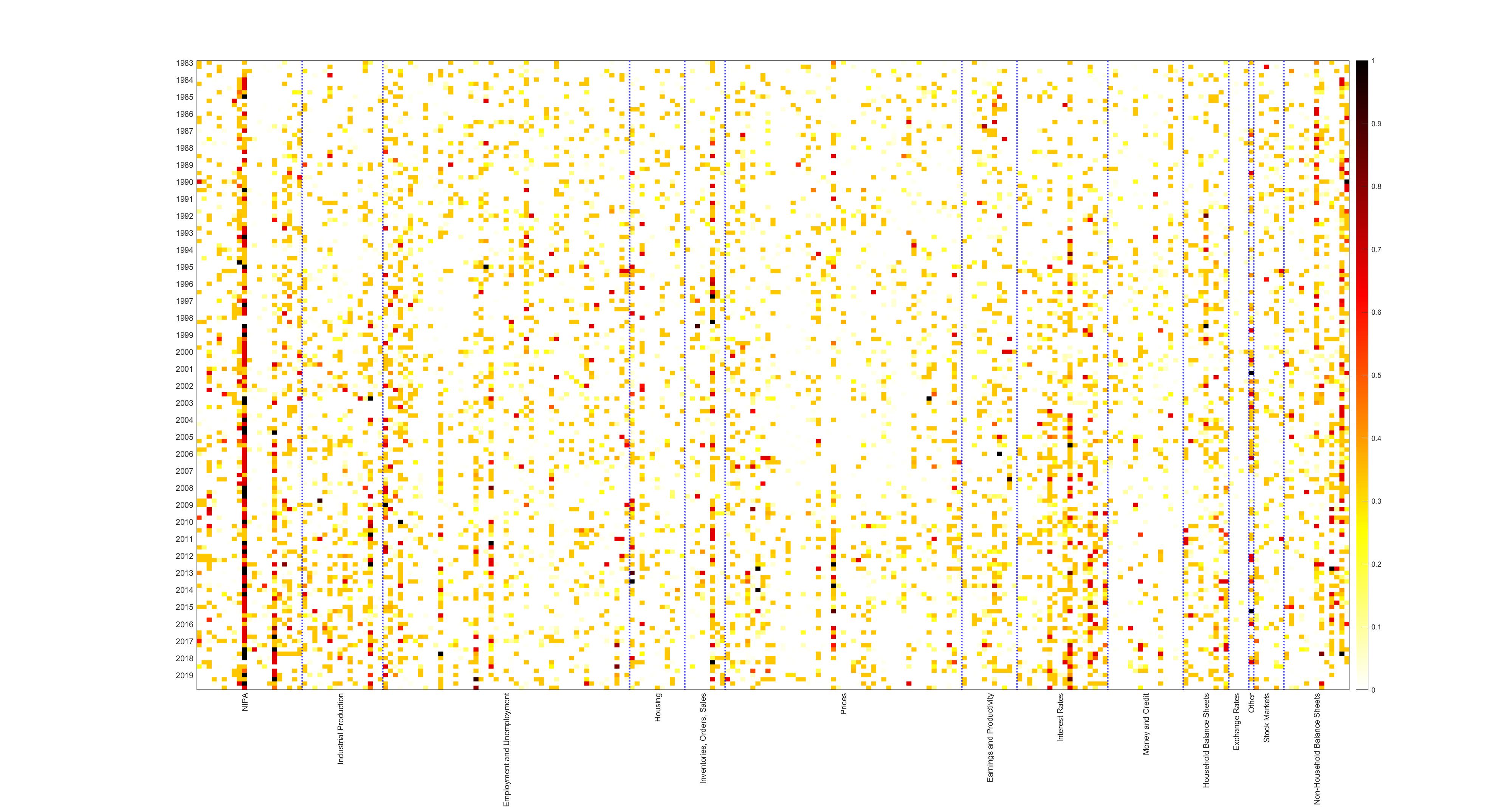}
    \end{subfigure}
    
    \begin{subfigure}[b]{0.5\textwidth}
        \centering
        \includegraphics[width=\linewidth]{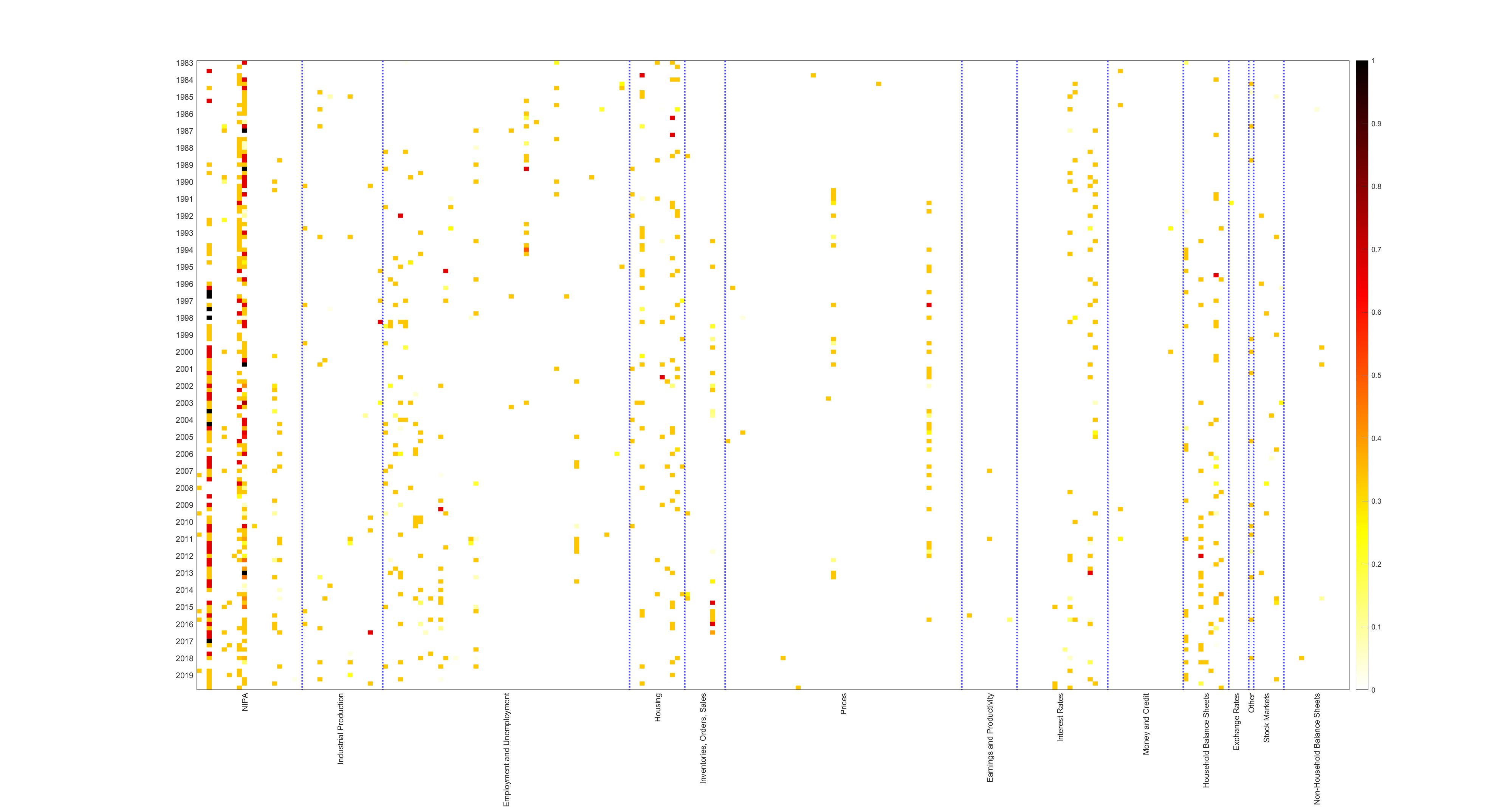}
    \end{subfigure}
    \begin{subfigure}[b]{0.5\textwidth}
        \centering
        \includegraphics[width=\linewidth]{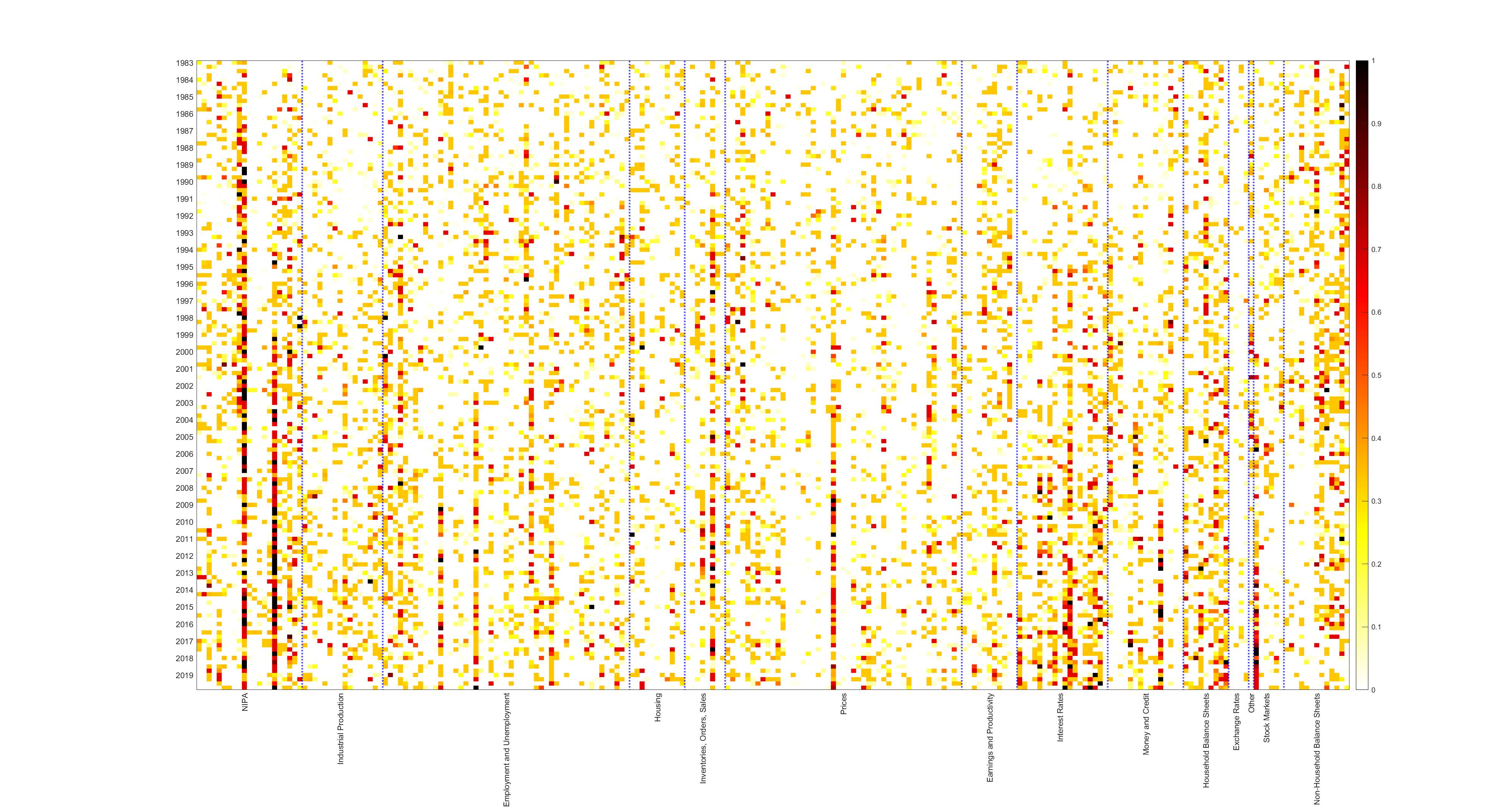}
    \end{subfigure}
    
    \begin{subfigure}[b]{0.5\textwidth}
        \centering
        \includegraphics[width=\textwidth]{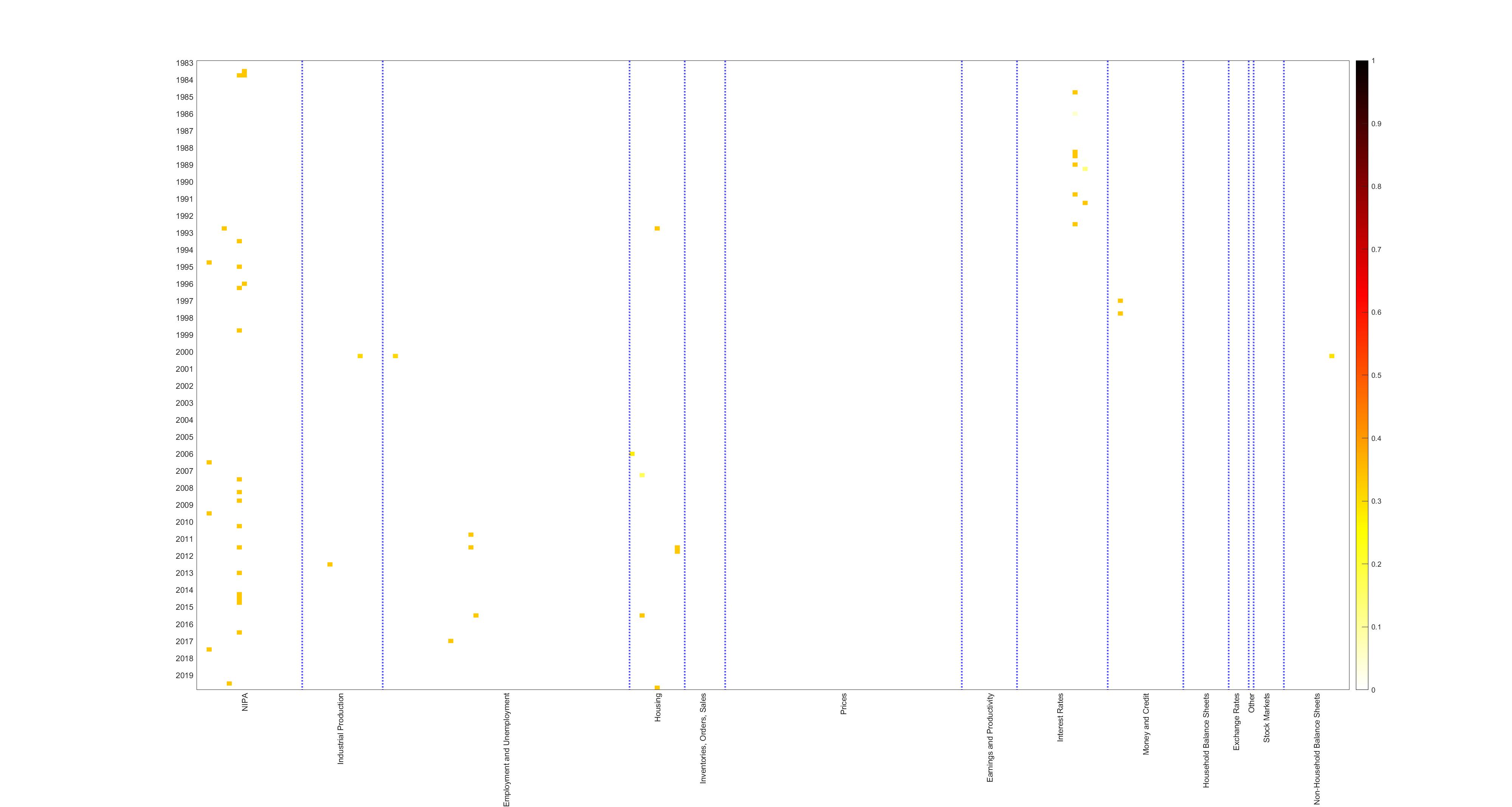}
    \end{subfigure}
    \begin{subfigure}[b]{0.5\textwidth}
        \centering
        \includegraphics[width=\textwidth]{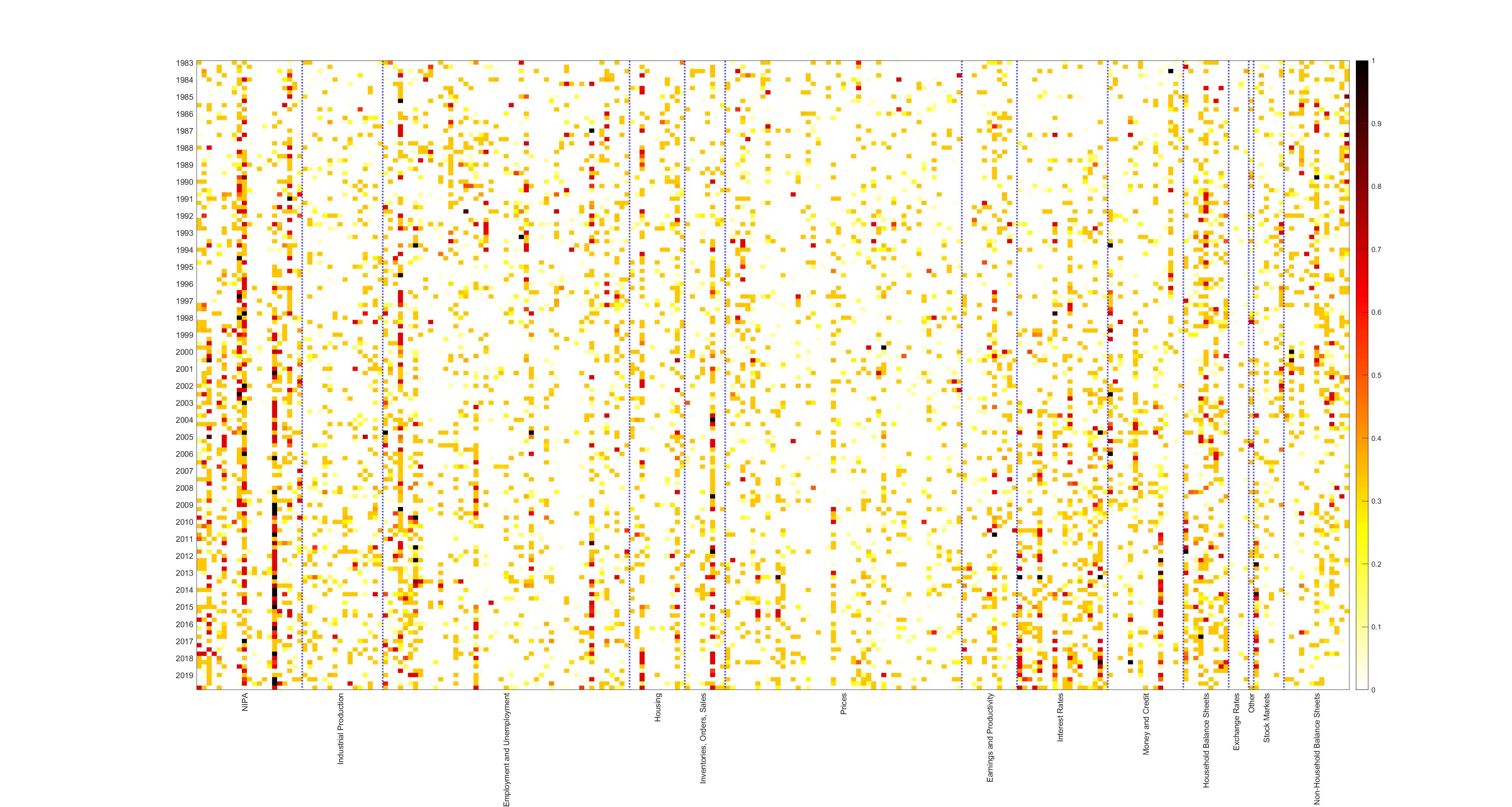}
    \end{subfigure}
    \caption{Average inclusion probability each year for the $LBQR_{BIC}$ (left) and $SSVSBQR$ (right) for their 3 left quantiles (top), 3 central quantiles (middle), and 3 right quantiles (bottom) for h=1}
    \label{fig:timevarying_other}
\end{figure}

While variable inclusion averaged across time revealed that there is support for quantile varying sparsity in the high dimensional GaR, figures (\ref{fig:timevarying}) and (\ref{fig:timevarying_other}) clearly show that variable selection has substantial time variation as well. 
Taking the $HSBQR_{BIC}$ as the benchmark, we can see in figure (\ref{fig:timevarying}) that the time variation in variable selection is in line with what one would expect given the economic history of the US. There are two periods that stand out at one-quarter ahead forecast: the financial crisis of 2008 and the mid to late 80's. Around the financial crisis the left tail increasingly includes employment and financial data such as interest rates and non-household balance sheet information. During the mid-to late 80's, a period which is often referred to as the Volcker period in the macro literature\footnote{When Paul Volcker became chairman of the Federal Reserve Board in 1979, the annual average inflation rate in the United States was 9\%. The Federal Reserves was able to bring inflation down to 4\% by the end of 1983. However, this disinflation came at a cost as during this period, the US experienced two recessions that are attributed to disinflationary monetary policy \citep{goodfriend2005incredible}.}, price variables have high inclusion probability. The middle and right tail share many common variable inclusion patterns over time, where employment and housing data impact the selected models during the 2000's and the financial crisis. 




Comparing to the $LBQR$ and $SSVSBQR$ in figure (\ref{fig:timevarying_other}), the contrasts in variable selection over time are even starker. The $LBQR_{BIC}$ produces very sparse and the $SSVSBQR$ models produce denser models, and the inclusion probabilities for both models are very erratic. Seldom is a variable included with high probability for either priors for more than a quarter which can be seen from the very dotted heatmaps. The $HSBQR$ models on the other hand produce far smoother inclusion patterns which indicates that variables with high inclusion probability impact forecasts for longer periods. While in figure (\ref{fig:timevarying_other}) we present $LBQR_{BIC}$, the same is true for $LBQR_{SAVS}$.

The general finding that there is time variation in the importance of variables for aggregate macroeconomic time-series has already been extensively studied for conditional mean models such as in \citet{koop2012forecasting}, \citet{billio2013time} and \citet{mcalinn2019dynamic}. Our findings motivate that similar methods be also considered for the Bayesian quantile regression moving forward. It is however not surprising that quantile regression models are more sensitive to varying variable importance, even with static parameters. Namely, as the time-series accumulates, crisis periods identify new information for the tails which the tick-loss function will heavily tilt the fitted quantiles toward. This explains the large increase in the importance of variable groups which help predict the 2000 and financial crisis, especially for the left tail and the median. 

\section{Conclusion} \label{sec: 5}
This paper has, in the spirit of \citet{hahn2015decoupling} and \citet{ray2018signal}, proposed to conduct variable selection for continuous shrinkage priors within the Bayesian Quantile Regression by decoupling shrinkage from sparsity as derived from a Bayesian decision theoretic perspective. The resultant easy to implement SAVS procedure for the BQR selects variables on a quantile specific basis where the degree of sparsity is estimated from the data via the quantile BIC in order to allow for heterogeneous levels of penalisation across the conditional distribution. 

Applying the proposed method to both simulated experiments and an empirical application, we have found that the qBIC augmented version, the $BQR_{BIC}$, retains or even improves the fit of the unsparsified posterior, while giving easy to interpret variable selection results.

Simulated experiments have tested a variety of sparsity patterns and error distributions, both for quantile constant and quantile varying coefficient vectors and have found that sparsification often significantly lowers bias, or at worst holds bias stable, while improving variable selection compared to the traditional Bayesian variable selection priors such as the SSVS. The proposed selection procedure via the qBIC outperforms the $BQR_{SAVS}$ especially in sparse DGPs and more generally in the tails where the extra flexibility of determining the penalisation from the data allows to capture the quantile profile. 

The high dimensional growth-at-risk application to the US verifies the findings of the Monte Carlo exercise, namely that the $BQR_{BIC}$ retains density forecast performance and even improves downside risk estimation. Compared to the benchmark 'vulnerable growth' model of \citet{adrian2019vulnerable}, the big data models provide significant gains in density forecasts. In particular, the proposed sparsification techniques applied to the $HSBQR$ provide well calibrated densities with much improved left tail performance. Variable selection results of the proposed methods revealed that there is substantial quantile varying sparsity for US macroeconomic data. As intuition would suggest variable groups which predict crises periods well, such as interest rate and housing data appear with higher inclusion probability in the left tail while right and middle quantiles tend to be affected by many overlapping variable groups.

Contrasting these results to variable selection of more commonly used conditional mean models, we find that mean models tend to select variable groups which have high inclusion probability not only in the median, but also the tails. This highlights the importance of quantile models as it allows us to decompose which variables impact different parts of the distribution. The general finding that there is time variation in the importance of variables for aggregate macroeconomic time-series seems to hold true for quantile models as well. We feel that it is a worthy endeavour to further research methods that allow for quantile as well as time variation for variable selection, to gain a better understanding of which variables are important for the different quantiles across time.

\pagebreak

\bibliographystyle{chicago}
\bibliography{text.bbl}

\pagebreak


\appendix 
\section{Appendix}
\subsection{Expected Loss function for the BQR} \label{sec:Appendix-SAVS-BQR}
In order to decouple shrinkage from sparsity, loss function (\ref{eq:loss_hd}) necessitates integration over the uncertainty in the latent predictions $\tilde{Y}$. This involves two integration steps: 1) over $\tilde{Y}$ conditional on $\theta$, and 2) over $\theta$, conditional on $\tilde{Y}$. The conditional predictive likelihood, based on in-sample values of X, is: 
\begin{equation}
    y_t | \beta, Z, \sigma \sim N(x_t'\beta + \xi z_t, \tau^2z_t\sigma)
\end{equation}
Note that for the sake of readability, the dependence on a given quantile is suppressed. 
  
\subsubsection*{Integration over $\tilde{Y}|\theta$}
\begin{equation}
    \begin{split}
        \mathcal{L}(\tilde{Y},\alpha) & = E_{\tilde{Y}|\bullet}[\tilde{Y},\alpha|\beta,\sigma,Z] \\
        & = \phi||\alpha||_0 + \int T^{-1}||X\alpha-\tilde{Y}||^2_2 \; p(\tilde{Y}|\beta,\sigma,Z)d\tilde{Y} \\
        & = \phi||\alpha||_0 + T^{-1}\alpha'X'X\alpha - T^{-1}2\alpha'X' \int \tilde{Y} \; p(\tilde{Y}|\beta,\sigma,Z)d\tilde{Y} + T^{-1} \int \tilde{Y}'\tilde{Y} \; p(\tilde{Y}|\beta,\sigma,Z)d\tilde{Y} \\
        & = \phi||\alpha||_0 + T^{-1}[\alpha'X'X\alpha - 2 \alpha X'(X\beta+ \xi Z)  + \Sigma_{\tilde{Y}} + ||X\beta+\xi Z||^2_2] \\
        & = \phi||\alpha||_0 + T^{-1}[||X\alpha-X\beta||^2_2 - 2 \alpha X'(X\beta+ \xi Z)  + \Sigma_{\tilde{Y}} + A ]
    \end{split}
\end{equation}
Where A is defined as $\xi^2Z'Z + 2\beta'X'\xi Z$, and the integral of $\tilde{Y}'\tilde{Y}$ follows from:
\begin{equation}
    \int \tilde{Y}'\tilde{Y} \; p(\tilde{Y}|\beta,\sigma,Z) = \sum_{t=1}^T\sigma^2_{\tilde{y}_t} + \sum_{t=1}^T(x_t'\beta + \xi z_t)^2,
\end{equation}
where $\sigma^2_{\tilde{y}_t} = \int (\tilde{y}_t-\tilde{\mu})^2p(\tilde{y_t}|\beta,\sigma,Z)$.

\subsubsection*{Integration over $\beta,\sigma,Z|\bullet$}

\begin{equation}
    \begin{split}
        \mathcal{L}(\alpha) & = E_{\beta,\sigma,Z|\bullet}[\mathcal{L(\beta,\sigma,Z,\alpha)}] \\ 
       & \phi||\alpha||_0 + T^{-1} \int ||X\alpha - X\beta||^2_2 \; p(\beta.\sigma,Z|\tilde{Y})d(\beta,\sigma,Z) \\
       & + T^{-1}\int \Sigma_{\tilde{y}} \; p(\beta.\sigma,Z|\tilde{Y})d(\beta,\sigma,Z) \\
       & + T^{-1} A \; p(\beta.\sigma,Z|\tilde{Y})d(\beta,\sigma,Z) + T{-1} \int \alpha'X'\xi Z \;  p(\beta.\sigma,Z|\tilde{Y})d(\beta,\sigma,Z)
    \end{split}
\end{equation}
Since all the terms involving $\Sigma_{\tilde{y}}$ and A don't involve $\alpha$, they do not affect the minimisation of the loss function and can therefore be dropped. 
 \begin{equation}
     \begin{split}
         & = \phi ||\alpha||_0 + T^{-1}[||X\alpha-X\beta||^2_2 - 2 \alpha'X'X\overline{\beta} + \overline{\beta}'X'X\overline{\beta} + tr(X'X\Sigma_{\overline{\beta}})] \\
        & - T^{-1}\alpha X'\xi\int Z \; p(\beta,\sigma,Z|\tilde{Y}) d (\beta,\sigma,Z) \\
         & = \phi||\alpha||_0 + T^{-1}||X\alpha- X\beta||^2_2 + T^{-1}tr(X'X\Sigma_{\overline{\beta}})-T^{-1}\alpha' X'\xi \overline{Z}.
     \end{split}
 \end{equation}
where $\overline{Z}_p = \Big( \frac{|y_1-x_1'\beta_p|}{\sqrt{\xi^2+2\tau^2}} + \frac{\sigma\tau^2}{\xi^2 + 2\tau^2}, \cdots, \frac{|y_T-x_T'\beta_p|}{\sqrt{\xi^2+2\tau^2}} + \frac{\sigma\tau^2}{\xi^2 + 2\tau^2} \Big)'$. Entries of $\overline{Z}_p$ follow from the definition of the expectation of the inverse Gaussian with location $\mu_t$ and scale $\lambda_t$: since $1/z_t \sim iG(\mu_t,\lambda_t)$, then $z_t \sim iG(\frac{1}{\mu_t +} \frac{1}{\lambda_t}, \frac{1}{\lambda_t\mu_t}+\frac{1}{\lambda_t^2})$ (see for example \cite{khare2012geometric}).

\subsection{Derivation of SAVS-BQR}
As described in section 2.4, instead of directly minimising (\ref{eq:intloss_adl}), we use $\ell_1$-norm as well as adaptive penalisation akin to \citet{zou2006adaptive}:
\begin{equation} \label{eq: BQR_loss}
    \mathcal{L}(\alpha) = \underset{\alpha}{\mathrm{argmin}} \{\frac{1}{2} ||X\overline{\beta}-X\alpha||^2_2 + \sum_{j=1}^K \phi_j |\alpha_j| - \alpha'X'\xi\overline{Z} \}.
\end{equation}

Notice that for the first term, we use the simple fact that $||Xa-Xb|| = ||Xb-Xa||$, and have introduced another factor $\frac{1}{2}$ as well as dropped $T^{-1}$ for notational convenience. We suppressed dependence of the latent quantities on quantile p for the same reason.

While the LARS algorithm (see \citet{friedman2001elements}) can be applied to (\ref{eq: BQR_loss}), we follow \citet{ray2018signal} in making use of the efficient coordinate descent algorithm introduced by \citet{friedman2007pathwise}. The coordinate descent algorithm reduces computational complexity by updating the entry in solution vector $\alpha_j$, conditional on all other $\alpha$, which is iterated until convergence. 

For a given state of the algorithm, $\tilde{\alpha}$, the objective function is recast as a function of the $j^{th}$ variable: 
\begin{equation} \label{eq:coord_single}
    \tilde{\mathcal{L}}(\alpha_j) = \{ \frac{1}{2}||X\beta-X_{-j}\tilde{\alpha}_{-j}-X_j\alpha_j ||^2_2 + \sum_{k\neq j}\phi_k|\alpha| - \tilde{\alpha}_{-j}'X'_{-j}\xi\overline{Z} - \alpha_j'X_j'\xi\overline{Z} \},
\end{equation}
where, likewise, $X_{-j}$ denotes columns in X which are not the $j^{th}$ column. Since the optimisation problem is thus broken down to a single covariate basis, the objective function (\ref{eq:coord_single}) is the classical thresholding problem. Taking first order conditions for any j: 
\begin{equation}
    \begin{split}
     \frac{\partial \tilde{L}_j(\alpha_j)}{\partial \alpha_j} & = X_j'X_j + X_j'(X_{-j}\tilde{\alpha}_{-j}-X\overline{\beta}) + \phi_j s_j + d_j = 0 \\
     \alpha_j & = \frac{1}{X_j'X_j} sign(X'_j\tilde{R}_j - d_j - \phi_j s_j) \\
     & = \frac{1}{X_j' X_j} sign(X_j'\tilde{R}_j-d_j)(|X_j'\tilde{R}_j-d_j|-\phi_j)_{+},
    \end{split}
\end{equation}
where $\tilde{R}_j$ is the partial residual vector between $X\overline{\beta}$ and $X_{-j}\tilde{\alpha}_j$ at a given state and $s_j$ is defined as the subgradient, $s_j \in \partial|\alpha_j|$. Notice that we collect $-X_{-j}\xi \overline{Z}$ into $d_j$. 

Now, since the coordinate descnet algorithm is stopped after the first iteration, as suggested in \citet{ray2018signal}, $R_j = \overline{\beta}_jX_j$, and therefore $X_jR_j = \overline{\beta}_j||X||^2_2$. Hence, by neglecting $d_j$ as suggested in section (3.2), the solution to theorem 1 is achieved. 

\subsection{Sampling Algorithms} \label{sec: sampling steps}
With conditional posteriors presented in section 2.2 at hand, we utilise standard Gibbs samplers to obtain draws from the posterior distributions. Collecting posterior hyperparameters for $\beta$, which will differ for each prior under investigation, in $\Lambda_{*}$, the dynamics of the of the Markov chain $\{\beta^m,\sigma^m,\Lambda_{*}^m,Z^m\}_{m=0}^{\infty}$ are implicitly defined through the following steps

    \begin{enumerate}
        \item Draw $z_t \sim \pi(.|\beta,\sigma,\Lambda_{*},\theta,\tau,Y)$ from $1/iG(\overline{c}_t,\overline{d}_t)$ for all t and stack to a T x 1 vector $Z_{n+1}$
        \item Draw $\sigma_{n+1} \sim \pi(.|\beta,\Lambda_{*},\theta,\tau,Y,Z_{n+1}) $ from $IG (\overline{a},\overline{b})$
        \item Draw $\beta_{n+1} \sim \pi(.|\sigma_{n+1},\Lambda_{*},\theta,\tau,Y,Z_{n+1})$ from $N(\overline{\beta},\overline{\Lambda}_*)$
        \item Draw $\Lambda_{n+1}$ according to each prior in section 2.2
        \item Iterate (1-4) until convergence is achieved
    \end{enumerate}
    
\noindent Note that for the horseshoe prior, we use a slice sampler to sample the elements of $\Lambda_*$, as suggested in \citet{kohns2020horseshoe}.     

\subsubsection{Fast BQR Sampler}
The most costly operation in the Gibbs sampler in (\ref{sec: sampling steps}), is the inversion of the possibly large dimensional posterior covariance of the regression coefficients, step 4. To speed up computation, we make use of the fast BQR sampler proposed in \citet{kohns2020horseshoe}: 

As derived section (\ref{eq:post_beta}), using the scale mixture representation in, the conditional posterior of $\beta$ given all other parameters can be written as:
\begin{equation} \label{proof1}
    \beta | \bullet \sim N(A^{-1}X'\Sigma y, A^{-1}), \quad A = (X'\Sigma X + \Lambda_{*}^{-1})
\end{equation}

Suppose, we want to sample from $N_K(\mu, \Sigma)$, where
\begin{equation} \label{proof2}
    \Sigma = (\Phi'\Phi + D)^{-1}, \quad \mu=\Sigma\Phi'\alpha.    
\end{equation}

Assume $D \in \mathbb{R}^{K \times K}$ is a positive definitive matrix, $\boldsymbol{\phi} \in \mathbb{R}^{T \times K}$, and $\alpha \in \mathbb{R}^{T \times 1}$. Then (\ref{proof1}) is a special case of of (\ref{proof2}) when setting $\Phi=\sqrt{\Sigma}X$, $D=\Lambda_{*}$ and $\alpha=\sqrt{\Sigma}y$. An exact algorithm to sample from (\ref{proof1}) is thus given by:

\newmdtheoremenv{theo}{Algorithm}
\begin{theo}
Fast HS-BQR sampler
\begin{enumerate}
    \item Sample independently $u \sim N(0,D)$ and $\delta \sim N(0,I_T)$
    \item Set $\xi=\Phi u + \delta$
    \item Solve $(\Phi D \Phi' + I_T)w=(\alpha - \xi)$
    \item Set $\theta = u + D\Phi'w$
\end{enumerate}
\end{theo}

\begin{flushleft}

\textbf{Proposition} \textit{Suppose $\theta$ is obtained through algorithm 1. Then $\theta\sim N(\mu, \Sigma)$}. \\
\end{flushleft}
For a proof, we refer to \citet{kohns2020horseshoe}.

\subsubsection{Conditional Mean Models} \label{sec:Appendix-MeanMod}
Assuming a Gaussian likelihood for linear model $y = X\beta + \epsilon$, $\epsilon \sim N(0,\sigma^2I_T)$, the prior for regression coefficients, $\beta$, is assumed to be multivariate norman, and the error variance, $\sigma^2$, inverse-Gamma: 
\begin{equation}
    \begin{split}
        p(\beta|V) & \sim N(0,V)) \\
        p(V) & \sim f \\
        p(\sigma^2) & \sim IG(\underline{a},\underline{b}). \\
    \end{split}
\end{equation}

\noindent The horseshoe, SSVS and lasso priors differ in how the hierarchical prior variance on $\beta$, $p(V)$, is specified. For all priors, we set $\underline{a}=\underline{b}=0.1$.

The conditional posteriors due to conditional conjugacy are: 
\begin{equation}
    \begin{split}
        \beta | \bullet & \sim N(\overline{\beta},\overline{V}) \\ 
        \sigma^2 | \bullet & \sim IG(\overline{a}/2, \overline{b}/2) \\
        \overline{V}|\bullet & \sim g,
    \end{split}
\end{equation}
where $\overline{a} = \underline{a} + T$ and $\overline{b} = \underline{b} + \frac{(y-X\beta)'(y-X\beta)}{T-K} + \overline{\beta}'[V + (X'X)^{-1}]^{-1}\overline{\beta}$. The posterior of the variance hyperparameters and of the regression parameters are prior specific and detailed below. These conditional posteriors are sampled in a standard Gibbs sampler with 5000 burnin draws and 5000 retained MCMC samples. 

\subsubsection*{Horseshoe}
For the horseshoe prior, set $V = \sigma^2 \nu^2 diag(\lambda^2_1, \cdots, \lambda^2_K)$, where
\begin{equation}
\begin{split}
    \nu & \sim C_+(0,1) \\
    \lambda_j & \sim C_+(0,1) \forall j.
\end{split}
\end{equation}

\noindent Then, by standard calculations (see \citet{bhattacharya2016fast})):

\begin{equation}
    \begin{split}
        \overline{\beta} & = \overline{V}^{-1}X'y, \\
        \overline{V} & = \sigma^2(X'X + V^{-1}) \\ 
    \end{split}
\end{equation}

\noindent Instead of computing the large dimensional inverse $A^{-1}$, we rely on a data augmentation technique introduced by \citet{bhattacharya2016fast}.

\subsubsection*{SSVS} 
For the SSVS prior, we use the same hierarchy as presented in (\ref{eq:ssvs_prior}), which is printed below for convenience:
\begin{equation} \label{eq:mean_ssvs_prior}
    \begin{split}
        \beta_{j,p} | \gamma_j,\lambda_j & \sim (1-\gamma_j)N(0,c\lambda_j^2) + \gamma_jN(0,\lambda_j^2) \forall j \in \{1,\cdots,K\} \\
        \lambda_j^2 & \sim IG(a_2,b_2) \\
        \gamma_j|\pi_0 & \sim Bern(\pi_0) \\
        \pi_0 & \sim B(a_3,b_3), 
    \end{split}
\end{equation}
where $V = diag(\lambda_1^2,\cdots,\lambda_K^2)$ for all j if $\gamma_j = 1$, and $diag(c\lambda_1^2,\cdots,c\lambda_k^2)$ otherwise. The conditional posteriors are standard and derived for example in \citet{george1993variable} and \citet{ishwaran2005spike}. The difference to the prior of \citet{george1993variable} lies in the additional prior on $\delta^2_j$ which is assumed to be inverse gamma. It can be shown that this implies a mixture of student-t distributions for $\beta_j$ marginally (Konrath et al., 2008):
\begin{equation}
    \begin{split}
        \overline{\beta} & = \overline{V}^{-1}X'y/\sigma^2 \\
        \overline{V} & = (X'X/\sigma^2 + V^{-1})^{-1}, \; V = diag(\lambda_j\gamma_j) \\
        \gamma_j | \bullet & \sim N(1-\pi_0)N(\beta_j|0,c\times\lambda^2_j)I_{\gamma_j = 0} + \pi_0N(\beta_j|0,\lambda_j^2)I_{\gamma_j=0} \\
        \pi_0 | \bullet & \sim Beta(a_3 + n_1, b_3 + K - n_1), \text{where} \; n_1 = \sum_jI_{\gamma_j=1}
    \end{split}
\end{equation}
To be agnostic about the degree of sparsity, we set $a_3=b_3=0$ as for the quantile SSVS model.

\subsubsection*{Lasso}
Similar to the SSVS, the lasso prior is the same as in (\ref{eq:b_lasso}):

\begin{equation} \label{eq:mean_b_lasso}
    \begin{split}
        \beta_j|\phi & \sim N(0,\lambda_j), \\
        \lambda_j | \phi & \sim exp(\frac{\phi}{2}) \\
        \phi & \sim G(a_1,b_1)
    \end{split}
\end{equation}

\noindent where $\Lambda_* = diag(\lambda_1,\cdots,\lambda_K)$. The conditional posteriors for the hyperparameters are standard:

\begin{equation} \label{eq:mean_post_lasso}
    \begin{split}
        p(\lambda^{-1}_j|\bullet) & \sim iG(\sqrt{\frac{\phi}{\beta_{j,p}^2}},\phi) \\
        p(\phi|\bullet) & \sim G(K+a_1, \frac{1}{2}\sum_{j=1}^K\lambda_j + b_1) \\
        \overline{\beta} & = \overline{V}^{-1}X'y/\sigma^2 \\
        \overline{V} & = (X'X/\sigma^2 + V^{-1})^{-1}, \; V = diag(\lambda_1, \cdots, \lambda_K) \\ 
    \end{split}
\end{equation}



\pagebreak

\subsection{Tables}

\begin{table}[h]
\centering
\resizebox{\textwidth}{!}{%
\begin{tabular}{lcccc|cccc}
               & MSFE     & LPDS      & CRPS     & qwCRPS  & MSFE     & LPDS     & CRPS     & qwCRPS \\ \hline \hline
               & \multicolumn{4}{c|}{h=1}                  & \multicolumn{4}{c}{h=2}                 \\
$ABG_{BQR}$    & 0.504    & -11.607   & 0.313    & 1.049   & 0.404    & -7.870   & 0.254    & 0.891  \\
$ABG_{BQR-Skt}$& 0.519    & -1.727    & 0.301    & 1.026   & 0.426    & -4.102   & 0.277    & 0.970  \\
$SSVSBQR$      & 0.539*   & -0.889**  & 0.320*** & 1.061   & 0.435**  & -0.607   & 0.221*** & 0.746**\\
$HSBQR$        & 0.542**  & -3.695    & 0.313*** & 0.952*  & 0.473**  & -5.101   & 0.236*** & 0.776**\\
$HSBQR_{BIC}$  & 0.522    & -3.020    & 0.300*** & 0.918** & 0.428**  & -5.775*  & 0.239*** & 0.779***\\
$HSBQR_{SAVS}$ & 0.537    & -2.838    & 0.304*** & 0.925*  & 0.416    & -3.959   & 0.259*** & 0.847  \\
$LBQR$         & 0.570    & -1.313    & 0.289*** & 0.949   & 0.480    & -0.760*  & 0.229*** & 0.748***\\
$LBQR_{BIC}$   & 0.544    & -1.316    & 0.289*** & 0.949   & 0.450    & -0.776*  & 0.230*** & 0.750***\\
$LBQR_{SAVS}$  & 0.544    & -2.021    & 0.301**  & 0.999   & 0.449    & -1.219*  & 0.245*** & 0.808  \\ \hline
               & \multicolumn{4}{c|}{h=3}                  & \multicolumn{4}{c}{h=4}                 \\
$ABG_{BQR}$    & 0.378    & -6.825    & 0.237    & 0.847   & 0.351    & -4.387   & 0.235    & 0.844  \\
$ABG_{BQR-Skt}$& 0.397    & -1.006    & 0.249    & 0.857   & 0.367    & -1.252   & 0.251    & 0.875  \\
$SSVSBQR$      & 0.431**  & -0.542    & 0.211*** & 0.723   & 0.392*   & -0.657*  & 0.214*** & 0.710  \\
$HSBQR$        & 0.437**  & -5.224    & 0.231*** & 0.777   & 0.392**  & -5.258*  & 0.224*** & 0.731  \\
$HSBQR_{BIC}$  & 0.400*   & -5.958**  & 0.230*** & 0.775   & 0.377*   & -4.815** & 0.220*** & 0.723  \\
$HSBQR_{SAVS}$ & 0.395    & -4.714    & 0.240*** & 0.810   & 0.380    & -3.568   & 0.216*** & 0.729  \\
$LBQR$         & 0.431*   & -0.620    & 0.222*** & 0.743** & 0.413*   & -0.667   & 0.213*** & 0.716**\\
$LBQR_{BIC}$   & 0.422*   & -0.630    & 0.223*** & 0.745** & 0.387*   & -0.675   & 0.214*** & 0.718**\\
$LBQR_{SAVS}$  & 0.421*   & -0.964    & 0.233*** & 0.797   & 0.386*   & -1.557   & 0.234*** & 0.804  \\ \hline
\end{tabular}
}
\caption{Forecast Evaluation Results (Skewed-t distribution)}
\label{tab:skew-t Forecast eval}
\end{table}

\begin{table}[]
\centering
\resizebox{\textwidth}{!}{%
\begin{tabular}{llccccc|ccccc|ccccc|ccccc}
\hline
            &                      & 0.05  & 0.25  & 0.5   & 0.75  & 0.95  & 0.05  & 0.25  & 0.5   & 0.75  & 0.95  & 0.05  & 0.25  & 0.5   & 0.75  & 0.95  & 0.05  & 0.25  & 0.5   & 0.75  & 0.95  \\ \hline \hline
            &                      & \multicolumn{5}{c}{$y_1$|}                & \multicolumn{5}{c}{$y_2$|}                & \multicolumn{5}{c}{$y_3$|}                & \multicolumn{5}{c}{$y_4$}                \\ \hline
            &                      & \multicolumn{20}{c}{Coefficient Bias}                                                                                                                               \\ \hline
\multicolumn{2}{l}{\textbf{T=100}} &       &       &       &       &       &       &       &       &       &       &       &       &       &       &       &       &       &       &       &       \\
\multicolumn{2}{l}{Sparse}         &       &       &       &       &       &       &       &       &       &       &       &       &       &       &       &       &       &       &       &       \\
            & $HSBQRZ$               & 0.121 & 0.092 & 0.086 & 0.098 & 0.126 & 0.172 & 0.099 & 0.088 & 0.097 & 0.176 & 0.156 & 0.098 & 0.077 & 0.094 & 0.148 & 0.128 & 0.095 & 0.079 & 0.094 & 0.127 \\
            & $HSBQRZ_{SAVS}$           & 0.110 & 0.072 & 0.064 & 0.081 & 0.116 & 0.165 & 0.081 & 0.066 & 0.080 & 0.169 & 0.146 & 0.089 & 0.062 & 0.086 & 0.138 & 0.119 & 0.085 & 0.063 & 0.085 & 0.119 \\
            & $HSBQRZ_{BIC}$            & 0.103 & 0.062 & 0.058 & 0.072 & 0.107 & 0.159 & 0.067 & 0.056 & 0.072 & 0.163 & 0.141 & 0.087 & 0.060 & 0.085 & 0.132 & 0.117 & 0.084 & 0.061 & 0.084 & 0.116 \\
\multicolumn{2}{l}{Block}          &       &       &       &       &       &       &       &       &       &       &       &       &       &       &       &       &       &       &       &       \\
            & $HSBQRZ$               & 0.250 & 0.247 & 0.252 & 0.254 & 0.244 & 0.260 & 0.240 & 0.244 & 0.250 & 0.272 & 0.289 & 0.293 & 0.291 & 0.285 & 0.277 & 0.257 & 0.266 & 0.268 & 0.263 & 0.256 \\
            & $HSBQRZ_{SAVS}$           & 0.252 & 0.242 & 0.246 & 0.248 & 0.243 & 0.262 & 0.236 & 0.239 & 0.245 & 0.272 & 0.290 & 0.287 & 0.283 & 0.279 & 0.276 & 0.260 & 0.262 & 0.262 & 0.258 & 0.256 \\
            & $HSBQRZ_{BIC}$            & 0.259 & 0.246 & 0.247 & 0.247 & 0.247 & 0.271 & 0.241 & 0.244 & 0.248 & 0.276 & 0.292 & 0.286 & 0.283 & 0.278 & 0.277 & 0.262 & 0.262 & 0.262 & 0.258 & 0.258 \\
\multicolumn{2}{l}{\textbf{T=500}} &       &       &       &       &       &       &       &       &       &       &       &       &       &       &       &       &       &       &       &       \\
\multicolumn{2}{l}{Sparse}         &       &       &       &       &       &       &       &       &       &       &       &       &       &       &       &       &       &       &       &       \\
            & $HSBQRZ$               & 0.099 & 0.073 & 0.062 & 0.073 & 0.103 & 0.152 & 0.074 & 0.061 & 0.075 & 0.152 & 0.140 & 0.066 & 0.048 & 0.067 & 0.135 & 0.113 & 0.065 & 0.047 & 0.066 & 0.111 \\
            & $HSBQRZ_{SAVS}$           & 0.086 & 0.053 & 0.038 & 0.054 & 0.089 & 0.145 & 0.057 & 0.037 & 0.059 & 0.144 & 0.127 & 0.058 & 0.026 & 0.058 & 0.123 & 0.100 & 0.056 & 0.026 & 0.057 & 0.098 \\
            & $HSBQRZ_{BIC}$            & 0.076 & 0.045 & 0.029 & 0.047 & 0.077 & 0.140 & 0.050 & 0.030 & 0.054 & 0.137 & 0.106 & 0.061 & 0.025 & 0.062 & 0.106 & 0.084 & 0.058 & 0.024 & 0.059 & 0.085 \\
\multicolumn{2}{l}{Block}          &       &       &       &       &       &       &       &       &       &       &       &       &       &       &       &       &       &       &       &       \\
            & $HSBQRZ$               & 0.143 & 0.101 & 0.094 & 0.103 & 0.144 & 0.174 & 0.104 & 0.091 & 0.103 & 0.176 & 0.187 & 0.098 & 0.082 & 0.096 & 0.184 & 0.156 & 0.101 & 0.085 & 0.099 & 0.154 \\
            & $HSBQRZ_{SAVS}$           & 0.139 & 0.089 & 0.081 & 0.092 & 0.138 & 0.173 & 0.093 & 0.077 & 0.092 & 0.171 & 0.185 & 0.091 & 0.072 & 0.087 & 0.181 & 0.152 & 0.092 & 0.074 & 0.091 & 0.150 \\
            & $HSBQRZ_{BIC}$            & 0.146 & 0.096 & 0.085 & 0.099 & 0.151 & 0.180 & 0.099 & 0.080 & 0.097 & 0.183 & 0.190 & 0.091 & 0.072 & 0.088 & 0.186 & 0.156 & 0.092 & 0.074 & 0.091 & 0.154 \\ \hline
            &                      & \multicolumn{20}{c}{MCC}                                                                                                                                      \\ \hline
\multicolumn{2}{l}{\textbf{T=100}} &       &       &       &       &       &       &       &       &       &       &       &       &       &       &       &       &       &       &       &       \\
\multicolumn{2}{l}{Sparse}         &       &       &       &       &       &       &       &       &       &       &       &       &       &       &       &       &       &       &       &       \\
            & $HSBQRZ_{SAVS}$           & 0.392 & 0.535 & 0.555 & 0.548 & 0.509 & 0.383 & 0.516 & 0.552 & 0.556 & 0.504 & 0.338 & 0.446 & 0.446 & 0.430 & 0.324 & 0.382 & 0.453 & 0.433 & 0.424 & 0.365 \\
            & $HSBQRZ_{BIC}$            & 0.555 & 0.725 & 0.780 & 0.772 & 0.713 & 0.561 & 0.722 & 0.793 & 0.791 & 0.707 & 0.423 & 0.496 & 0.540 & 0.464 & 0.405 & 0.446 & 0.485 & 0.523 & 0.448 & 0.426 \\
\multicolumn{2}{l}{Block}          &       &       &       &       &       &       &       &       &       &       &       &       &       &       &       &       &       &       &       &       \\
            & $HSBQRZ_{SAVS}$           & 0.386 & 0.412 & 0.420 & 0.411 & 0.393 & 0.405 & 0.429 & 0.439 & 0.421 & 0.398 & 0.336 & 0.342 & 0.344 & 0.351 & 0.345 & 0.384 & 0.392 & 0.386 & 0.394 & 0.382 \\
            & $HSBQRZ_{BIC}$            & 0.420 & 0.445 & 0.457 & 0.452 & 0.435 & 0.435 & 0.457 & 0.471 & 0.462 & 0.441 & 0.372 & 0.371 & 0.350 & 0.376 & 0.383 & 0.413 & 0.405 & 0.372 & 0.405 & 0.411 \\
\multicolumn{2}{l}{\textbf{T=500}} &       &       &       &       &       &       &       &       &       &       &       &       &       &       &       &       &       &       &       &       \\
\multicolumn{2}{l}{Sparse}         &       &       &       &       &       &       &       &       &       &       &       &       &       &       &       &       &       &       &       &       \\
            & $HSBQRZ_{SAVS}$           & 0.397 & 0.540 & 0.567 & 0.543 & 0.462 & 0.389 & 0.568 & 0.576 & 0.574 & 0.471 & 0.362 & 0.679 & 0.641 & 0.653 & 0.408 & 0.398 & 0.668 & 0.648 & 0.654 & 0.434 \\
            & $HSBQRZ_{BIC}$            & 0.681 & 0.898 & 0.904 & 0.898 & 0.840 & 0.702 & 0.905 & 0.906 & 0.903 & 0.850 & 0.659 & 0.783 & 0.836 & 0.792 & 0.665 & 0.709 & 0.785 & 0.827 & 0.785 & 0.697 \\
\multicolumn{2}{l}{Block}          &       &       &       &       &       &       &       &       &       &       &       &       &       &       &       &       &       &       &       &       \\
            & $HSBQRZ_{SAVS}$           & 0.691 & 0.804 & 0.810 & 0.807 & 0.691 & 0.707 & 0.809 & 0.820 & 0.815 & 0.710 & 0.613 & 0.837 & 0.865 & 0.850 & 0.617 & 0.676 & 0.821 & 0.855 & 0.844 & 0.677 \\
            & $HSBQRZ_{BIC}$            & 0.779 & 0.946 & 0.947 & 0.943 & 0.828 & 0.781 & 0.946 & 0.951 & 0.952 & 0.839 & 0.714 & 0.889 & 0.907 & 0.899 & 0.717 & 0.767 & 0.861 & 0.883 & 0.875 & 0.765 \\ \hline
            &                      & \multicolumn{20}{c}{Hit rate}                                                                                                                                 \\ \hline
\multicolumn{2}{l}{\textbf{T=100}} &       &       &       &       &       &       &       &       &       &       &       &       &       &       &       &       &       &       &       &       \\
\multicolumn{2}{l}{Sparse}         &       &       &       &       &       &       &       &       &       &       &       &       &       &       &       &       &       &       &       &       \\
            & $HSBQRZ_{SAVS}$           & 0.578 & 0.817 & 0.842 & 0.835 & 0.768 & 0.567 & 0.790 & 0.828 & 0.822 & 0.762 & 0.538 & 0.608 & 0.612 & 0.565 & 0.511 & 0.540 & 0.618 & 0.608 & 0.571 & 0.518 \\
            & $HSBQRZ_{BIC}$            & 0.481 & 0.640 & 0.707 & 0.703 & 0.690 & 0.495 & 0.642 & 0.728 & 0.724 & 0.684 & 0.479 & 0.546 & 0.484 & 0.507 & 0.460 & 0.501 & 0.586 & 0.509 & 0.536 & 0.481 \\
\multicolumn{2}{l}{Block}          &       &       &       &       &       &       &       &       &       &       &       &       &       &       &       &       &       &       &       &       \\
            & $HSBQRZ_{SAVS}$           & 0.554 & 0.652 & 0.685 & 0.671 & 0.583 & 0.570 & 0.665 & 0.691 & 0.672 & 0.583 & 0.549 & 0.631 & 0.650 & 0.639 & 0.556 & 0.566 & 0.649 & 0.667 & 0.655 & 0.567 \\
            & $HSBQRZ_{BIC}$            & 0.358 & 0.414 & 0.443 & 0.439 & 0.388 & 0.368 & 0.423 & 0.444 & 0.436 & 0.388 & 0.454 & 0.570 & 0.632 & 0.581 & 0.461 & 0.499 & 0.624 & 0.686 & 0.632 & 0.498 \\
\multicolumn{2}{l}{\textbf{T=500}}      &       &       &       &       &       &       &       &       &       &       &       &       &       &       &       &       &       &       &       &       \\
\multicolumn{2}{l}{Sparse}         &       &       &       &       &       &       &       &       &       &       &       &       &       &       &       &       &       &       &       &       \\
            & $HSBQRZ_{SAVS}$           & 0.769 & 0.956 & 0.966 & 0.951 & 0.924 & 0.745 & 0.950 & 0.957 & 0.963 & 0.916 & 0.818 & 0.935 & 0.949 & 0.920 & 0.830 & 0.834 & 0.940 & 0.956 & 0.927 & 0.845 \\
            & $HSBQRZ_{BIC}$            & 0.651 & 0.831 & 0.853 & 0.843 & 0.763 & 0.632 & 0.846 & 0.862 & 0.848 & 0.786 & 0.656 & 0.889 & 0.887 & 0.868 & 0.653 & 0.658 & 0.905 & 0.926 & 0.891 & 0.655 \\
\multicolumn{2}{l}{Block}          &       &       &       &       &       &       &       &       &       &       &       &       &       &       &       &       &       &       &       &       \\
            & $HSBQRZ_{SAVS}$           & 0.963 & 0.999 & 0.998 & 0.998 & 0.969 & 0.966 & 0.998 & 1.000 & 0.998 & 0.972 & 0.932 & 0.986 & 0.995 & 0.989 & 0.929 & 0.954 & 0.988 & 0.995 & 0.990 & 0.953 \\
            & $HSBQRZ_{BIC}$            & 0.905 & 0.982 & 0.989 & 0.980 & 0.875 & 0.918 & 0.983 & 0.994 & 0.985 & 0.876 & 0.852 & 0.982 & 0.993 & 0.985 & 0.854 & 0.915 & 0.986 & 0.994 & 0.988 & 0.913 \\ \hline
\end{tabular}%
}
\caption{Nuisance Parameter correction results for the HSBQR}
\label{tab:HSBQRZ}
\end{table}


\end{document}